\documentclass{jfm}
\usepackage{graphicx}
\usepackage{epstopdf, epsfig}
\usepackage{natbib}
\usepackage{amsmath, amsfonts, amsbsy, amssymb, array, float,psfrag,subfigure}
\usepackage{color}

\providecommand\bnabla{\boldsymbol{\nabla}}
\providecommand\bcdot{\boldsymbol{\cdot}}
\definecolor{dgreen}{rgb}{0.0,0.545,0.0}

\providecommand\eg{e.g.}

\providecommand{\bs}{\boldsymbol}

\providecommand{\EQ}{\begin{equation}}
\providecommand{\EN}{\end{equation}}
\providecommand{\EQA}{\begin{eqnarray}}
\providecommand{\ENA}{\end{eqnarray}}

\providecommand\bomega{\boldsymbol{\omega}}
\providecommand{\<}{\langle}
\renewcommand{\>}{\rangle}

\shorttitle{Evolution of material surfaces in transitional channel flow}
\shortauthor{Y. Zhao, Y. Yang and S. Chen}

\title{Evolution of material surfaces in the temporal transition in channel flow}

\author{Yaomin Zhao\aff{1},
  Yue Yang\aff{1,2}
  \corresp{\email{yyg@pku.edu.cn}},
 \and Shiyi Chen\aff{3,1,2}}

\affiliation{\aff{1}State Key Laboratory for Turbulence and Complex Systems, College of Engineering,\break Peking University, Beijing 100871, China
\aff{2}Center for Applied Physics and Technology, Peking University, Beijing 100871, China
\aff{3}Department of Mechanics and Aerospace Engineering, South University of Science and Technology of China, Shenzhen 518055, China}

\begin{document}

\maketitle

\begin{abstract}
We report a Lagrangian study on the evolution of material surfaces in the Klebanoff-type temporal transitional channel flow. Based on the Eulerian velocity field from the direct numerical simulation,
a backward-particle-tracking method is applied to solve the transport equation of the Lagrangian scalar field, and then the iso-surfaces of the Lagrangian field can be extracted as material surfaces in the evolution. Three critical issues for Lagrangian investigations on the evolution of coherent structures using material surfaces are addressed. First, the initial scalar field is uniquely determined based on proposed criteria, so that the initial material surfaces can be approximated as vortex surfaces,
and keep invariant in the initial laminar state. Second, the evolution of typical material surfaces initially from different wall distances is presented, and then the influential material surface with the maximum deformation is identified. Large vorticity variations with the maximum curvature growth of vortex lines are also observed on this surface. Moreover, crucial events in the transition can be characterized in a Lagrangian approach by conditional statistics on the material surfaces. Finally, the influential material surface, which is initially a vortex surface, is demonstrated as a surrogate of the vortex surface before significant topological changes of vortical structures. Therefore, this material surface can be used to elucidate the continuous temporal evolution of vortical structures in transitional wall-bounded flows in a Lagrangian perspective. The evolution of the influential material surface is divided into three stages: the formation of a triangular bulge from an initially disturbed streamwise-spanwise sheet, rolling up of the vortex sheet near the bulge ridges with the vorticity intensification, and the generation and evolution of signature hairpin-like structures with self-induced dynamics of vortex filaments.

\end{abstract}

\begin{keywords}
Boundary layer structure, Turbulent boundary layers, Turbulent transition.
\end{keywords}


\section{Introduction}
The laminar-turbulent transition has been one of the most fundamental and challenging problems in turbulence research for decades. In wall-bounded shear flows, there are two different classes of transitions, the {classical} transition and the bypass transition  {\citep[]{Morkovin1984}}. In the  {classical} transition, the Klebanoff-type (referred to as K-type below) transition \citep[]{Klebanoff1962} has been extensively studied  {\citep[see e.g.][]{Rist1995,Borodulin2002a,Bake2002}}. Other types of transitions, {such as the New-type \citep[]{Kachanov1977,Kachanov1994} or Herbert-type \citep[]{Herbert1984}} transition, and the oblique transition \citep[]{Fasel1993}, show different spatial arrangements of the similar elementary vortical structures of the K-type transition \citep[]{Bake2000,Borodulin2006}.
{In the present study, we focus on the K-type transition owing to its fundamental and practical importance}.

In the literature, the whole transition process is often divided into different conceptual stages  {\citep[see e.g.][]{Herbert1988,Kachanov1994}: (1) receptivity; (2) linear instability; (3) weakly-nonlinear instability;
(4) late or essentially nonlinear stage; (5) final laminar flow breakdown and transition to turbulence \citep[]{Borodulin2007,Guo2010}. 
In the receptivity stage, the instability modes are excited by various external perturbations.}
In the  {linear instability} stage, the amplitudes of the initial disturbances are very small compared to the mean flow velocity, so the flow can be described by the linear perturbation theory with the Orr-Sommerfeld equation.
 {After the amplification of the instability modes in the linear stage, the flow enters the weakly nonlinear stage, in which the insability modes start to interact with each other \citep[see \eg][]{Borodulin2002b}.
The growth of the spanwise modulation of disturbances resulting in the `peak-valley spilliting' (see \citet[]{Klebanoff1962} and \citet[]{Gilbert1990}) can be interpreted by the secondary instability theory \citep[]{Herbert1988}.}

{Compared with the earlier stages, the late nonlinear stages are less investigated.
In the late stages, the transition is mainly characterized by the intensive vortical structures.}
The high-shear layer is induced by the so called `$\Lambda$-shaped' vortical structures with the three-dimensional characteristics {\citep[]{Borodulin2002a,Guo2010}}, and then the hairpin-like structures  {or ring-like vortices} are generated.
In this process, the `spike' signal, indicating the streamwise velocity dramatically decreasing at some point in the flow,
is considered as a result of the velocity induced by the hairpin-like structure  {\citep[see][]{Borodulin1988, Rist1995,Borodulin2002a}}.
 {Subsequently, secondary hairpin-like structures are generated at the upstream of the primary one \citep[]{Borodulin2002a}. The self-similar dynamical procedure is also described as the formation of structure packets \citep[]{Zhou1999, Adrian2007}.}
 {After the non-linear interaction of the vortical structures, the flow breaks down and finally evolves into the fully developed turbulent state.}

Previous studies about the K-type transition suggest that the formation and evolution of coherent structures play an important role in flow dynamics,
and various identification methods have been developed to investigate the coherent structures.
Flow visualization techniques such as smoke and hydrogen bubbles were employed in experiments to identify coherent flow structures in fully developed and transitional turbulent boundary layer flows  {\citep[\eg][]{Head1981,Lee2008,Guo2010}}.
The identified flow structures can be tracked in a Lagrangian sense to show the structural evolution.
Nevertheless, these methods face difficulties of diffusion or following property of the tracers,
and they are not easy to be used for quantified analysis \citep[see][]{Robinson1991}.

Based on the high resolution velocity fields provided by numerical simulations and the particle image velocimetry (PIV) technique,
several Eulerian vortex identification methods have been introduced to study coherent structures in turbulence: (1) vortex lines and iso-surface of the vorticity magnitude \citep[\eg][]{Moin1985,Kim1987}; (2) criteria based on the Eulerian velocity gradient tensor $\bnabla\boldsymbol{u}$, e.g., the $Q$-criterion \citep[]{Hunt1988},
the $\Delta$-criterion \citep[]{Chong1990}, the $\lambda_2$-criterion \citep[]{Jeong1995}, and the $\lambda_{ci}$-criterion \citep[]{Zhou1999}.
It is noted that a weakness of the Eulerian identification methods is on the tracking of a uniquely defined structure.
Since the Eulerian methods are based on instantaneous velocity fields, the mapping in the evolution of vortical structures among different times has to be determined by heuristic algorithms  {from time-resolved experimental \citep[]{Lehew2013} and direct numerical simulation (DNS) data \citep[]{Duran2014}}.
Additionally, the iso-contour level of the iso-surface extracted from the vortex identification criteria is often subjectively selected, so it has to be varied for different times in visualizations.

Compared with the Eulerian methods, the Lagrangian structure identification methods appear to be more natural to describe the continuous evolution of coherent structures, and a few of Lagrangian-type identification methods were proposed. \citet[]{Haller2001} used the ridge of the direct Lyapunov exponent (DLE) to define hyperbolic attracting `Lagrangian coherent structures' (LCS) in three-dimensional flow.
This method was applied to study the evolutionary geometry of an isolated hairpin-like vortical structure in fully developed channel turbulence \citep[]{Green2007}.
Recently, another type of LCS, the elliptic LCS was developed to identify the boundary of vortices and was applied in two-dimensional flows and simple three-dimensional flow \citep[see][]{Blazevski2014,Haller2015}, but it has not been used in transitional wall flows.

 {In the studies of the coherent structures, there are some controversies on the existence and importance of the $\Lambda$-like and hairpin-like vortical structures.
For transitional flows, on one hand, it has been discussed extensively \citep[see \eg][]{Herbert1988,Kleiser1991,Kachanov1994} that the `hairpin-like vortices' evolving from transitional `$\Lambda$-shaped vortices' can be considered as the dominant structures in the late transitional stages \citep[\eg][]{Borodulin2002a,Wu2009,Schlatter2014}, though the mechanism in their evolution processes is still not completely clear.
On the other hand, alternative views of the vortical structures such as `vortex furrows' rather than `hairpin vortices' in the late stage in the K-type transition is described in \citet[]{Bernard2011,Bernard2013} by tracking an ensemble of vortex filaments in a Lagrangian sense.
Hence, it is worth to explore the essence of the coherent structures and their dynamical evolution process in transitional flows.} 
More importantly, it is desired to elucidate the  {late} transition by characterizing the dynamical evolution of a small number of representative structures \citep[\eg][]{Pope2000}.
The major obstacle, however, still exists owing to the lack of methods to describe the dynamical evolution of uniquely defined flow structures.

Based on the tracking of material surfaces, the Lagrangian scalar field, whose iso-surface is a material surface, and the multi-scale geometric analysis have been successfully applied in isotropic turbulence \cite[]{yang2010multi}, simple transitional flows
such as Taylor--Green and Kida--Pelz flows \citep[]{yang2010lagrangian}, and fully developed channel flows \citep[]{yang2011geometric}. This method can present the temporal evolution of material surfaces with quantitative multi-scale and multi-directional statistical geometry.

In the present study, this Lagrangian method is extended to the laminar-turbulent transition to investigate the evolution of material surfaces in transitional flow.
If a material surface is defined at the initial time, then it can be uniquely defined and tracked as a Lagrangian object in the evolution,
and quantitative statistics can be obtained and analyzed on the evolving material surface.

On the other hand, there are several critical issues that should be addressed for the study of  structural evolution using material surfaces in transitional flows.
First, the initial Lagrangian scalar field in isotropic turbulence can be arbitrary \citep{yang2010multi}, but it should be determined uniquely via a rational approach in transitional wall flow owing to the preferred direction of the mean velocity and shear.
Second, given an initial Lagrangian field,  {although an infinite number of material surfaces can be tracked, only a very small number of the material surfaces that have distinguished geometric changes or dynamics in the evolution should be identified and characterized.} These surfaces may be referred as `influential material surfaces' \citep{Haller2015}.
Finally, the evolution of the passive material surfaces only provide kinematics of structures, but it is possible to show essential vortex dynamics for physically interesting initial conditions and within a finite time period \citep[see][]{yang2010lagrangian}.
Thus the connection between the evolutionary geometry of material surfaces and the vortex dynamics needs to be clarified.

By tracking the identified influential material surface, the relation between the Lagrangian statistics on the surfaces and Eulerian flow statistics will be investigated.
Furthermore, we will characterize crucial events in the transition, such as ejections/sweeps and the formation of hairpin-like structures, using the evolution of material surfaces.
In this approach, we can provide a Lagrangian perspective to elucidate the flow dynamics in the transition.

The outline of this paper is as follows. The overview of numerical simulations with the backward-particle-tracking method for the Lagrangian scalar field is illustrated in \S\,\ref{sec:2}, with discussions on the criteria to determine the initial scalar field.
The qualitative visualization for the evolution of material surfaces and Lagrangian fields during the  {late transition} stages is shown in \S\,\ref{sec:3}.
In \S\,\ref{sec_influential}, the influential material surface which plays an important role in evolution is identified,
and the relation between Lagrangian statistics on surfaces and Eulerian statistics is also discussed.
Then the important stages and corresponding dynamical mechanisms in the evolution of the influential material surface are elucidated in \S\,\ref{sec:5}.
Some conclusions are drawn in \S\,\ref{sec:conclusion}.

\section{Simulation overview}\label{sec:2}

\subsection{Direct numerical simulation of temporal transitional channel flow}

\begin{figure}
\begin{center}
 \includegraphics[width=0.7\textwidth]{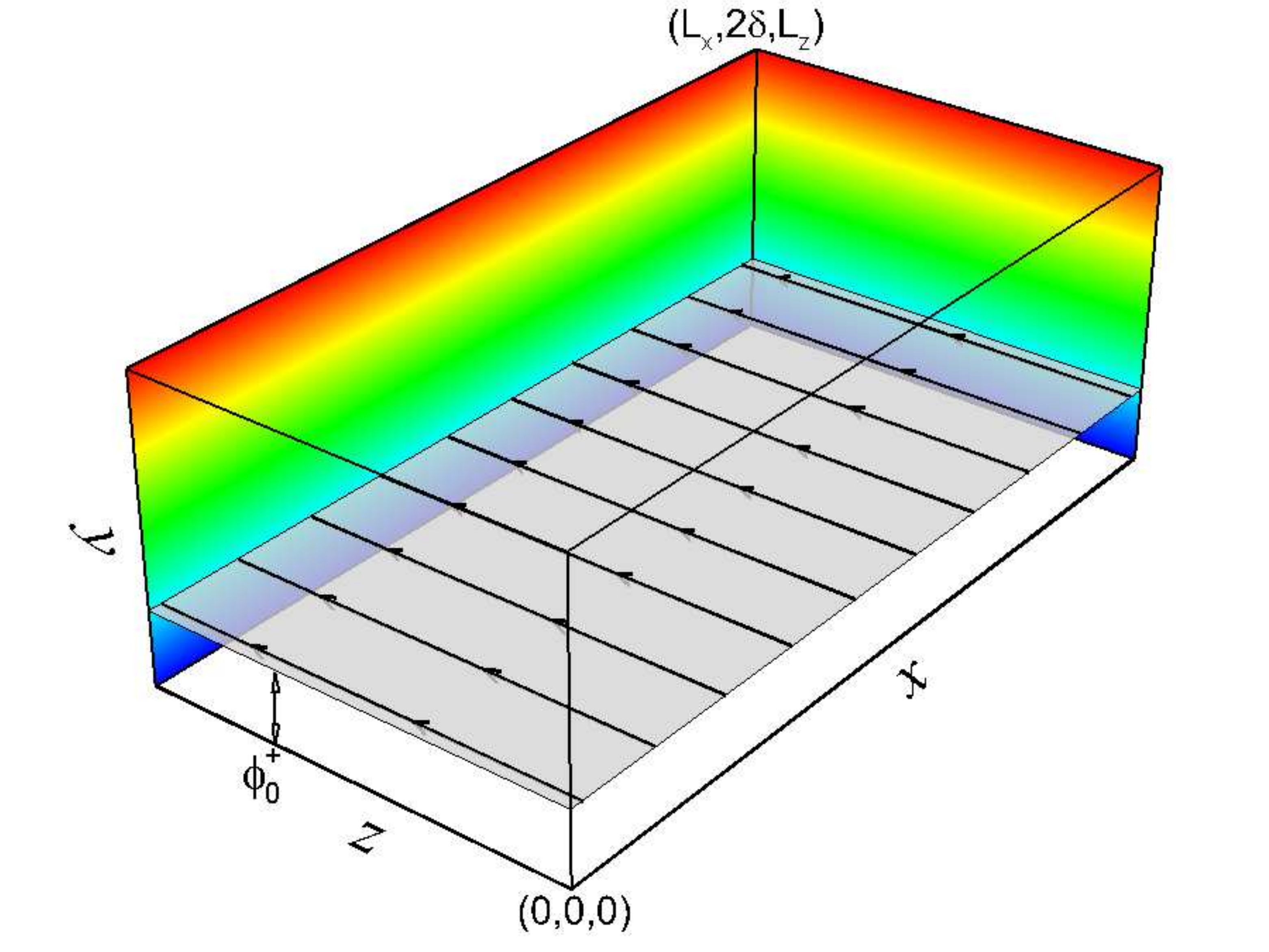}
  \caption{(Colour online) Diagram of the computational domain.\protect\\}
\label{fig:1}
\end{center}
\end{figure}

A diagram of the computation domain of channel flow is shown in figure~\ref{fig:1}.
The sides of the domain in the streamwise $x$-, the wall-normal $y$- and the spanwise $z$-directions
are $L_x$, $L_y$ and $L_z$, respectively. The three-dimensional incompressible Navier-Stokes equations for the velocity $\bs u=(u,v,w)$ are non-dimensionalized
by the channel half-height $\delta$ and the bulk velocity $U_b$ as
\begin{equation}
\left. \begin{array}{ll}
\displaystyle\
    \frac{\partial{\boldsymbol{u}}}{\partial{t}}+\boldsymbol{u}\bcdot\bnabla\boldsymbol{u}
    =-\nabla p+\boldsymbol{f}+\frac{1}{Re_b}\nabla^2\boldsymbol{u},\\[8pt]
\displaystyle\
    \bnabla\bcdot\boldsymbol{u}=0,
 \end{array}\right\}
  \label{NS}
\end{equation}
where $p$ is the pressure, $\nu$ is the kinetic viscosity, $Re_b=U_b\delta/\nu$ denotes the Reynolds number with $U_b=\int_{0}^{2\delta}u dy/L_y$.
The channel half-height is set as $\delta=1$, and the flow is driven by a time-dependent external force $\boldsymbol{f}(t)$ to maintain a constant flow flux with $U_b\approx1$ in the streamwise direction.

For DNS, (\ref{NS}) is solved by the Fourier--Chebyshev pseudo-spectral method \citep[]{Kim1987}.
The non-slip conditions are applied at the walls at $y=0$ and $y=2\delta$,
and the periodic boundary conditions are applied in the streamwise and spanwise directions.
The nonlinear term is dealiased using the two-thirds truncation method \citep[]{Canuto1988}.
The low-storage third-order semi-implicit Runge-Kutta method \citep[]{Spalart1991} is applied for the temporal integration.
The number of grids $N_x$, $N_y$ and $N_z$ in all the directions and other parameters of the DNS are listed in table~\ref{tab:Re}.
The wall-friction Reynolds number $Re_\tau=u_\tau\delta/\nu$ in this table is calculated after the flow reaches the fully developed turbulent state,
where $u_\tau=\sqrt{\tau_w/\rho}$ is the wall friction velocity with the wall shear stress $\tau_w$ and the density $\rho$.
The grid sizes $\Delta x$ and $\Delta z$ in the streamwise and the spanwise directions are scaled in wall units as $\Delta x^+=\Delta x/\delta_\nu$ and $\Delta z^+=\Delta z/\delta_\nu$
with the viscous length scale $\delta_\nu=\delta u_\tau/\nu$ for the fully developed turbulent state. In the wall-normal direction, $\Delta y^+_{w}$ denotes the distance of the first grid point from the wall scaled by $\delta_\nu$.

\begin{table}
  \begin{center}
\def~{\hphantom{0}}
  \begin{tabular}{lcccccccccc}
       $Re_\tau$ & $L_x$ & $L_y$ & $L_z$ & $N_x$ & $N_y$ & $N_z$ & $\Delta x^+$ & $\Delta y^+_{w}$  &  $\Delta z^+$\\[3pt]
         207.8   & 5.61  & 2  & 2.99     & 192   & 192   & 192   & 6.07            &   0.0278    &  3.23 \\
       \end{tabular}
  \caption{Summary of DNS parameters.}
  \label{tab:Re}
  \end{center}
\end{table}

The initial condition is set as the laminar Poiseuille flow with three Tollmien--Schlichting (TS) waves to trigger the K-type transition. The velocity profile of the Poiseuille flow is
\begin{equation}
\left. \begin{array}{ll}
\displaystyle\
    u=U_0(1-(1-y)^2),\\[8pt]
\displaystyle\
    v=0,\\[8pt]
\displaystyle\
    w=0,
 \end{array}\right\}
  \label{Poiseuille}
\end{equation}
where $U_0=1.5$ is the laminar velocity on the centerline of the channel.
The initial disturbances \citep[see][]{Sandham1992, Schlatter2004}
\begin{equation}
\boldsymbol{u}'_{0}=A_1\boldsymbol{u}_{2\textrm{D}}(y)\exp[i(\alpha x)]+A_2\boldsymbol{u}_{3\textrm{D}}(y)\exp[i(\alpha x+\beta z)]
\label{Eq:TS}
\end{equation}
consist of a two-dimensional TS wave with the amplitude $A_1=3.5\%$ and a pair of superimposed oblique three-dimensional waves with the amplitude $A_2=0.09\%$, which are similar to the simulations in \citet[]{Gilbert1990}. In (\ref{Eq:TS}), $\boldsymbol{u}_{2\textrm{D}}(y)$ and $\boldsymbol{u}_{3\textrm{D}}(y)$ are the eigenfunctions of the two-dimensional TS wave and the three-dimensional TS waves, respectively.
They are obtained by solving the Orr-Sommerfeld equations using a fourth-order compact difference method \citep[]{Malik1990}.
 {With different temporal frequencies,} the streamwise wavenumber is $\alpha=1.12$ and the spanwise wavenumber is $\beta=\pm2.1$ for the TS waves.
The numerical solver and initial disturbances used in this DNS were verified in \citet[]{Zhao2014}.

\begin{figure}
\psfrag{X}{$t$}  \psfrag{Y}{$Re_\tau$}
\begin{center}
  \includegraphics[width=0.65\textwidth]{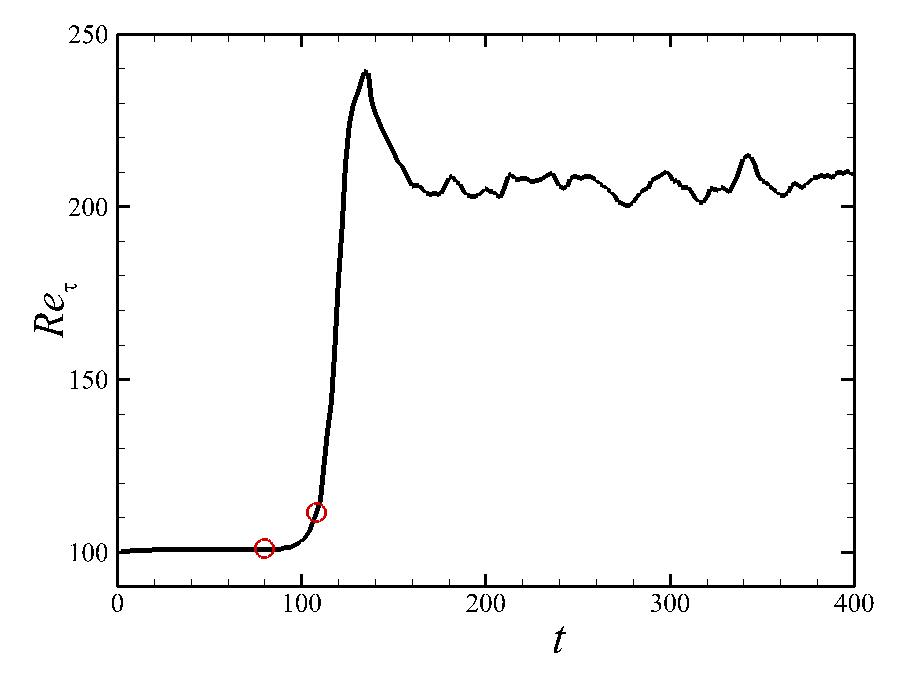}
  \caption{Temporal evolution of the wall-friction Reynolds number $Re_\tau$ in the K-type temporal transitional channel flow.
  The initial and the ending tracking times are marked by red circles.\protect\\}
\label{fig:ret}
\end{center}
\end{figure}

The flow gradually evolves from the laminar state into the fully developed turbulent state with the imposed initial disturbances.
The temporal evolution of the wall-friction Reynolds number $Re_\tau$ is shown in figure~\ref{fig:ret}.
The transition in this calculation can be roughly divided into three  {periods}.
First, the initial disturbances gradually amplify at $t<80$, with the integral flow quantity such as $Re_\tau$ remains at the laminar value.
After the slow amplification of the disturbances, the integral quantities change significantly in a short period, which implies the onset of the  {late stages of} transition at $80\leq t\leq108$.
The flow structures undergo the first significant topological changes around $t=110$ (shown in \S\,\ref{sec:3}) followed by a series of  {interactions} of vortical structures, and finally the flow develops into the fully turbulent state around $t=200$.
The evolution of flow structures in the  {beginning of the late} transitional stages for $80\leq t \leq 108$ will be studied in \S\S\,\ref{sec:3}--\ref{sec:5} in detail.

\subsection{Backward-particle-tracking method for the Lagrangian field}\label{sec:backtracking}

The transport equation for the Lagrangian scalar field $\phi(\boldsymbol{x},t)$ reads
\begin{equation}
\frac{\partial{\phi}}{\partial{t}}+\boldsymbol{u}\bcdot\bnabla\phi=0,
  \label{Eq:phi}
\end{equation}
and iso-surfaces of the scalar field are material surfaces in the temporal evolution. A set of ordinary differential equations (ODEs) are converted from (\ref{Eq:phi}) to calculate trajectories of fluid particles as
\begin{equation}
\frac{\partial{\boldsymbol{X}(\boldsymbol{x}_0,t_0| t)}}{\partial{t}}=\boldsymbol{V}(\boldsymbol{x}_0,t_0| t)=\boldsymbol{u}(\boldsymbol{X}(\boldsymbol{x}_0,t_0| t),t),
  \label{Eq:lag}
\end{equation}
where $\boldsymbol{X}(\boldsymbol{x}_0,t_0| t)$ is the location at time $t$ of the particle
which was located at $\boldsymbol{x}_0$ at the initial time $t_0$
and $\boldsymbol{V}(\boldsymbol{x}_0,t_0| t)$ is the particle velocity at time $t$.

A backward-particle-tracking method is implemented to solve (\ref{Eq:lag}).
More details about this numerical scheme can be referred to \cite{yang2010multi} and \cite{yang2011geometric}.
Firstly, the particles are placed on the uniform grids $N_x^L\times N_y^L\times N_z^L=768\times768\times768$ at a particular time $t$.
Here, the superscript $L$ denotes the grids used for the Lagrangian scalar field.
Then, these particles are tracked backwards to the initial time $t_0$ with the Eulerian velocity field $\bs u$ saved on disk from DNS.
After the backward tracking, the initial locations of the particles $\boldsymbol{x}_0$ can be obtained,
and the corresponding flow map is
\begin{equation}
   F_t^{t_0}(\bs X):\boldsymbol{X}(\boldsymbol{x}_0,t_0| t)\mapsto\boldsymbol{x}_0,~~~t\ge t_0.
  \label{Eq:mapping}
\end{equation}
Then the Lagrangian field $\phi(\boldsymbol{x},t)$ at any given time $t$ can be obtained as
\begin{equation}
\phi(\boldsymbol{x},t)=\phi(F_t^{t_0}\left(\boldsymbol{X}\right),t_0)=\phi_0
  \label{Eq:lag1}
\end{equation}
from (\ref{Eq:phi}) and (\ref{Eq:mapping}) with the initial Lagrangian field $\phi_0\equiv\phi(\boldsymbol{x}_0,t_0)$.

\subsection{Determination of the initial Lagrangian field}\label{sec:initial_scalar}
Although the initial Lagrangian field $\phi_0=f(x,y,z)$ can be an arbitrary three-dimensional scalar field in a general case,
several criteria are proposed to uniquely determine $\phi_0$ that is suitable for  {this particular laminar-turbulent transition problem}.

\begin{figure}
\centering \subfigure{
\begin{minipage}[c]{0.7\textwidth}
 \includegraphics[width=1\textwidth]{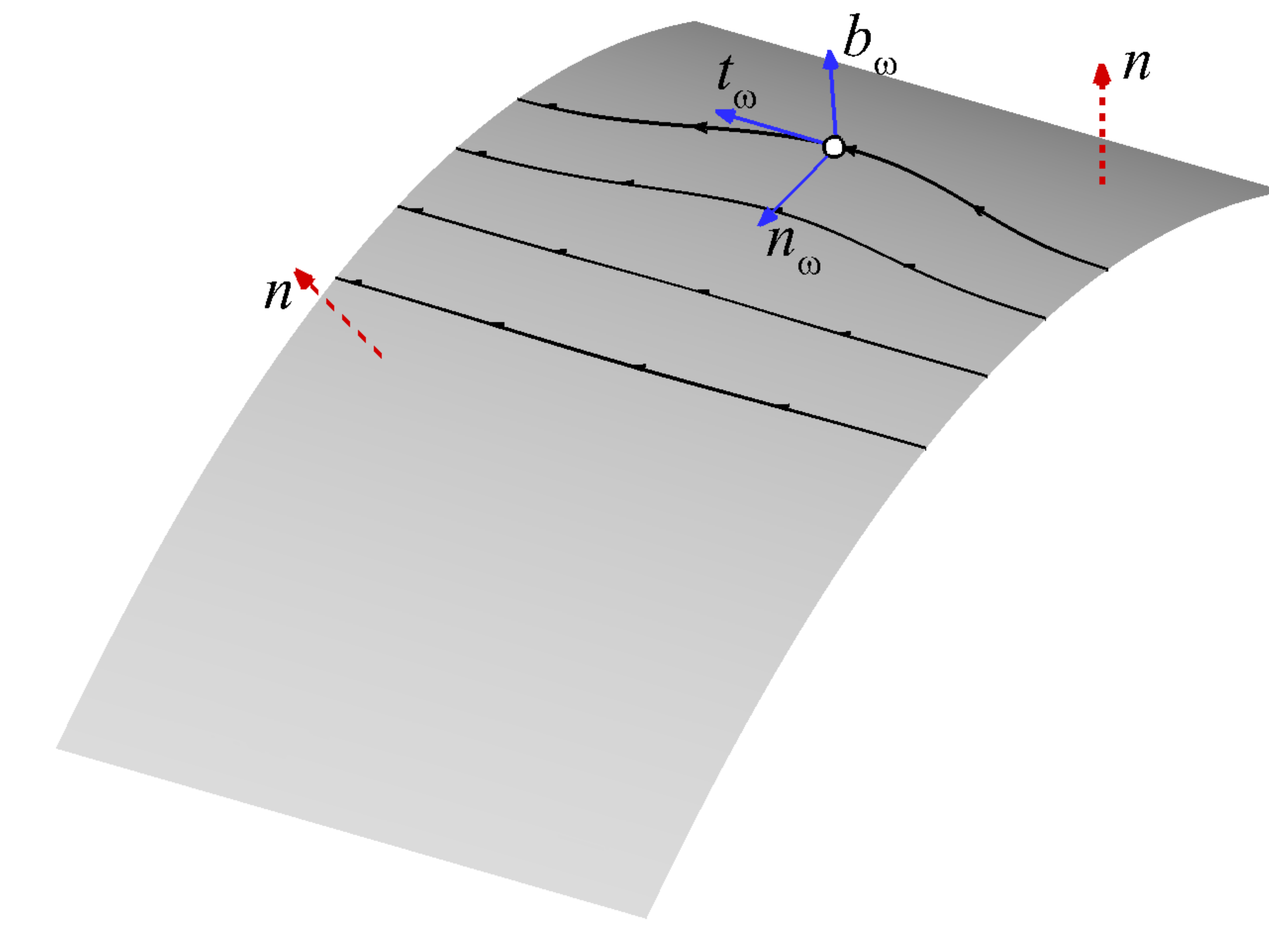}
\end{minipage}}%
  \caption{(Colour online) Diagram of the geometry of a material surface and typical vortex lines attached on the surface.
  The dashed vector $\boldsymbol{n}$ is the normal of the surface,
  and solid vectors are the tangential vector $\boldsymbol{t}_\omega$,
  the normal vector $\boldsymbol{n}_\omega$ and the bi-normal vector $\boldsymbol{b}_\omega$ of a vortex line.}
\label{fig:diagram_n}
\end{figure}

First, in order to investigate the evolution of vortical structures, the iso-surfaces of $\phi_0$ should be vortex surfaces \citep{yang2010lagrangian}.
The vortex surface is a surface composed of vortex lines, which implies that the local vorticity vector is tangent at every point on the
surface.
As shown in figure~\ref{fig:diagram_n}, the unit vector $\boldsymbol{n}=(n_x, n_y, n_z)\equiv\bnabla\phi/g$ is the normal of the material surface,
 {where $g\equiv(\bnabla\phi\bcdot\bnabla\phi)^{1/2}$ is the magnitude of the scalar gradient}.
Thus it requires that the vorticity $\bs \omega\equiv\nabla\times\bs u$  is perpendicular to $\boldsymbol{n}$ in the whole field.
In the initial state, the criterion can be written as
\begin{equation}
\boldsymbol{\omega}_0\bcdot\bnabla\phi_0=0,
  \label{criteria-1}
\end{equation}
where the subscript $0$ denotes a quantity at the initial time.
Considering the base flow without the initial disturbances, the laminar Poiseuille flow (\ref{Poiseuille}) or any other simple shear flow as $\boldsymbol{u}_0=(h(y),0,0)$, where $h(y)$ is an arbitrary smooth function of $y$, the vorticity only has one component in the spanwise direction as $\boldsymbol{\omega}_0=(0, 0, \omega_z)$.
Thus, $\phi_0$ is reduced to a two-component scalar field as $\phi_0=f(x,y)$.

Second, the growth of disturbances in the natural transition is slow in a long period when the flow state is nearly laminar.
To focus on the evolution of material surfaces in the transition and avoid the deformation under the mean shear, $\phi_0$ is expected to stay invariant as $\partial{\phi_0}/\partial{t}=0$ in a laminar shear flow $\boldsymbol{u}_0=(h(y),0,0)$.
With the transport equation (\ref{Eq:phi}) of $\phi$, we obtain
\begin{equation}
\boldsymbol{u}_0\bcdot\bnabla\phi_0=0,
  \label{criteria-2}
\end{equation}
so $\phi_0$ can be reduced as $\phi_0=f(y)$. The iso-surfaces of $\phi_0$ are streamwise-spanwise planes at different wall distances in figure~\ref{fig:1}, and they are both vortex and stream surfaces in the laminar state. In contrast, an arbitrary initial surface with an inclination angle with the streamwise direction is not a stream surface, and it can be persistently stretched by the mean shear. Eventually it evolves to nearly streamwise-spanwise planes within a finite time.

Finally, $\phi_0$ is required to be non-constant and as smooth as possible.  {Based} on a polynomial expansion of $\phi_0=f(y)$, we can assume $\phi_0=C_0+C_1 y$ where $C_0$ and $C_1$ are constants.
Without loss of generality, the initial scalar field is set as $\phi_0=y$ with $C_0=0$ and $C_1=1$.

\subsection{Tracking interval}\label{sec:fragments}
The imposed growing disturbances such as (\ref{Eq:TS}) can break the temporal invariance of the scalar field.
The initial tracing time $t_0=80$ is selected when the amplitudes of the initial disturbances are small enough, so that every initial material surface can be approximated as a vortex sheet composed of vortex lines.
The choice of $t_0=80$, instead of the theoretical value $t_0=0$, can avoid a very long particle tracking time for $t\leq80$ and greatly reduce the computational cost with negligible deviations. We find that the statistical results are not sensitive to the value of $t_0$ for $t_0\leq80$ from numerical experiments.

The evolution and dynamics of material surfaces can be studied from the evolution of material surfaces in the  {late} transitional stages for $80\leq t\leq108$ except for a few of qualitative visualizations at $t=112$.
It is noted that the particle-backward-tracking method has no numerical diffusion. Although it ensures the geometrical and topological conservation properties for material surfaces, the fragmentation for the under-resolved material surfaces can occur for long times \citep[see][]{yang2010multi}.
To avoid the effect of numerical artifacts caused by the fragmentation on statistical results, we select the upper limit of the tracking period as $t=108$.
During this period, satisfactorily resolved material surfaces can be obtained by using the high resolution grids $N_x^L\times N_y^L\times N_z^L$ for the Lagrangian field.
 {Futhermore, iteration strategies with the local grid refinement can be considered in the further study to reduce the computational cost.}


\section{Qualitative visualizations}
\label{sec:3}
\subsection{Evolution of material surfaces}
\label{sec:L-surface}

In the transition process,  all the material surfaces evolve from the streamwise-spanwise planes at different distances from the wall at the initial tracing time $t_0$, but their evolutions can be different.
The typical material surfaces studied in the present work are listed in table~\ref{tab:material}. They are labelled by the Lagrangian scalar field scaled by $\delta$ or $\delta_\nu$ that corresponds to the laminar state or the fully developed turbulent state, respectively.
The evolution of the material surfaces in the  {late transition} at $80\leq t\leq112$ will be discussed from $\phi=0.024$ to $\phi=0.722$.
\begin{table}
  \begin{center}
\def~{\hphantom{0}}
  \begin{tabular}{lccccc}
       $\phi=\phi_0/\delta$   & 0.024  & 0.217  & 0.433     & 0.722\\[3pt]
       $\phi^+=\phi_0/\delta_\nu$ & 5 & 45 & 90 & 150 \\
       \end{tabular}
  \caption{Typical material surfaces labelled by scaled $\phi_0=y$.}
  \label{tab:material}
  \end{center}
\end{table}

First of all, the material surface initially very close to the wall with $\phi=0.024$ is only slightly wrinkled in the evolution. Since the non-slip condition is applied at the wall and the amplitudes of the initial disturbances in this region are very small, the geometry of the surface has no obvious change.

\begin{figure}
\centering \subfigure{
\begin{minipage}[c]{0.5\textwidth}
\includegraphics[width=2.6in]{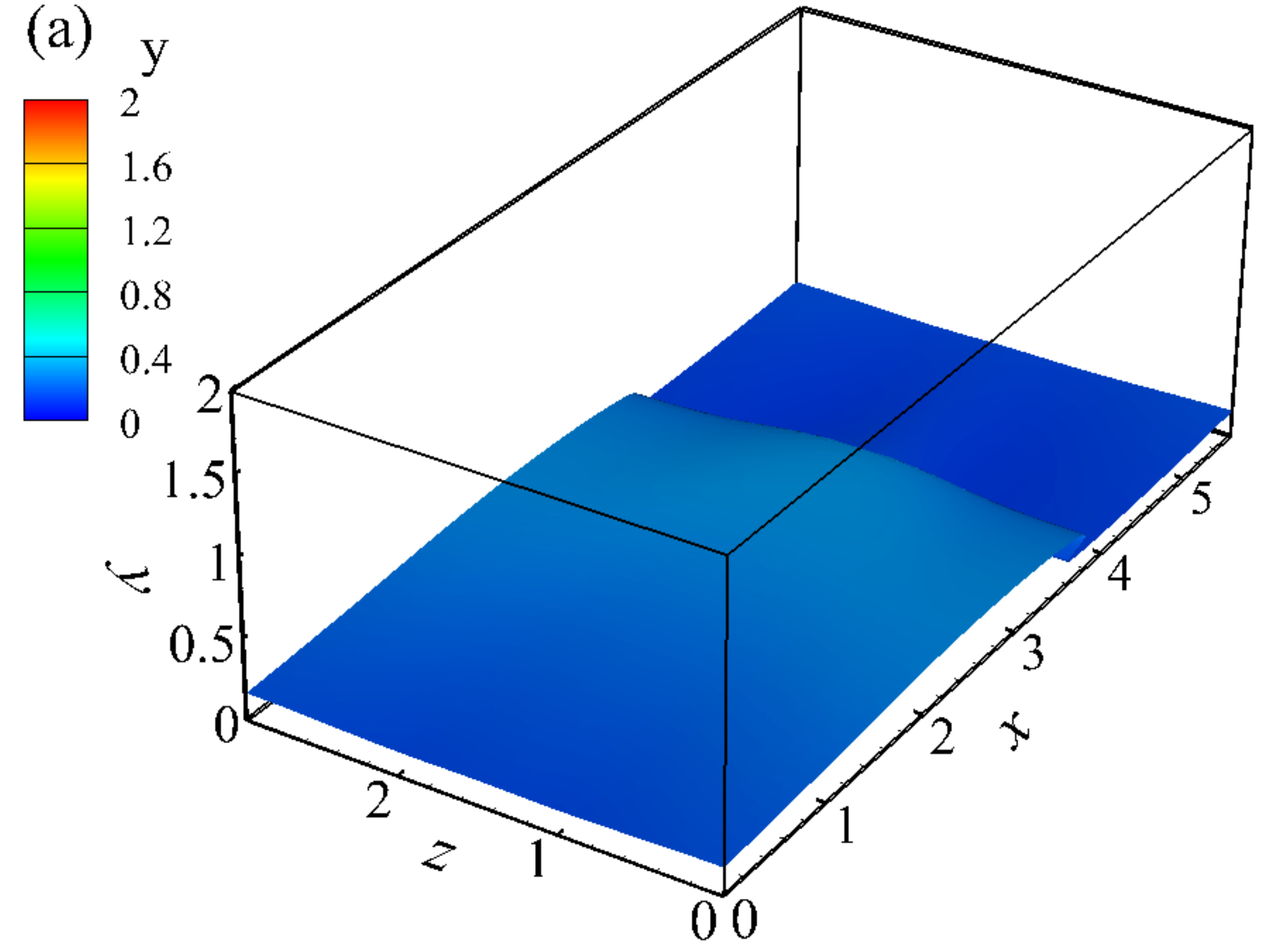}
\end{minipage}}%
\centering \subfigure{
\begin{minipage}[c]{0.5\textwidth}
\includegraphics[width=2.6in]{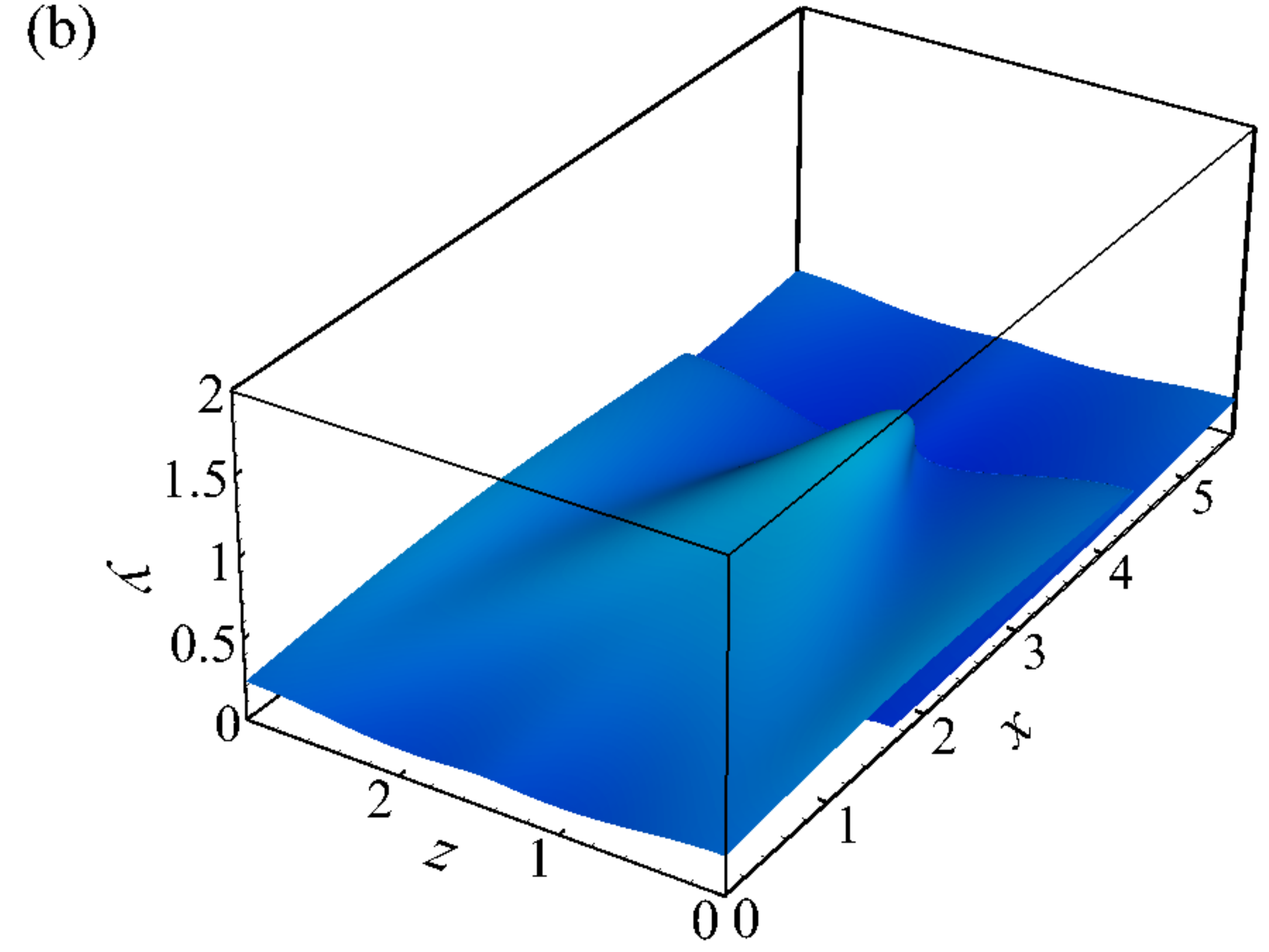}
\end{minipage}}%

\centering \subfigure{
\begin{minipage}[c]{0.5\textwidth}
\includegraphics[width=2.6in]{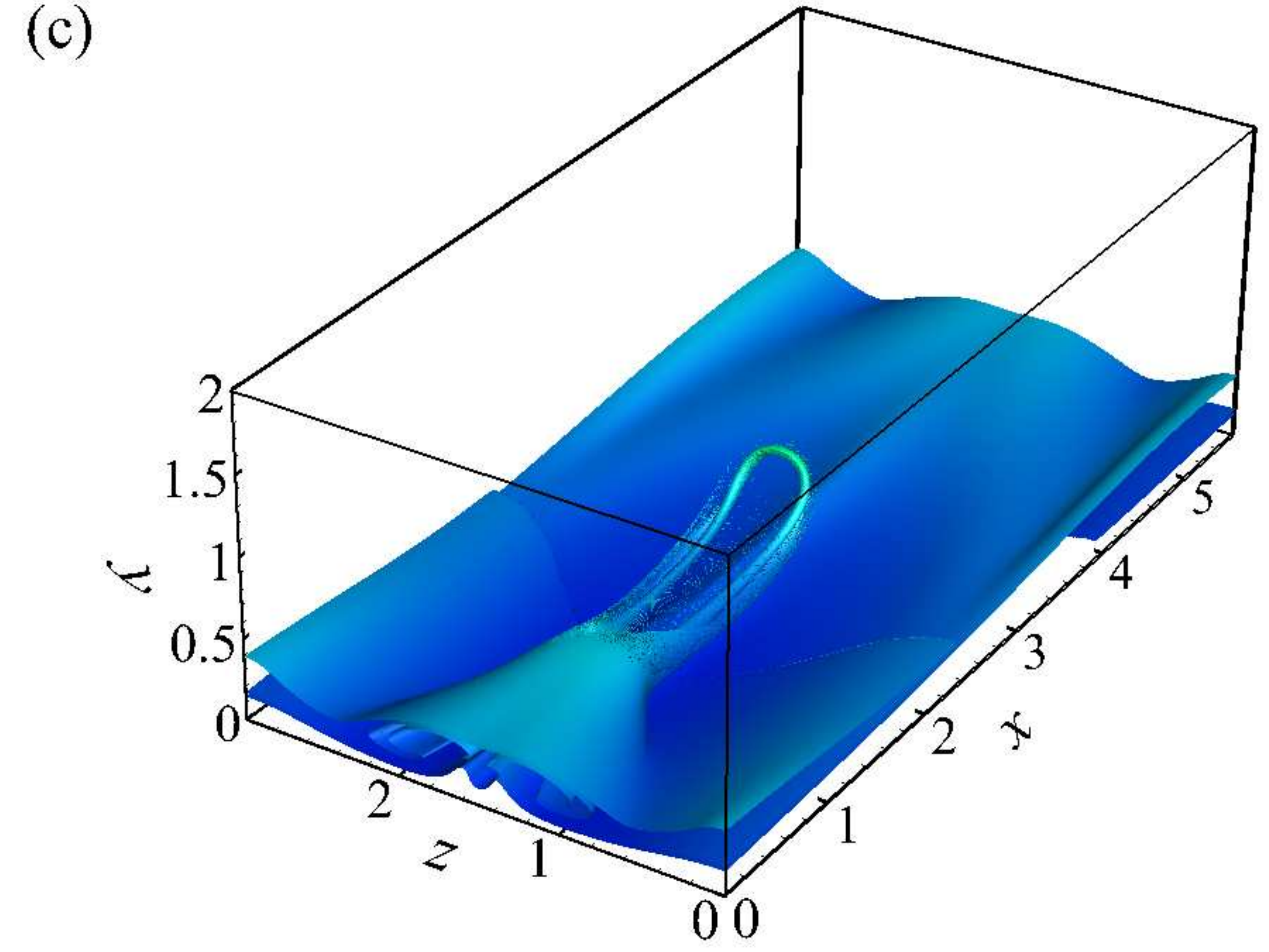}
\end{minipage}}%
\centering \subfigure{
\begin{minipage}[c]{0.5\textwidth}
\includegraphics[width=2.6in]{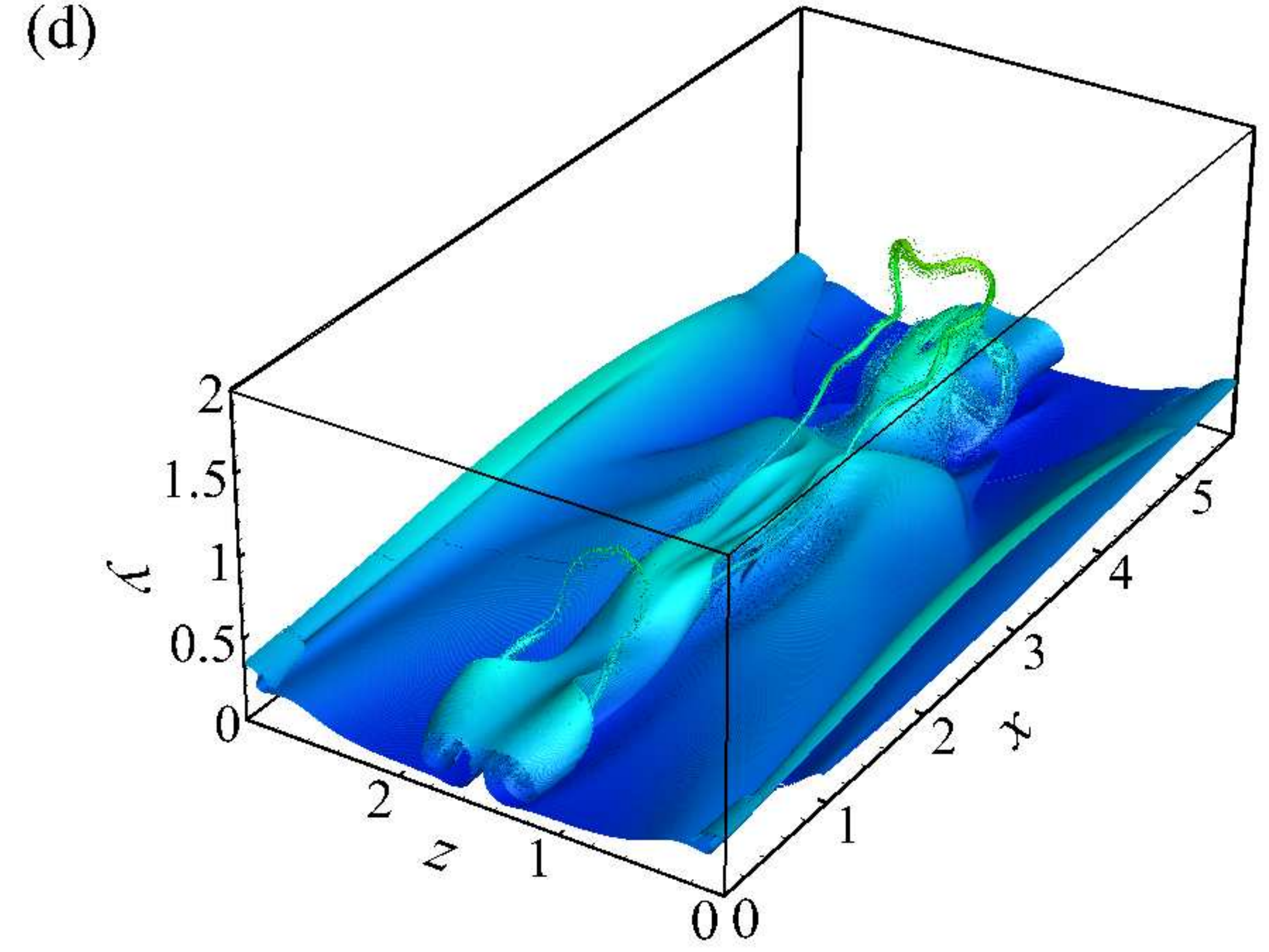}
\end{minipage}}%

  \caption{(Colour online) Temporal evolution of the material surface with $\phi=0.217$ or $\phi^+=45$.
  The surface is colour-coded by the wall-normal coordinate $y$. (a) $t=90$, (b) $t=100$, (c) $t=106$, (d) $t=112$.  {The surfaces at $t=106$ and $t=112$ are partially under-resolved.} \protect\\}
\label{fig:all-2}
\end{figure}

On the contrary, the material surface of $\phi=0.217$, which is more distant from the wall than that of $\phi=0.024$ at the initial time,  shows significant deformation in the evolution in figure~\ref{fig:all-2}.
The planar surface is first folded and then rolled up at $t=90$ in figure~\ref{fig:all-2}(a).
A triangle-shaped bulge is generated around $t=100$ in figure~\ref{fig:all-2}(b), and then gradually evolves into a typical hairpin-shaped structure in figure~\ref{fig:all-2}(c).
Finally, the `head' of the hairpin-like structure is lifted and the two `legs' are stretched in figure~\ref{fig:all-2}(d).
 {
This stretching process was also observed by tracking an ensemble of material lines initially parallel to the wall in DNS of a transitional boundary layer in \citet[]{Rist1995}, but the evolution of the material surface can display more details, such as the rolling-up of tube-like structures, than the scattered lines, and provide quantitative results based on the conditional statistics on the surfaces.}
It is noted that there are some fragments near the head and legs of the hairpin-like structure in figures~\ref{fig:all-2}(c) and (d). Since these parts of the surface endure persistent stretching and rolling during the evolution,
they are partially under-resolved for the iso-surface visualization.
Although the fragmentation can be resolved by using higher resolutions for the scalar field as discussed in \S\,\ref{sec:fragments}, the current resolution is enough to show the skeleton of the hairpin-like structures.

\begin{figure}
\centering \subfigure{
\begin{minipage}[c]{0.5\textwidth}
\includegraphics[width=2.6in]{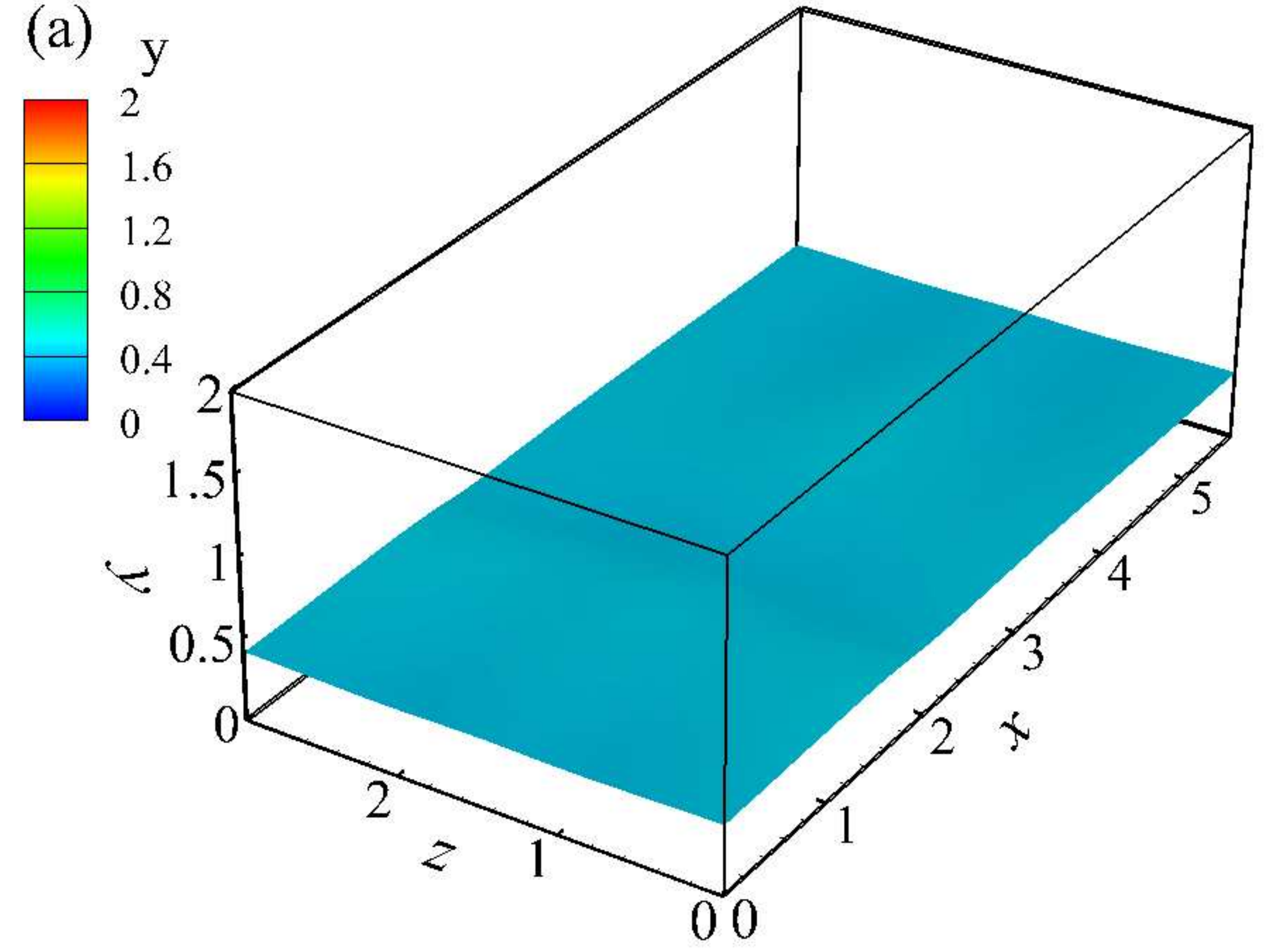}
\end{minipage}}%
\centering \subfigure{
\begin{minipage}[c]{0.5\textwidth}
\includegraphics[width=2.6in]{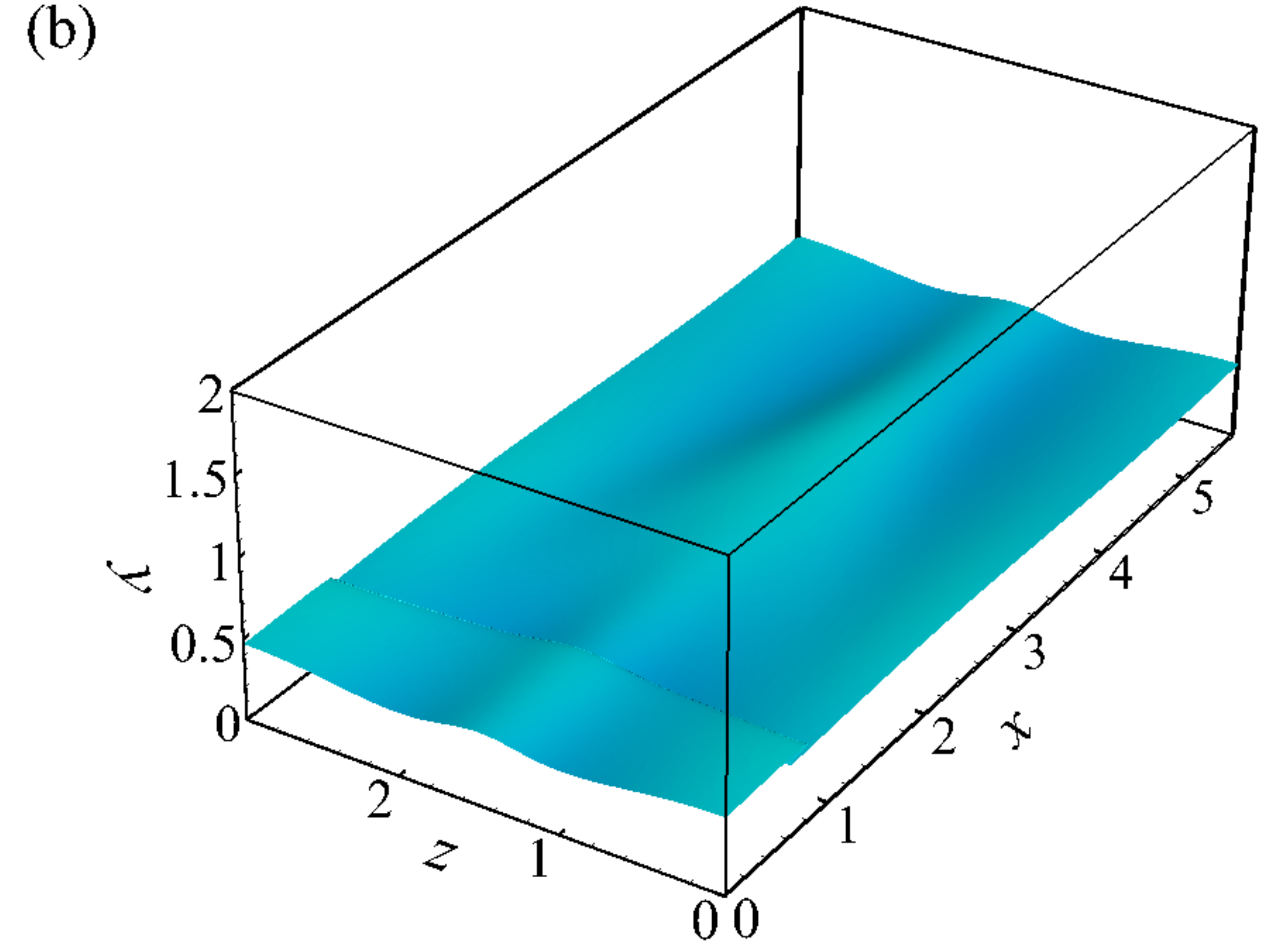}
\end{minipage}}%

\centering \subfigure{
\begin{minipage}[c]{0.5\textwidth}
\includegraphics[width=2.6in]{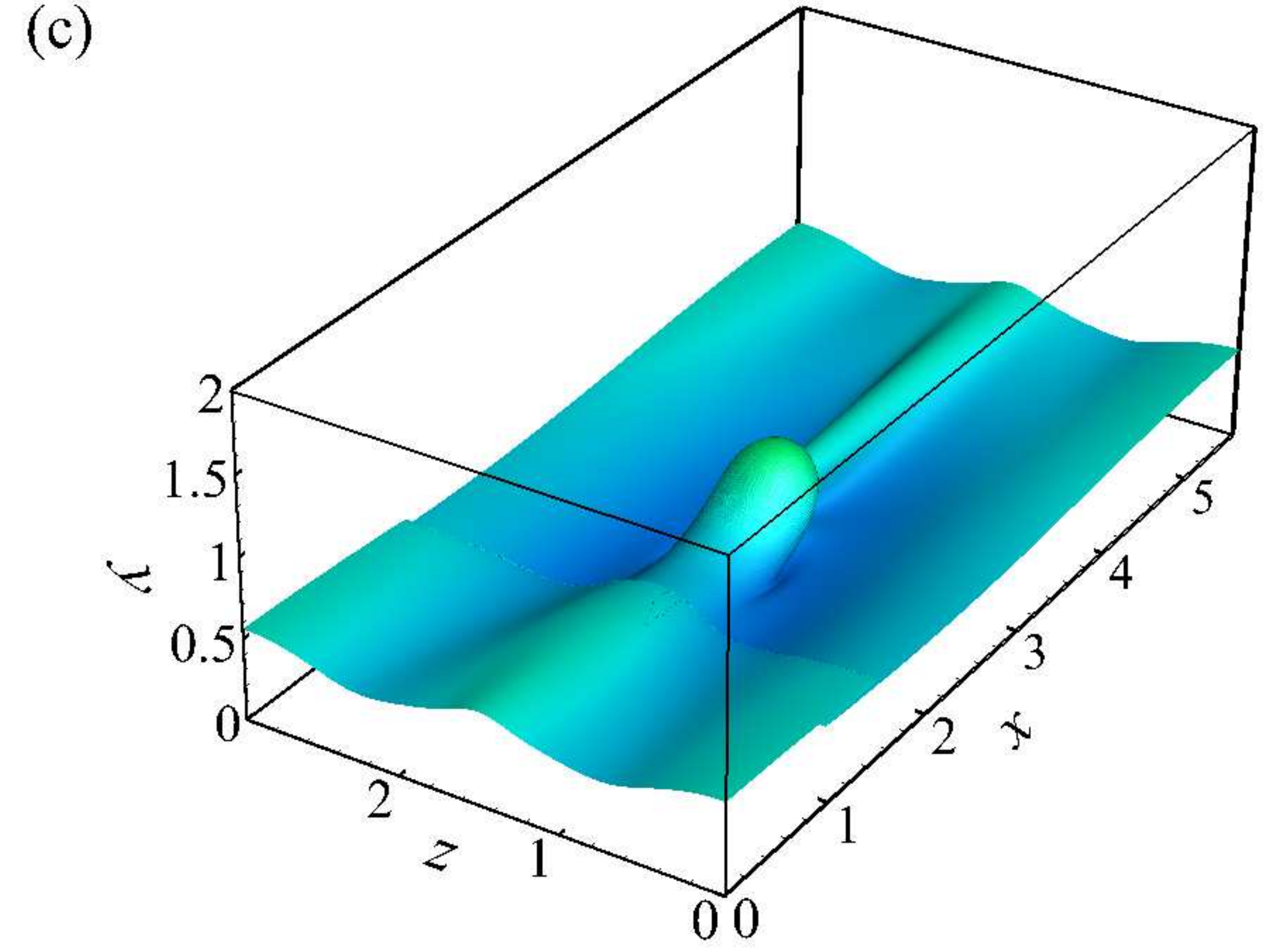}
\end{minipage}}%
\centering \subfigure{
\begin{minipage}[c]{0.5\textwidth}
\includegraphics[width=2.6in]{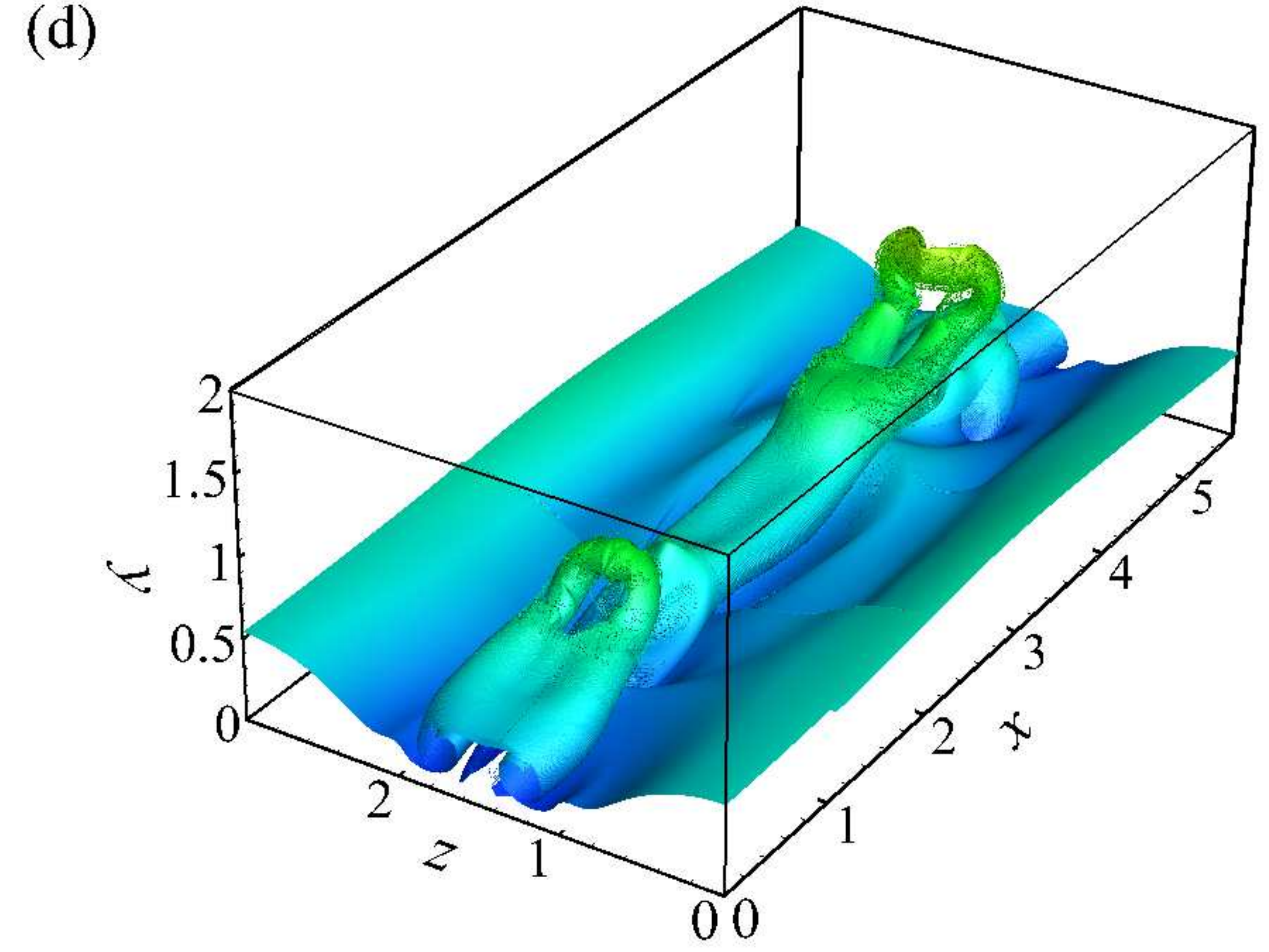}
\end{minipage}}%

  \caption{(Colour online) Temporal evolution of the material surface with $\phi=0.433$ or $\phi^+=90$.
  The surface is colour-coded by the wall-normal coordinate $y$. (a) $t=90$, (b) $t=100$, (c) $t=106$, (d) $t=112$.  {The surface at $t=112$ is partially under-resolved.}\protect\\}
\label{fig:all-3}
\end{figure}

The evolution of the surface of $\phi=0.433$ is shown in figure~\ref{fig:all-3}.
This process is mainly influenced by the lift of the hairpin-like structure as that in figure~\ref{fig:all-2}.
Before the lift, this surface is only slightly disturbed in figures \ref{fig:all-3}(a) and \ref{fig:all-3}(b).
A thumb-shaped bulge in figure~\ref{fig:all-3}(c) at $t=106$ is above the hairpin-like structure in figure~\ref{fig:all-2}(c).
The material surface of $\phi=0.433$ warps around the structure of $\phi=0.217$ and also rolls into a hairpin-shaped structure in figure~\ref{fig:all-3}(d).
 {A similar rolling-up process was observed from the evolution of a vortex layer from an initially disturbed sheet in \citet[]{Moin1986}, but the tracking object represented by scattered vortex lines was initially set in the mean-velocity profile of a fully developed channel flow rather than a transitional flow.
}
Finally, the secondary hairpin-shaped structure is generated behind the primary one in figures \ref{fig:all-2}(d) and \ref{fig:all-3}(d).
Furthermore, the interactions between different hairpin-like structures will be addressed in \S\,\ref{sec:2d}, and the dynamics in the evolution of the hairpin-like structures will be discussed in \S\,\ref{sec:5} in detail.

 {For the evolution of the material surface that is initially remote from the wall, the surface of $\phi=0.722$ is almost undisturbed until $t=106$ (not shown). Then it is influenced by the rapid elevation of the hairpin-like structure as shown in figure~\ref{fig:all-2}, and the rolling-up process is similar to the surface of $\phi=0.433$.
}

\subsection{Evolution of Lagrangian fields on the streamwise and wall-normal plane cut}
\label{sec:2d}

The evolution of the Lagrangian scalar field on the streamwise and wall-normal ($x$--$y$) plane at $z=L_z/2$ is used to study evolving material surfaces,
because the most rapidly growing fluctuations, the tip of the triangular bulge and the head of the hairpin-like structures occur on this `peak plane' \citep[]{Klebanoff1962,Kleiser1991}.
The number of two-dimensional grids for the Lagrangian field on this plane is increased as $N_x^L\times N_y^L=1536\times1536$ to achieve the visualization with high resolution.

\begin{figure}
\centering \subfigure{
\begin{minipage}[c]{1\textwidth}
\begin{center}
\includegraphics[width=4.8in]{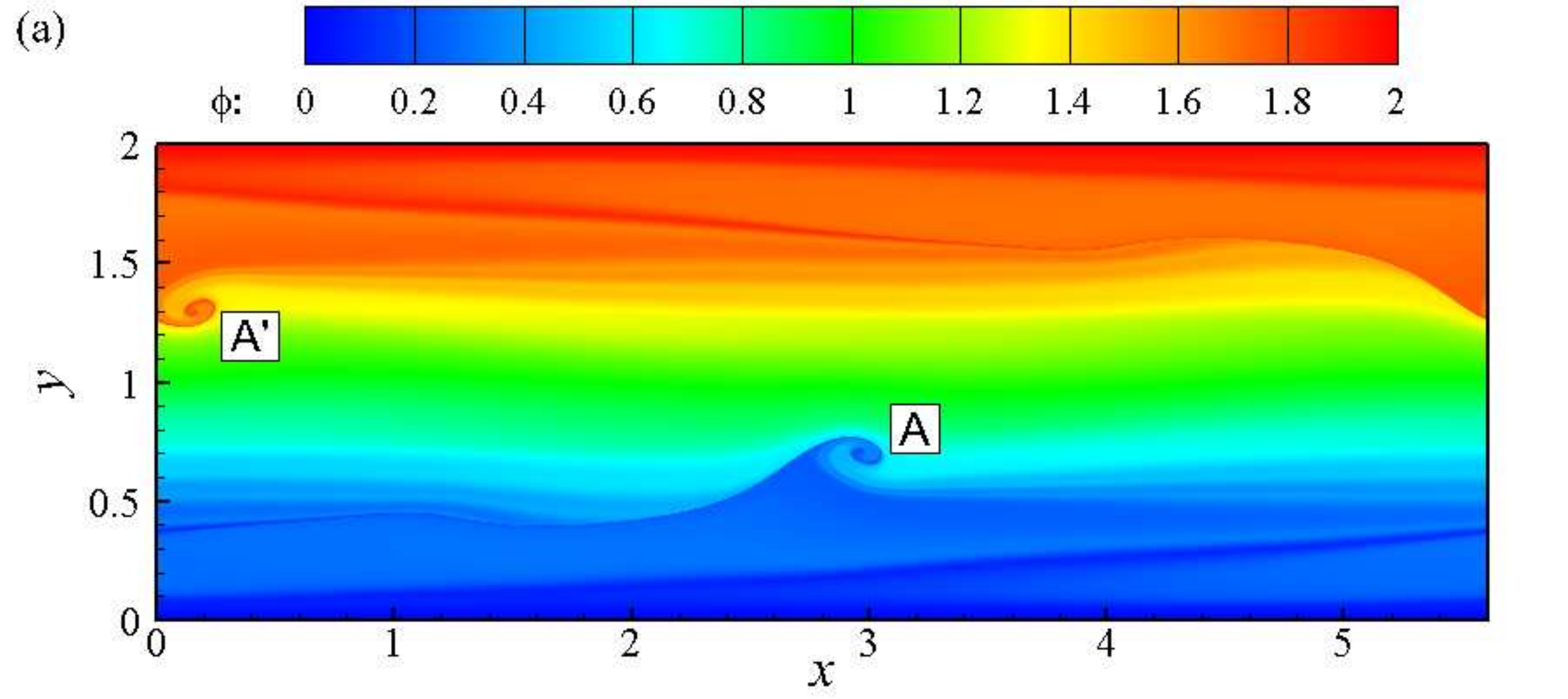}
\end{center}
\end{minipage}}
\centering \subfigure{
\begin{minipage}[c]{1\textwidth}
\begin{center}
\includegraphics[width=4.8in]{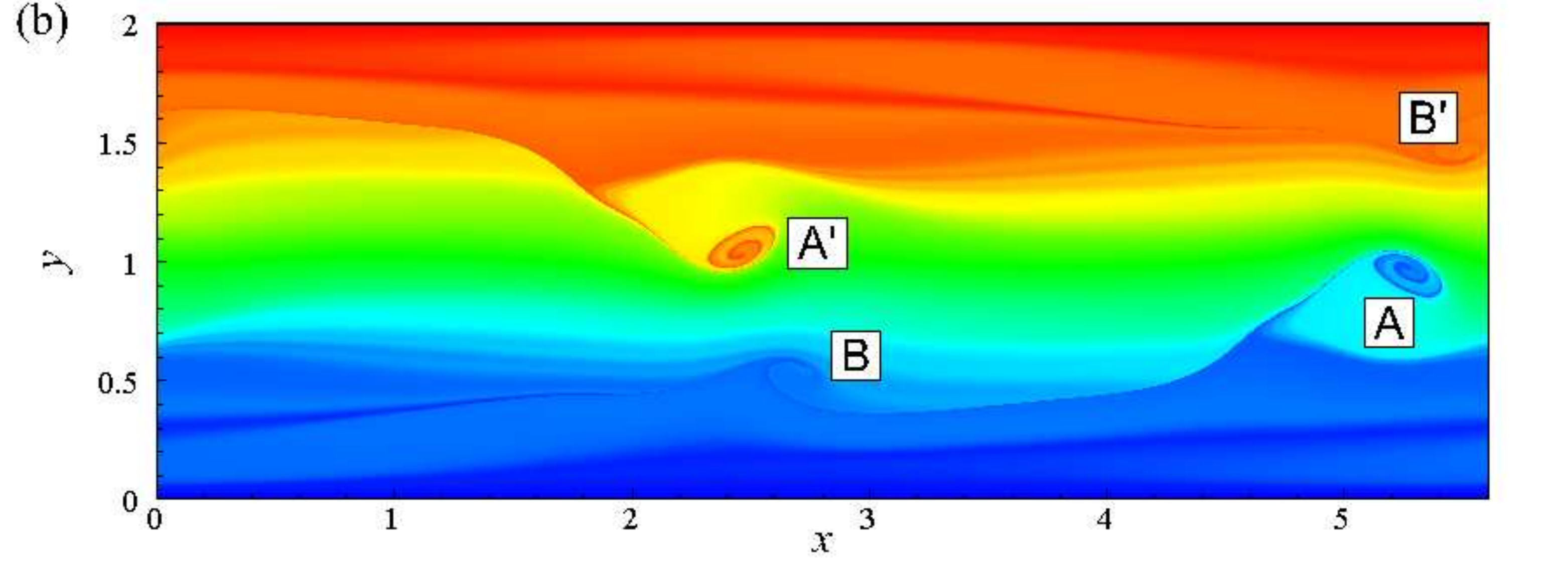}
\end{center}
\end{minipage}}
\centering \subfigure{
\begin{minipage}[c]{1\textwidth}
\begin{center}
\includegraphics[width=4.8in]{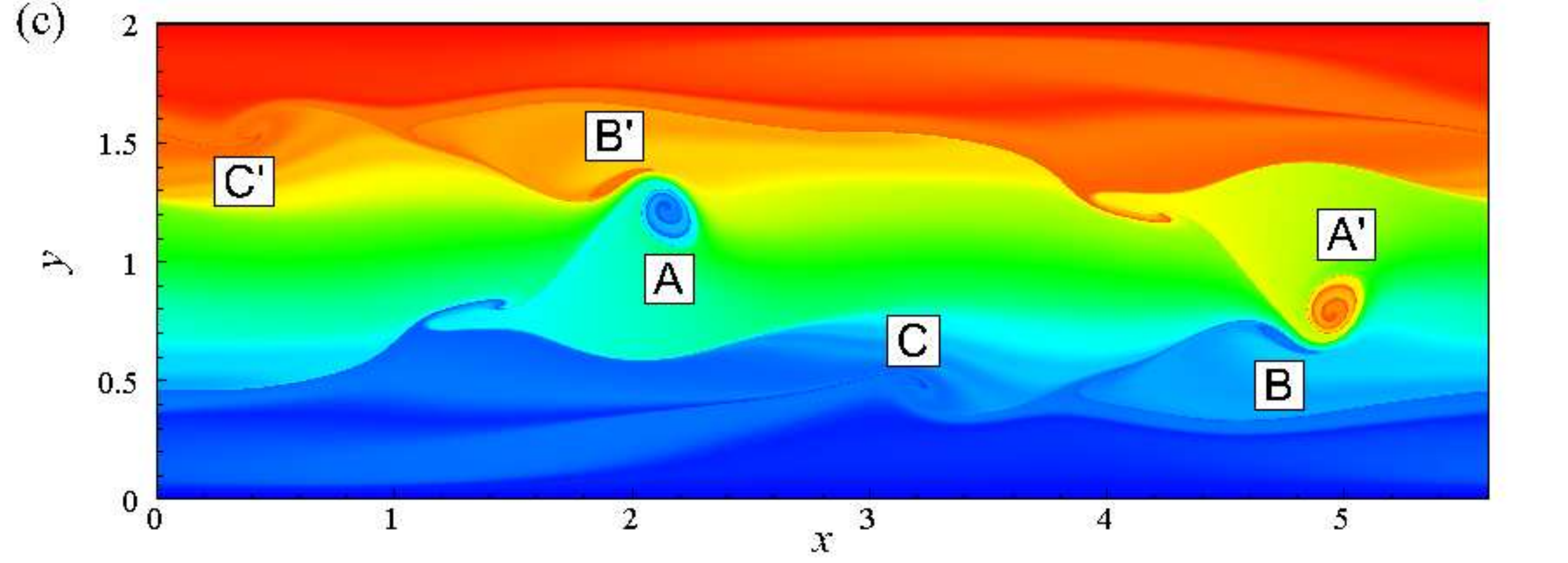}
\end{center}
\end{minipage}}

\centering \subfigure{
\begin{minipage}[c]{1\textwidth}
\begin{center}
\includegraphics[width=4.8in]{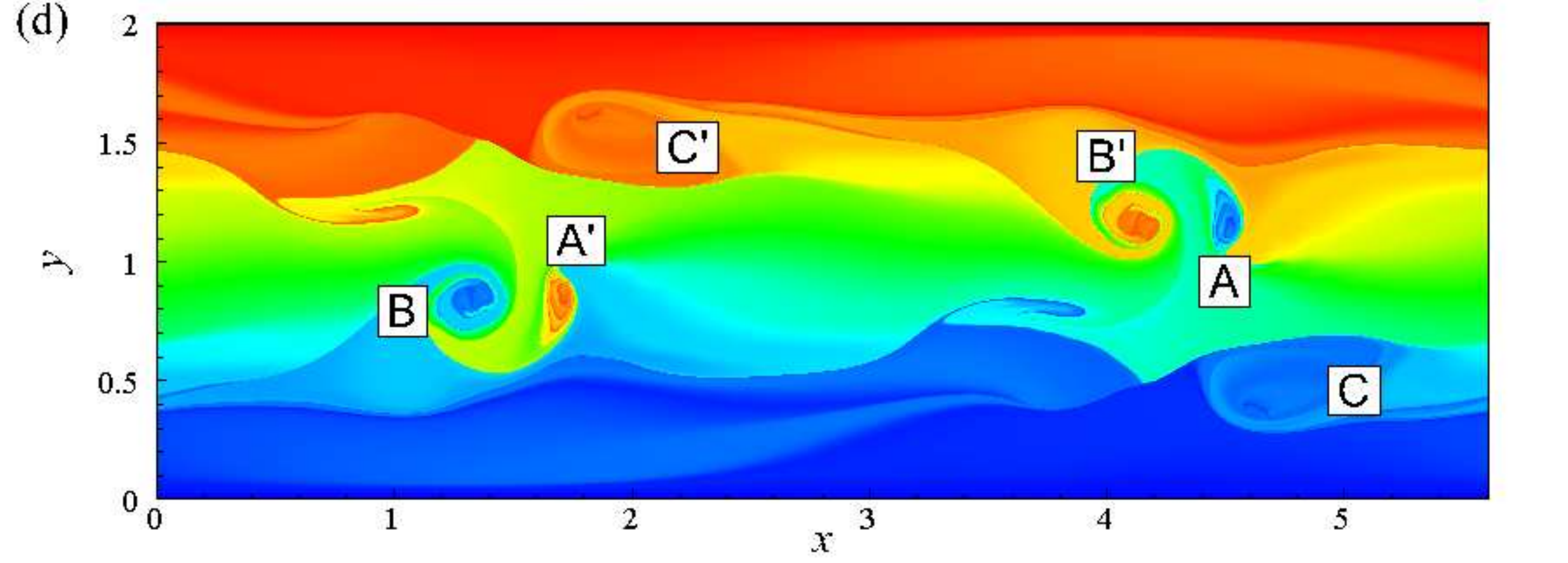}
\end{center}
\end{minipage}}

  \caption{(Colour online) Temporal evolution of the Lagrangian scalar field on the spanwise and wall-normal plane-cut at  {the peak position} $z=L_z/2$.
  The heads of the hairpin-like structures from the lower half of the channel are marked by $A$, $B$ and $C$,
  and those from the upper half are marked by $A'$, $B'$ and $C'$.
  (a) $t=106$, (b) $t=108$, (c) $t=110$, (d) $t=112$.\protect\\}
\label{fig:2d}
\end{figure}

The temporal evolution of the Lagrangian field on the $x$--$y$ plane at  {the peak position} $z=L_z/2$ in the  {late transitional} flow is shown in figure~\ref{fig:2d}. The scalar is colour-coded by $0\leq\phi\leq2$.
Corresponding to the structures in figure~\ref{fig:all-2}(b),
the heads of primary hairpin-like structures marked as $A$ and $A'$ are stretched out from the triangular bulges at $t=106$ in figure~\ref{fig:2d}(a), and they are then lifted in figure~\ref{fig:2d}(b).
 {The spiral patterns with large scalar gradients indicate swirling motions near the heads of hairpin-like structures.}

Subsequently the secondary hairpin-like structures marked as $B$ and $B'$ at $t=108$ form in the upstream of the primary ones.
Both structures continue lifting, and at the mean time the tertiary hairpin-like structures marked as $C$ and $C'$ are generated in figure~\ref{fig:2d}(c).
It is noted that the tertiary hairpin-like structures cannot fully develop in the transitional channel flow, because they are suppressed by structures $A$ and $A'$ in figure~\ref{fig:2d}(d).
With the periodic boundary condition in the streamwise direction,
the structure $A$ reenters the domain from the left boundary and then catch $C$. Furthermore, since the induced velocity of the vortical structure $A$ at the location of $C$ is downward,
the lift of $C$ is also suppressed. It is noted that the structures marked as $A'$, $B'$ and $C'$ can be considered as the images of $A$, $B$ and $C$ with corresponding phase differences in the streamwise direction owing to the reflectional symmetry with the mid-plane of the channel flow.

It can be inferred from figure~\ref{fig:2d}(d) that the hairpin-like structures from upper and bottom halves of the channel have strong interactions,  {which was observed and called as the `cross-channel interactions' in \citet[]{Sandham1992}}.
In particular, the structure $A$ appears to collide with $B'$ at $110\leq t\leq112$ and they are entangled with each other. This  { structure} interaction or reconnection causes  {flow to break down into the turbulent state}, because a vortex surface can have violent topological changes, and then evolve into small-scale, tube-like structures with stretching and twisting processes \citep{yang2011vsf}. At this point, the evolution of material surfaces and vortex surfaces with the same initial condition can have essential differences, which will be addressed in \S\,\ref{sec:approx_vortex_surface}.

\section{Identification and characterization of influential material surfaces}\label{sec_influential}
As presented in \S\,\ref{sec:3}, the material surfaces with different initial wall distances show notable different behaviors in evolution.
The material surface very close to the wall has no obvious deformation. In contrast, the surface of $\phi=0.217$ or $\phi^+=45$ shows significant deformation to form the signature hairpin-like structure, and the elevation of this structure also affects the deformation of neighbouring surfaces such as that of $\phi=0.433$ or $\phi^+=90$.
Therefore, it is natural to identify the distinguished material surfaces that can generate coherent flow patterns and have an impact on the behaviors of other material surfaces. These surfaces are expected to provide representative kinematics on the evolution of the Lagrangian scalar field.

In the LCS theory \citep[see][]{Haller2015}, the distinguished or influential material surfaces are identified as locally maximal repelling, attracting, or shearing impact on neighboring fluid elements.
Rather than using the LCS framework rooted in the general theory of nonlinear dynamics, we identify and characterize the material surfaces that have distinguished geometric changes or kinematics based on conditional means of the properties that are important in transitional wall flow over different evolving material surfaces.
Presently, we call the surface with the maximum conditional mean as a general term `influential material surface' that was used based on different criteria in the LCS context by \citet[]{Haller2015}.

\subsection{Deformation of material surfaces}\label{sec:geometry_surface}
In order to identify influential material surfaces, we firstly analyze the geometric deformation of different material surfaces.
The level of deformation of the material surfaces can be quantified by the statistics for the magnitude of the scalar gradient $g$, which is critical in scalar transport and is straightforward to be obtained in numerical or experimental data set.
The transport equation of $g$ can be derived from (\ref{NS}) and (\ref{Eq:phi}) as
\begin{equation}
   \frac{\mathrm{D}g}{\mathrm{D}t}=-(\boldsymbol{n}\bcdot\mathsfbi{S}\bcdot\boldsymbol{n})g,
  \label{Eq:grad_phi}
\end{equation}
where ${\mathrm{D}}/{\mathrm{D}t}\equiv{\partial}/{\partial t}+\boldsymbol{u}\bcdot\bnabla$ is the material derivative, $\mathsfbi{S}=(\partial u_i/\partial x_j+\partial u_j/\partial x_i)/2$ is the rate-of-strain tensor.
From (\ref{Eq:grad_phi}), the evolution of $g$ is mainly controlled by the alignment of the scalar gradient and the most compressive strain direction
of the rate-of-strain tensor \citep[see][]{Brethouwer2003}.

\begin{figure}
\centering \subfigure{
\begin{minipage}[c]{0.6\textwidth}
\psfrag{X}{$\phi$}\psfrag{Y}{$\<g|\phi\>$}
\psfrag{A}{$t=108$}\psfrag{B}{$t=104$}\psfrag{C}{$t=100$}\psfrag{D}{$t=90$}
\includegraphics[width=3.2in]{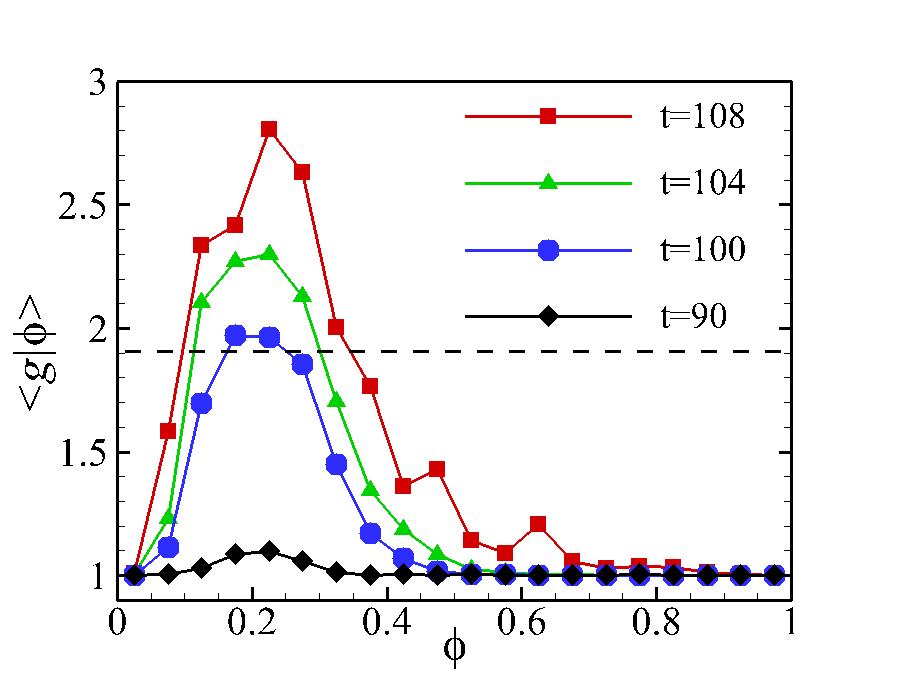}
\end{minipage}}%
  \caption{(Colour online)  Averaged $\langle g|\phi\rangle$ conditioned on different material surfaces. The dashed line denotes the level of $0.5\<g|\phi\>_{max}$ at $t=108$.
}
\label{fig:grad_p}
\end{figure}

For the initial field $\phi_0=y$ with $g=1$ in the whole field at the initial time, $\phi$ is time invariant in the laminar state (\ref{Poiseuille}).
The imposed disturbances cause the evolution of $\phi$ and $g$.
In figure~\ref{fig:grad_p}, the conditional averages $\<g|\phi\>$ for all the surfaces monotonically increase in the  {late transition}, where $\langle\cdot\rangle$ denotes the volume average.
This growth owing to the local straining motion can be explained from the statistical nature of the transport equation (\ref{Eq:grad_phi}), which was discussed in \citet[]{yang2010multi}.

Meanwhile, the growth of $\<g|\phi\>$ corresponds to the evolution of the structures presented in \S\,\ref{sec:3}.
In figure~\ref{fig:grad_p}, we can see that the surfaces with $0.1<\phi<0.35$ have significant deformation with $\<g|\phi\> > 0.5 \<g|\phi\>_{max}$ at $t=108$.
In particular, the surface of $\phi=0.217$ (also see table~\ref{tab:material}) shown in figure~\ref{fig:all-2} is very close to the peaks of $\<g|\phi\>$ at $t=106$ and $t=108$.
It is interesting that this surface with the most intensive deformation
shows the generation of the signature hairpin-like structure.
Furthermore, the range of $\phi$ with significant deformations is broaden with time in figure~\ref{fig:grad_p},  {which implies that the elevation of the hairpin-like structures near $\phi=0.217$ can have an impact on the motion of neighboring material surfaces.}

As an additional note, the scalar gradient magnitude $g$ is essentially similar to the DLE field that was used to identify the hyperbolic LCS in \citet[]{Green2007}, which is discussed in \S\,\ref{sec:LCS}.
Compared with the contour lines of the DLE field on two-dimensional plane-cuts in \citet[]{Green2007} rather than the explicit identification of LCSs as parameterized
material surfaces \citep[]{Haller2015}, figures~\ref{fig:all-2} and \ref{fig:all-3} display the evolution of uniquely defined material surfaces in three-dimensional space in the transition.

\begin{figure}
\centering \subfigure{
\psfrag{X}{$t$}\psfrag{Y}{$\<|n_x|| \phi\>$}
\psfrag{A}{$\phi=0.024$}\psfrag{b}{$\phi=0.217$}\psfrag{c}{$\phi=0.433$}\psfrag{d}{$\phi=0.722$}\psfrag{e}{average}
\begin{minipage}[c]{0.48\textwidth}
\includegraphics[width=2.6in]{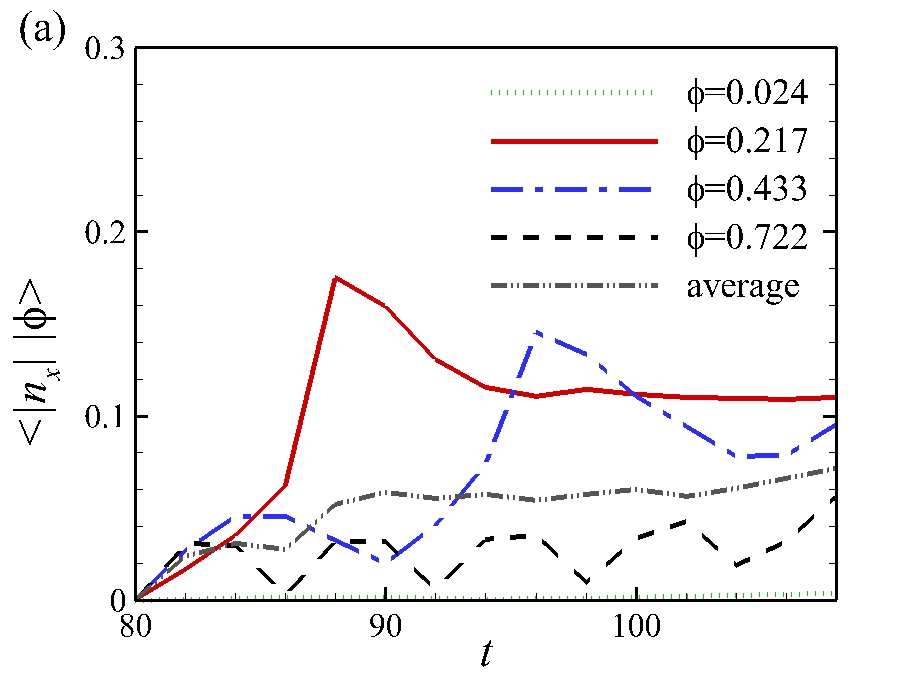}
\end{minipage}}
\centering \subfigure{
\psfrag{X}{$t$}\psfrag{Y}{$\<|n_y|| \phi\>$}
\psfrag{a}{$\phi=0.024$}\psfrag{B}{$\phi=0.217$}\psfrag{c}{$\phi=0.433$}\psfrag{d}{$\phi=0.722$}\psfrag{e}{average}
\begin{minipage}[c]{0.48\textwidth}
\includegraphics[width=2.6in]{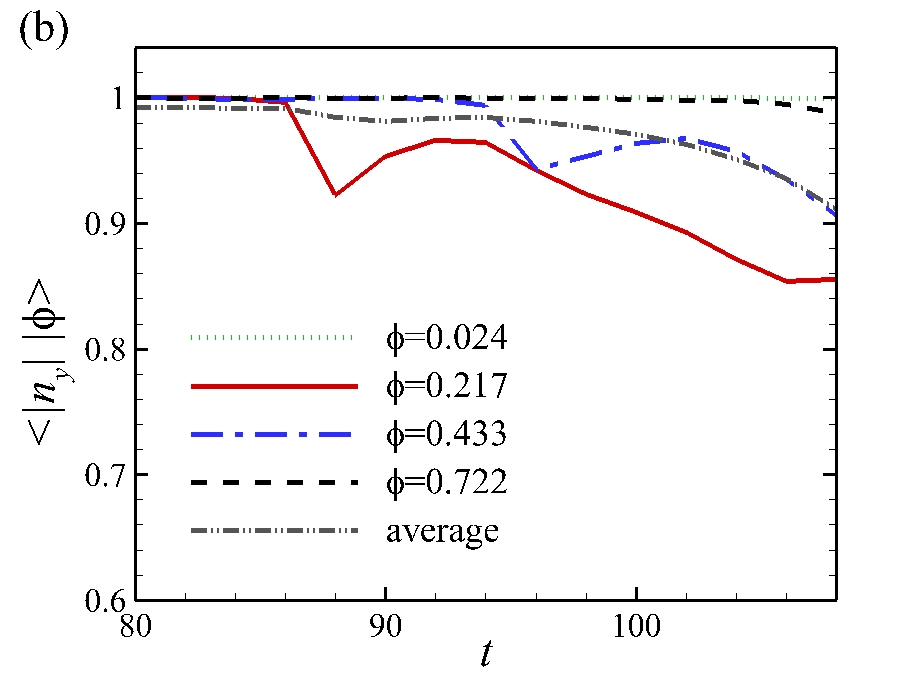}
\end{minipage}}

\centering \subfigure{
\psfrag{X}{$t$}\psfrag{Y}{$\<|n_z|| \phi\>$}
\psfrag{a}{$\phi=0.024$}\psfrag{b}{$\phi=0.217$}\psfrag{C}{$\phi=0.433$}\psfrag{d}{$\phi=0.722$}\psfrag{e}{average}
\begin{minipage}[c]{0.48\textwidth}
\includegraphics[width=2.6in]{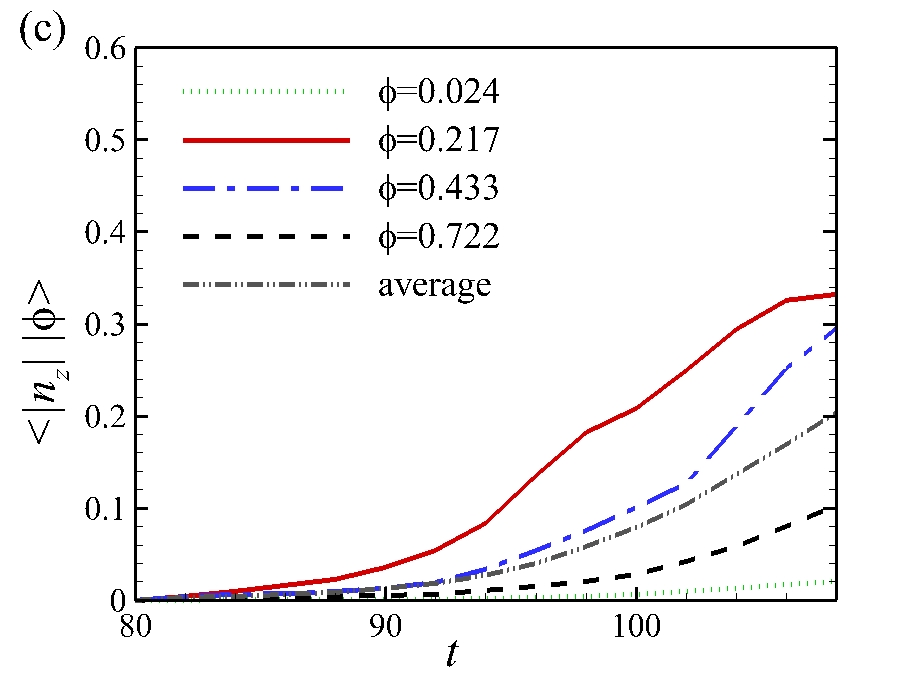}
\end{minipage}}

  \caption{(Colour online) Temporal evolution of the conditional average of normal components for material surfaces with different initial wall distances,
  and the volume averaged normal components,  {(a) absolute value of the streamwise component $|n_x|$; (b) absolute value of the wall-normal component $|n_y|$; (c) absolute value of the spanwise component $|n_z|$.}}
\label{fig:n}
\end{figure}

The conditional averaged $\boldsymbol{n}$ is used to characterize the geometry of the evolving material surfaces.
As shown in figure~\ref{fig:n}, the influential material surface of $\phi=0.217$ with the maximum deformation is selected to study the conditional averages of the absolute values of normal components $n_x$, $n_y$ and $n_z$, and other surfaces initially located at different wall distances with $\phi=0.024, 0.433$ and 0.722 are also presented for comparison.
With $\bs n=(0,1,0)$ for the initial planar surfaces, the averaged $|n_y|$ decreases from $|n_y(t_0)|=1$ at $t=t_0$, while both $|n_x|$ and $|n_z|$ grow from $|n_x(t_0)|=|n_z(t_0)|=0$.

In the evolution, the conditional average for the surface of $\phi=0.217$ is much earlier and more disturbed than those from other initial wall distances, which is similar to $\<g\>$.
In particular, $|n_x|$ amplifies much earlier than $|n_z|$ before the triangular bulge is generated in figure~\ref{fig:all-2}(b),
so the Lagrangian field is essentially two-component in terms of $x$ and $y$ before $t=85$ only with a slight increase of $|n_z|$.

Additionally, the generation of the triangular bulge quantified by the increase of $|n_x|$ and $|n_z|$ corresponds to the crucial event usually mentioned as `peak-valley splitting' \citep[]{Klebanoff1962,Kleiser1991} in transitional wall flows.
A qualitative analysis about the formation mechanism of the triangular bulge is presented in \S\,\ref{sec:triangular}.
It is noted that the amplitude of the initial two-dimensional TS wave is much larger than those of the three-dimensional TS waves in (\ref{Eq:TS}),
and the two-dimensional TS wave amplifies earlier than the three-dimensional ones \citep[]{Kleiser1991}.
Therefore, the flow is approximately quasi two-component until three-dimensional disturbances are amplified by the secondary instability \citep[]{Herbert1988}.
This transition can be represented by the formation of the triangular bulge for the most deformed surface with the evolution of $|n_x|$ and $|n_z|$.
Thus, the peak-valley splitting can be illustrated in a quantitative Lagrangian approach.

\subsection{Lagrangian elevation and descent}\label{sec:elevation}
The analysis presented in \S\,\ref{sec:geometry_surface} is on the evolutionary geometry of the material surfaces themselves, which is also related to other underlying flow physics such as momentum transport.
The contours presented in figure~\ref{fig:2d} show that the Lagrangian scalar is transported violently in the wall-normal direction.
Since the streamwise flow velocity is increasing from the wall at the initial state, the scalar transport, indicating the blending of rapid and slow motions from different wall distances, can be related to momentum transport in wall turbulence.

We define the wall-normal Lagrangian displacement as
\begin{equation}
\Delta Y(\bs x_0, t_0 | t)=Y(\bs x_0, t_0 | t)-y(\bs x_0, t_0 )=Y(\bs x_0, t_0 | t)-\phi(\bs X(\bs x_0, t_0 | t),t),
\label{Eq:DZ}
\end{equation}
where $Y$ is the wall-normal location of a fluid particle on a material surface at time $t$.
The displacement $\Delta Y$ quantifies the scalar transport in the wall-normal direction within a time interval of interest, and we define Lagrangian events `elevation' with $\Delta Y>0$ and `descent' with $\Delta Y<0$.

We can prove that the overall elevation/descent of a material surface is caused by the mean shear and vortical structures.
A material surface $\phi=\phi_0$ can be presented as a graph function $y_\phi=h(x,z)$, where $y_\phi$ is the coordinate of the surface in the wall-normal direction.
Since there is no mass flux through a material surface and the boundaries are periodic in the streamwise and spanwise directions, the volume of the fluid enclosed by two material surfaces, including $y_\phi$ and the rigid wall, and four periodic boundaries is conserved for constant-density flows.
Hence, $\<\Delta Y|\phi\>$ is always vanishing in a channel flow from the integration of $h(x,z)$ over the $x$-$z$ plane with a constant area, but only if $h(x,z)$ is a single-valued function.
In other words, the variation of $\<\Delta Y|\phi\>$ implies that $h(x,z)$ becomes multi-valued for the surface. It corresponds to folding or rolling up of material surfaces caused by the mean shear and vortical structures as shown in figures~\ref{fig:all-2} and \ref{fig:all-3}.

\begin{figure}
\centering \subfigure{
\begin{minipage}[c]{0.5\textwidth}
\psfrag{X}{$\phi$}\psfrag{Y}{$\<\Delta Y|\phi\>$}
\psfrag{A}{$t=108$}\psfrag{B}{$t=100$}
\includegraphics[width=2.6in]{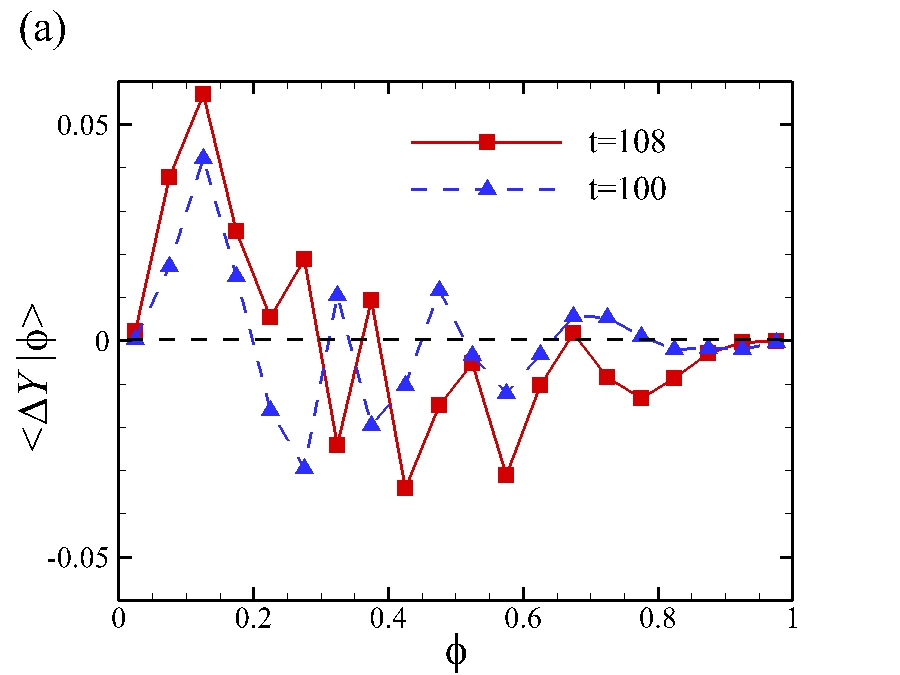}
\end{minipage}}%
\centering \subfigure{
\begin{minipage}[c]{0.5\textwidth}
\psfrag{X}{$\phi$}\psfrag{Y}{$\<\Delta u|\phi\>$}
\psfrag{A}{$t=108$}\psfrag{B}{$t=100$}
\includegraphics[width=2.6in]{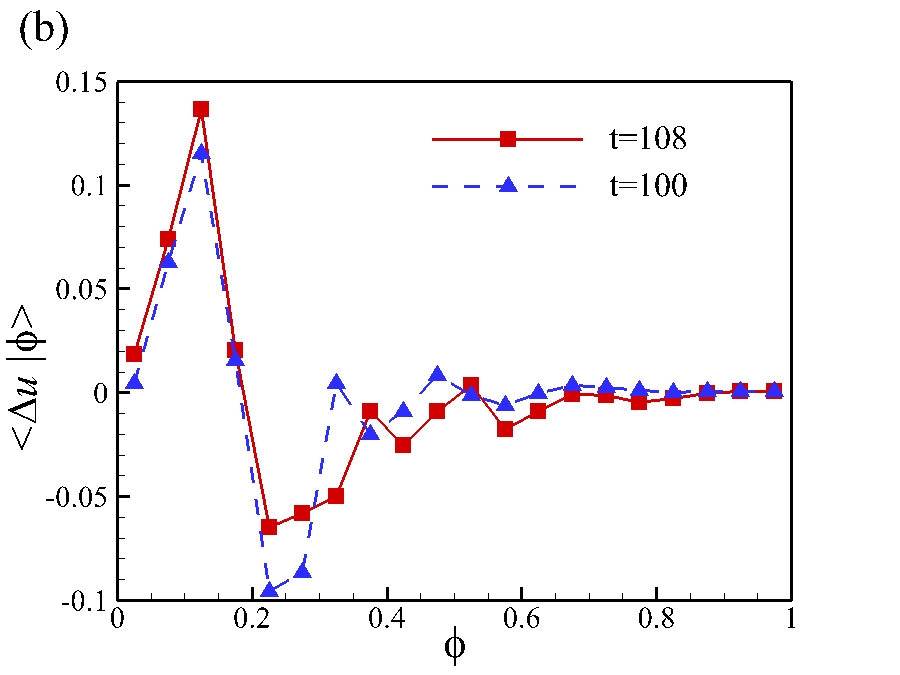}
\end{minipage}}%

  \caption{(Colour online) (a) Lagrangian wall-normal displacement $\langle \Delta Y | \phi\rangle$ conditioned on different material surfaces;
  (b) Lagrangian streamwise velocity difference $\langle \Delta u | \phi\rangle$  averaged on material surfaces.
}
\label{fig:delta_Z}
\end{figure}

As shown in figure~\ref{fig:delta_Z}(a), generally the fluid from the near-wall region is elevated from the wall, and the fluid close to the mid-plane descends toward the wall.
From the visualizations in figures~\ref{fig:all-2} and \ref{fig:2d}, the elevation is mainly caused by the hairpin-like structures with the intensive folding and rolling up of material surfaces.
Thus the elevation events are strong, and a peak of $\<\Delta Y|\phi\>$ appears near $\phi=0.125$.
In contrast, few of coherent patterns moving toward the wall are observed, and relatively weak descent events occur in a broad range of $0.3<\phi<0.9$.
We can also observe that the elevation of the region near $\phi=0.125$ is significantly intensified in the temporal evolution, which appears to be related to the elevation of the hairpin-like vortical structure in figure~\ref{fig:all-2}.

The elevation and descent events are closely related to ejections and sweeps. 
The ejection event and the sweep event indicate low-speed fluid moving away from the wall and high-speed fluid moving toward the wall, respectively \citep[]{Corino1969,Willmarth1972}.
To investigate the significance of  ejections and sweeps in turbulence production, these events are often studied by the Eulerian quadrant analysis of $u'v'$, where $u'$ and $v'$ are fluctuating velocities in the streamwise and wall-normal directions.
Both the ejection event in the second quadrant (Q2) and the sweep event in the fourth quadrant (Q4) are of importance of producing turbulent energy \citep[]{Adrian2007}.



 { 
Futhermore, the elevation and descent events can be directly observed in experiments using Lagrangian-type tracers such as hydrogen bubbles \citep[see e.g.][]{Hama1960,Hama1963,Lee2007,Lee2008,Guo2010}.
Since the experimental visualizations are based on the concentration of hydrogen bubbles that are initially released in a series of lines parallel or perpendicular to the wall, the bubbles are similar to material lines and their concentration can be described by the Lagrangian field $\phi$ with a particular initial condition as banded Heaviside functions on a plane. Compared with experimental visualizations, the visualization of $\phi$ with a smooth initial condition can provide the evolution of the entire collection of material surfaces and lines, and corresponding quantifications can be easily obtained from the conditional statistics in the scalar field.
}

\begin{figure}
\begin{center}
\includegraphics[width=0.95\textwidth]{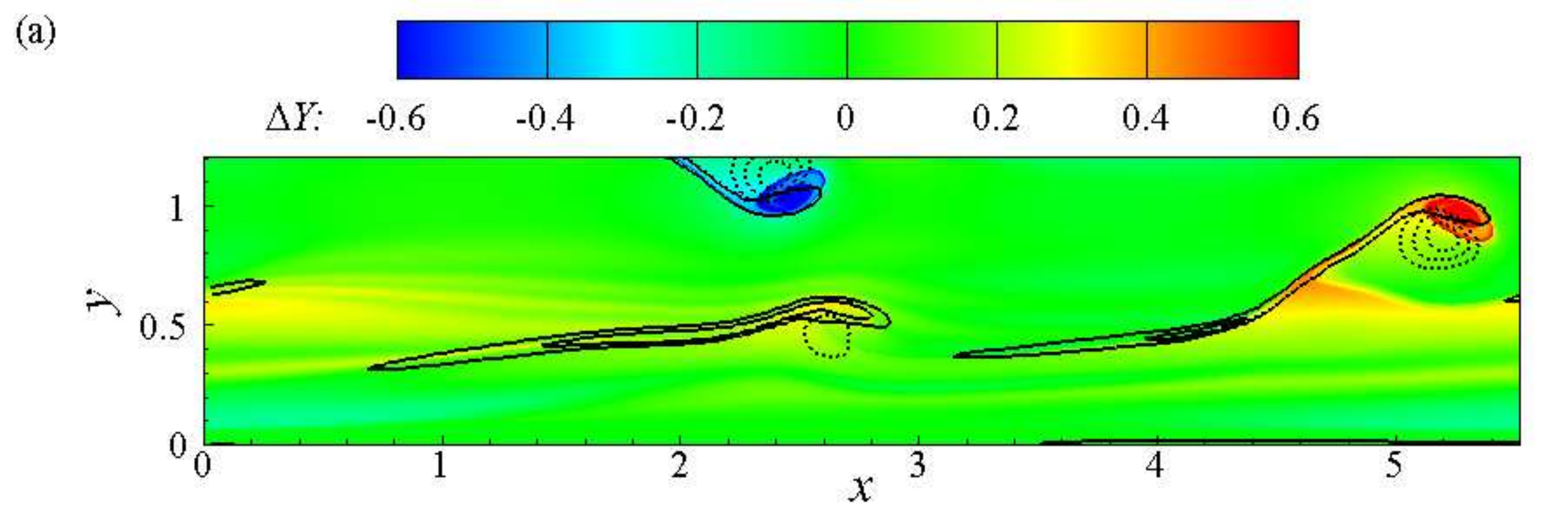}

\includegraphics[width=0.95\textwidth]{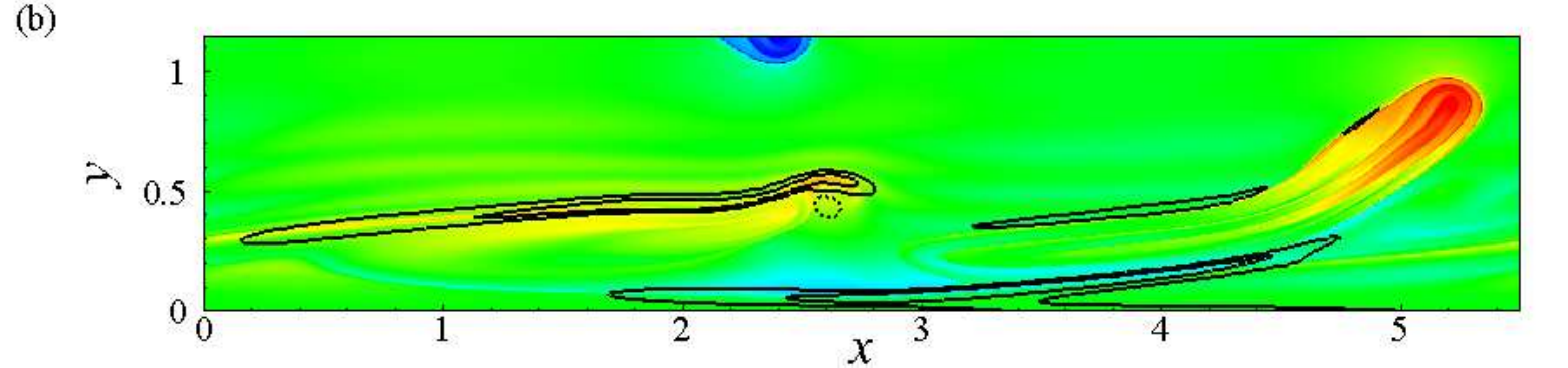}
  \caption{(Colour online)  Lagrangian wall-normal displacement $\Delta Y$ on the spanwise and wall-normal plane-cut at $t=108$, (a) at  {the peak position $z=L_z/2$; (b) $0.17\delta$ away from the peak position, the plane cut is on the `hairpin legs'.} 
  The solid lines represents contour lines of shear $\partial{u}/\partial{y}$, and the dashed lines represents contour lines of $u'v'$.}
\label{fig:2d-com}
\end{center}
\end{figure}

The distribution of $\Delta Y$ and the contour lines of large negative $u'v'$ with $v'>0$ on the  {peak plane} and another plane slightly aside from the peak position are shown in figure~\ref{fig:2d-com}, where $\Delta Y>0$ indicates the elevation and $u'v'<0$ represents the Q2 event in the lower half of channel.
In figure~\ref{fig:2d-com}(a), the region with the intensive Q2 event is mainly beneath the strong elevation region, which demonstrates that the elevation of low-speed fluid is related to the Q2 event.
 {The most intensive elevation is near the `head' of the primary hairpin-like structure (the dark red region around $x=5.2$) and the second most intensive elevation is near the triangular bulge and the secondary hairpin-like structure (red region from $x=1.0$ to $x=2.8$). This quantitatively support the observation that the ejection event caused by hairpin-like vortical structures is much stronger than that caused by the $\Lambda$-shaped structures in \citet[]{Guo2010}.

It is interesting that the elevation near the `head' of the hairpin-like structures is highly concentrated, and the nearby fluids tend to follow the motion of the `head'.
In contrast, the descent events are scattered and they are not as strong as the localized elevations in figure~\ref{fig:2d-com}, and they are quantified as the broad range of $\<\Delta Y |\phi\>< 0$ in figure~\ref{fig:delta_Z}(a).
}

In figure~\ref{fig:2d-com}(a), the strong elevation region is covered by the high-shear layer represented by the contour lines of large $\partial u/\partial y$.
It is noted that the instability of the high-shear layer is crucial in the transition process \citep[see \eg][]{Klebanoff1962,Gilbert1990}.
 {The quantified visualization in figure~\ref{fig:2d-com}(a) provides a direct proof for the relation between the high-shear layer and the elevation event.}
The high-shear layer can be generated between the elevated low-speed fluid and the surrounding high-speed fluid.
 {In figure~\ref{fig:2d-com}(b), there exists a high-shear layer close to the wall on the planes aside from the peak position, which accompanies the descent events near the `legs' of the primary hairpin-like structure.}

\begin{figure}
\centering \subfigure{
\begin{minipage}[c]{0.5\textwidth}
\psfrag{X}{$\Delta Y/\phi_0$}\psfrag{Y}{$\Delta u/U_0$}
\includegraphics[width=2.6in]{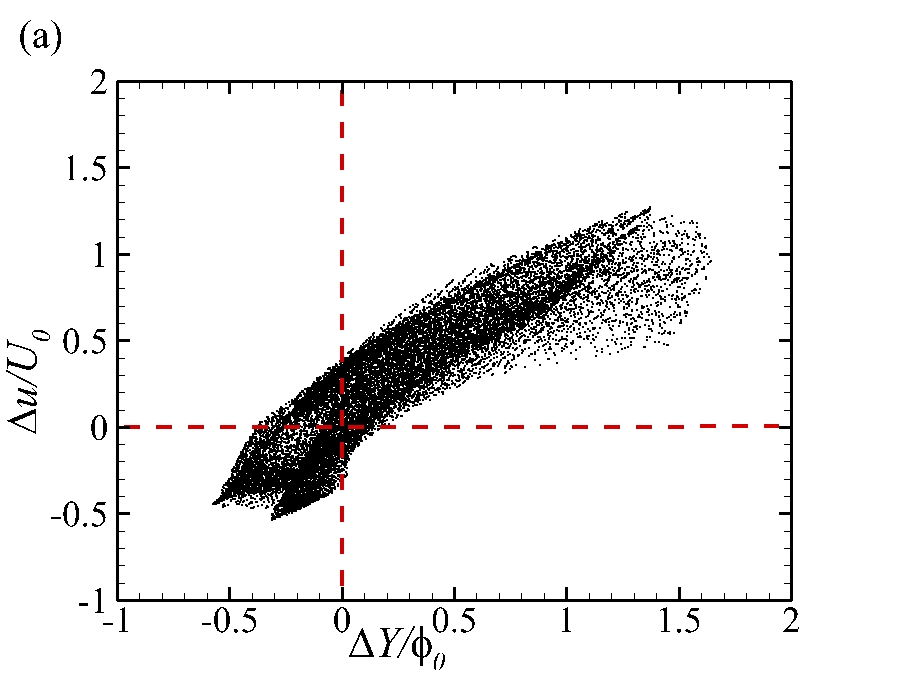}
\end{minipage}}%
\centering \subfigure{
\begin{minipage}[c]{0.5\textwidth}
\psfrag{X}{$\Delta Y/\phi_0$}\psfrag{Y}{$\Delta u/U_0$}
\includegraphics[width=2.6in]{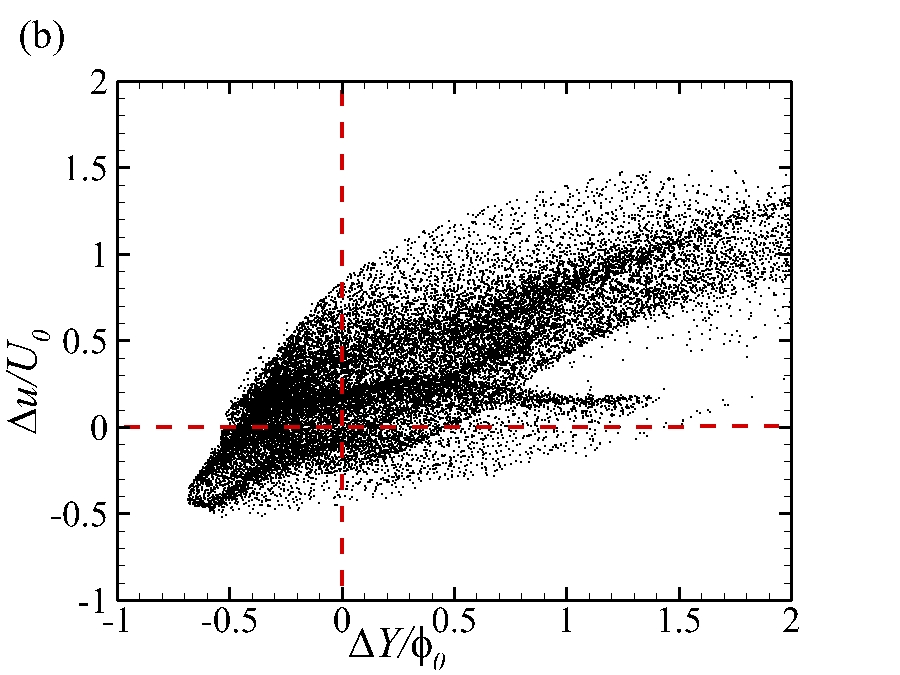}
\end{minipage}}%

  \caption{(Colour online) Scatter plots for Lagrangian displacement  $\Delta Y/\phi_0$ and Lagrangian streamwise velocity difference $\Delta u/U_0$ for the most elevated material surface with $\phi=0.125$, (a) $t=100, \rho_{Yu}=0.896$; (b)$t=108, \rho_{Yu}=0.780$.
}
\label{fig:scatter}
\end{figure}

The Lagrangian elevation and descent events are relevant to momentum transport.
 {\citet[]{Guo2010} found that the Lagrangian acceleration and deceleration are qualitatively related with the ejections and sweeps from the visualization of decelerating fluid streaks identified by the motion of hydrogen bubbles near vortical structures.}
We quantify the momentum transport using the Lagrangian streamwise velocity difference as
\begin{equation}
\Delta u(\bs X(\bs x_0, t_0 | t),t)=u(\bs X(\bs x_0, t_0 | t),t)-u(\bs x_0, t_0 ),
\label{Eq:Du}
\end{equation}
which represents the fluid acceleration/deceleration in the streamwise direction.
In figure~\ref{fig:delta_Z}, the profile of $\langle \Delta u | \phi\rangle$ is qualitatively similar to $\langle \Delta Y | \phi\rangle$, in particular, the most elevated material surface of $\phi=0.125$ is most accelerated.
The correlations of $\Delta Y/\phi_0$ and $\Delta u/U_0$ for the material surface $\phi=0.125$ at $t=100$ and $t=108$ are shown as scatter plots in figure~\ref{fig:scatter}.
We can see that $\Delta Y$ and $\Delta u$ are positively correlated.
In the scatter plots, the elevated fluid or the descending fluid tend to be accelerated or decelerated, respectively.
The correlation coefficient $\rho_{Yu}=0.896$ is high at $t=100$, and decreases with time because the mean velocity is no longer monotonically increased with $y$ during the transition.


\subsection{Vorticity growth and distribution}\label{sec:vor_distribution}
It is widely accepted that the vortical structures play key roles in the momentum and scalar transport \citep[]{Adrian2007,Wallace2012},
so the evolutionary geometry of material surfaces is affected by the mean shear and the vortices.
On the other hand, the Lagrangian evolution of the surfaces can also characterize the evolution of vortical structures themselves (discussed in \S\,\ref{sec:approx_vortex_surface}) and it appears to more natural to capture the growth of vorticity than Eulerian methods.

\begin{figure}
\centering \subfigure{
\begin{minipage}[c]{0.5\textwidth}
\psfrag{X}{$\phi$}\psfrag{Y}{$\<\Delta_L|\bomega||\phi\>$}
\psfrag{A}{$t=108$}\psfrag{B}{$t=100$}\psfrag{C}{$t=108$}\psfrag{D}{$t=100$}
\includegraphics[width=2.6in]{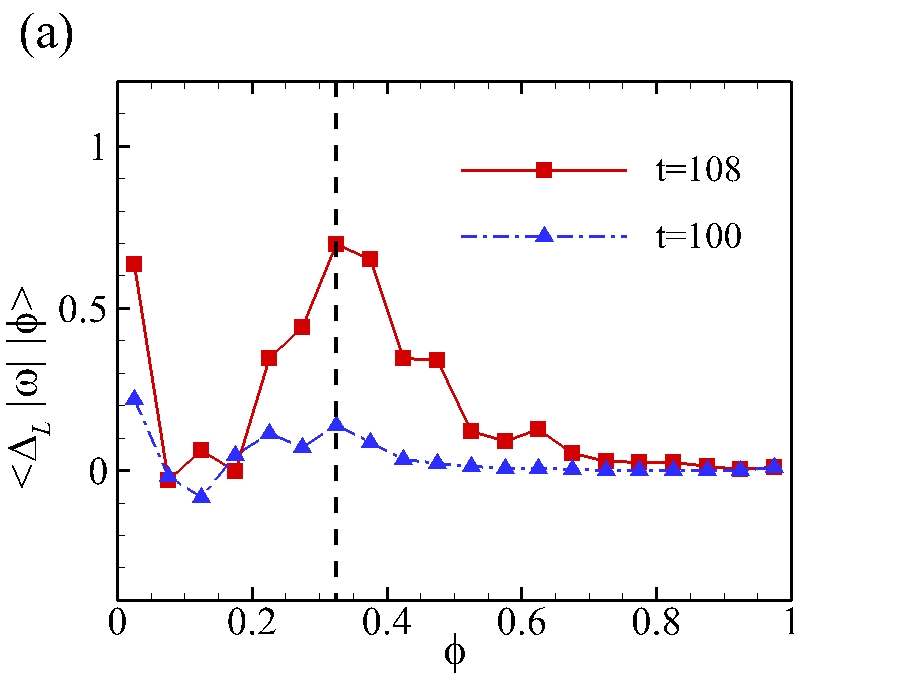}
\end{minipage}}%
\centering \subfigure{
\begin{minipage}[c]{0.5\textwidth}
\psfrag{X}{$z$}\psfrag{Y}{$\<\Delta_E|\bomega||y\>$}
\psfrag{A}{$t=108$}\psfrag{B}{$t=100$}\psfrag{C}{$t=108$}\psfrag{D}{$t=100$}
\includegraphics[width=2.6in]{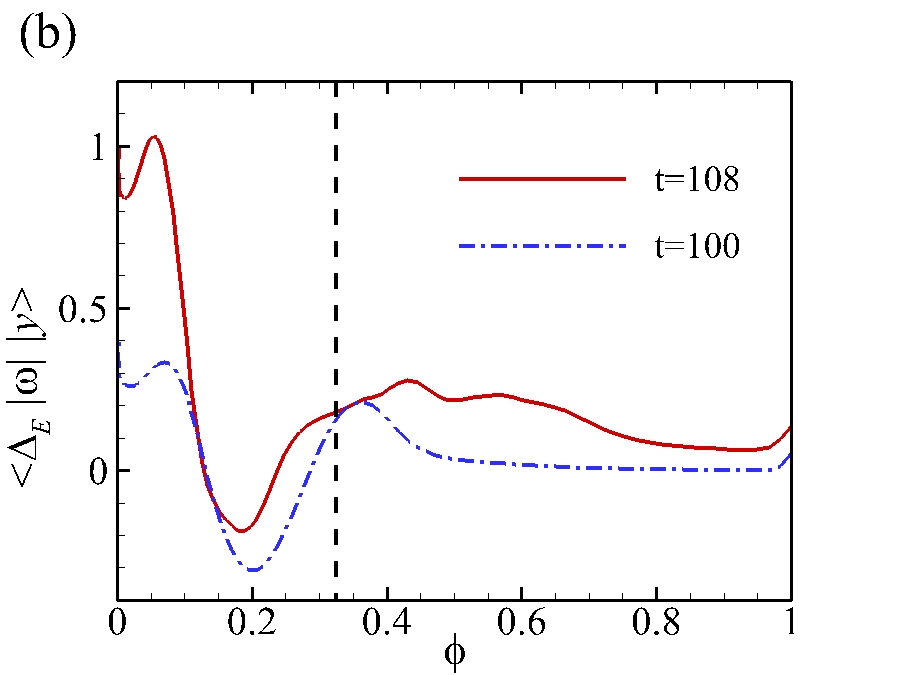}
\end{minipage}}%

  \caption{(Colour online) Change of vorticity magnitude, (a) averaged on material surfaces $\<\Delta_L|\bomega||\phi\>$; (b) averaged on on Eulerian planes $\<\Delta_E|\bomega||y\>$. The location of the peak of Lagrangian vorticity accumulations is indicated by the dashed line.
}
\label{fig:sta_w}
\end{figure}

The conditional averages of the change of the vorticity magnitude are calculated over both material surfaces with different initial wall distances and fixed Eulerian $x$--$z$ planes at different $y$-coordinates at $t=100$ and $t=108$.
The averaged Lagrangian difference $\Delta_L|\bomega|\equiv|\bomega(\bs X(\boldsymbol{x}_0,t_0| t),t)|-|\bomega(\boldsymbol{x}_0,t_0)|$ conditioned on $\phi$ and Eulerian difference $\Delta_E|\bomega|\equiv|\bomega(\bs x,t)|-|\bomega(\bs x,t_0)|$ conditioned on $y$ are shown in figure~\ref{fig:sta_w} for comparison.
Since the material surfaces close to the wall are hardly disturbed, both $\<\Delta_L|\bomega||\phi\>$ and $\<\Delta_E|\bomega||y\>$ show the similar vorticity accumulation in the near-wall region. The growth of $|\bomega|$ corresponds to the increase of the wall shear stress $\tau_w=\rho \nu \partial{u}/ \partial{y}$ at $y=0$, which is usually observed in the laminar-turbulent transition.

Unlike the Eulerian result, the Lagrangian vorticity accumulation has an obvious peak near $\phi=0.325$ at $t=108$. We analyze the terms in the governing equation of $|\bomega|$
\begin{equation}
   \frac{\mathrm{D}|\boldsymbol{\omega}|}{\mathrm{D}t}=
   (\boldsymbol{t}_\omega\bcdot\mathsfbi{S}\bcdot\boldsymbol{t}_\omega)|\boldsymbol{\omega}|
   +\nu\boldsymbol{t}_\omega\bcdot\bnabla^2\boldsymbol{\omega}
  \label{Eq:mag_w}
\end{equation}
to investigate the vorticity transport, where $\boldsymbol{t}_\omega=\boldsymbol{\omega}/|\boldsymbol{\omega}|$ is the unit vector for vorticity.
In the right hand side, the transport of $|\bomega|$ is governed by the stretching term $(\boldsymbol{t}_\omega\bcdot\mathsfbi{S}\bcdot\boldsymbol{t}_\omega)|\boldsymbol{\omega}|$ and the viscous diffusion term $\nu\boldsymbol{t}_\omega\bcdot\bnabla^2\boldsymbol{\omega}$.  {Although the two terms involve non-local effects on the trajectory of a fluid particle, the qualitative relation between the different mechanisms and the growth of $|\bomega|$ can be explained by the conditional statistics.}
In the near-wall region, the vorticity is generated through the boundary vorticity flux \citep[]{Lighthill1963,Wu2005}, and is diffused away from the high vorticity region by the viscous term.

\begin{figure}
\begin{center}
\includegraphics[width=0.95\textwidth]{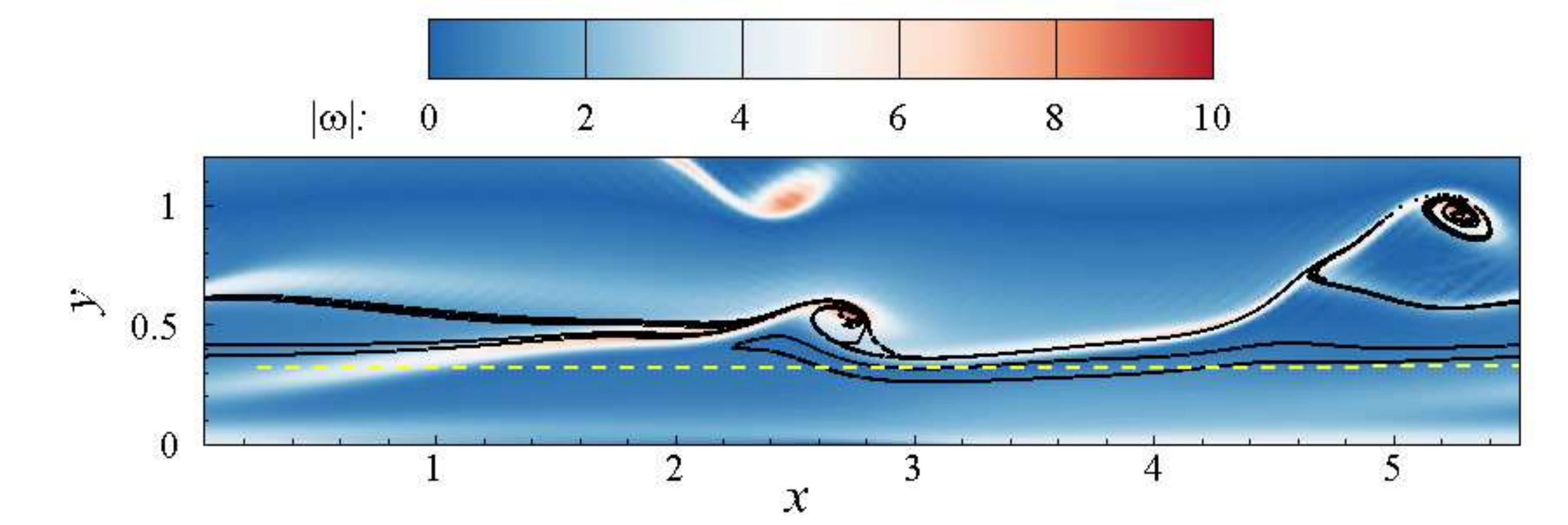}
  \caption{(Colour online)  Vorticity distribution on the spanwise and wall-normal plane-cut at  {the peak position} $z=L_z/2$ at $t=108$. The black solid lines represents the material surfaces around $\phi=0.325$, and the yellow dashed line represents the Eulerian plane at $y=0.325$.}
\label{fig:2d-ww}
\end{center}
\end{figure}

The major mechanism for the vorticity intensification remote from the wall is through the vorticity stretching term $(\boldsymbol{t}_\omega\bcdot\mathsfbi{S}\bcdot\boldsymbol{t}_\omega)|\boldsymbol{\omega}|$.
The distribution of $|\bomega|$ on the streamwise and wall-normal plane-cut at  {the peak position} $z=L_z/2$ at $t=108$ is presented in figure~\ref{fig:2d-ww}.
We can see that the vorticity concentrates in ribbon-like regions, and the distribution appears to resemble the high-shear layers in figure~\ref{fig:2d-com}(a),
because the shear $\partial u/\partial y$ is the dominating component in $\omega_z$.
Several material surfaces around $\phi=0.325$ with the maximum Lagrangian vorticity growth $\Delta_L|\bomega|$ are warped in the high vorticity region in figure~\ref{fig:2d-ww}, while their Eulerian counterpart, the plane at $y=0.325$, does not capture this strong vorticity growth and the corresponding vortex stretching mechanism in figure~\ref{fig:sta_w}.

In order to shed light on the vorticity growth and the evolutionary geometry of vortex lines, the vorticity components averaged on material surfaces are analyzed.
Both the streamwise and the wall-normal components of vorticity are strongly intensified in the evolution.
The peak of $|\omega_x|$ and $|\omega_y|$ for the surfaces with $\phi\approx0.2$ at $t=100$, which are close to the most deformed material surface, suggests that the triangular bulge in the evolution of the material surface with $\phi=0.217$ is related to the inclined quasi-streamwise vortices.
In addition, the migration of the peak of $|\omega_y|$ from $\phi=0.2$ at $t=100$ to $\phi=0.375$ at $t=108$ appears to be relevant to the elevation of the hairpin-like structures in figure~\ref{fig:all-3}.
Further explanations on the mechanism of the intensification of $|\omega_x|$ and $|\omega_y|$ are discussed in \S\,\ref{sec:stage2}.
Since the dominant term of $\omega_z$ is the shear $\partial{u}/\partial{y}$, the change of the spanwise vorticity $\Delta_L|\omega_z|\equiv|\omega_z(\bs X(\boldsymbol{x}_0,t_0| t),t)|-|\omega_z(\boldsymbol{x}_0,t_0)|$ averaged over material surfaces is related with the momentum transport as indicated in figure~\ref{fig:delta_Z}, and it decreases significantly around $\phi=0.2$. 

\begin{figure}
\centering \subfigure{
\begin{minipage}[c]{0.48\textwidth}
\psfrag{X}{$\phi$}\psfrag{Y}{$\<|\omega_x||\phi\>$}
\psfrag{A}{$t=108$}\psfrag{B}{$t=100$}
\includegraphics[width=2.6in]{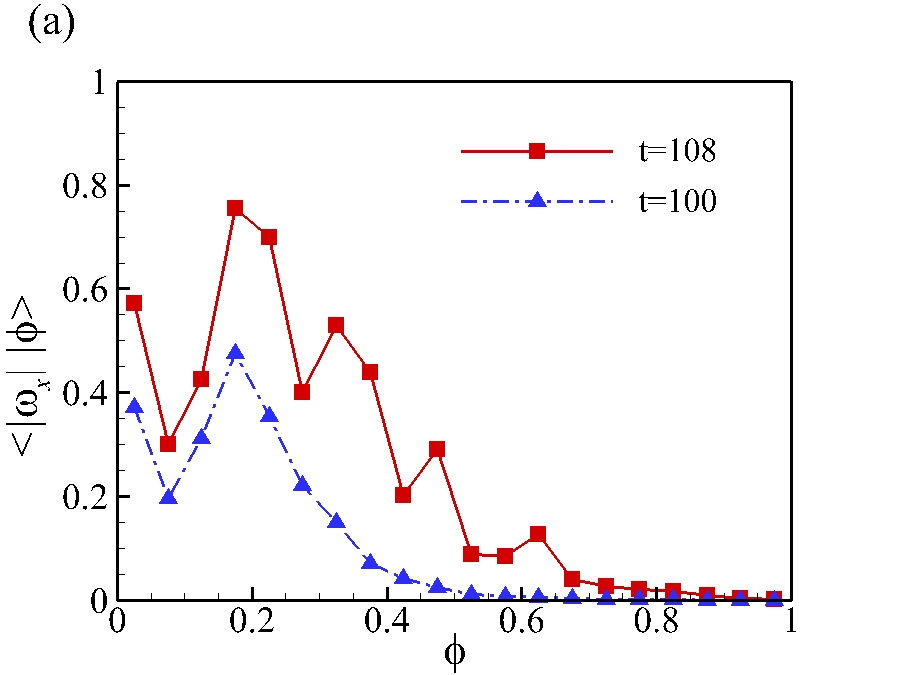}
\end{minipage}}%
\centering \subfigure{
\begin{minipage}[c]{0.48\textwidth}
\psfrag{X}{$\phi$}\psfrag{Y}{$\<|\omega_y||\phi\>$}
\psfrag{A}{$t=108$}\psfrag{B}{$t=100$}
\includegraphics[width=2.6in]{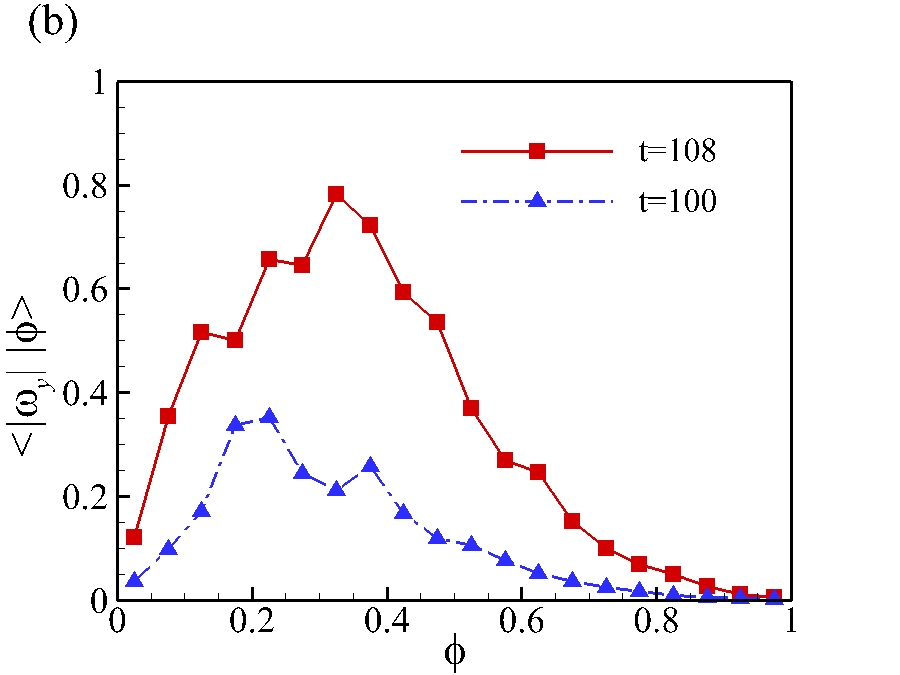}
\end{minipage}}%

\centering \subfigure{
\begin{minipage}[c]{0.48\textwidth}
\psfrag{X}{$\phi$}\psfrag{Y}{$\<\Delta_L|\omega_z||\phi\>$}
\psfrag{A}{$t=108$}\psfrag{B}{$t=100$}
\includegraphics[width=2.6in]{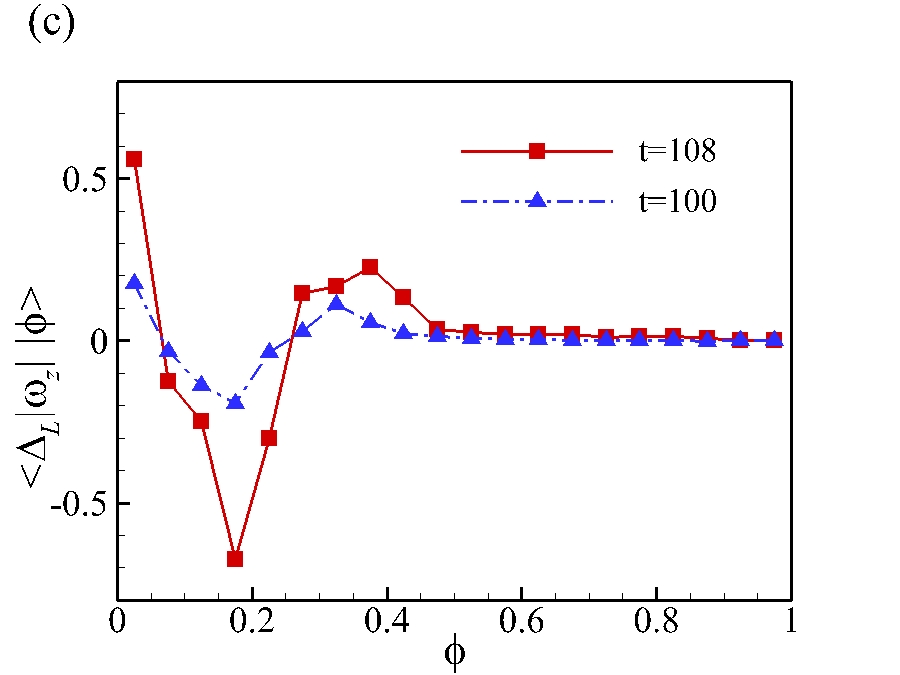}
\end{minipage}}%
\centering \subfigure{
\begin{minipage}[c]{0.48\textwidth}
\psfrag{X}{$\phi$}\psfrag{Y}{$\<\kappa|\phi\>$}
\psfrag{A}{$t=108$}\psfrag{B}{$t=100$}
\includegraphics[width=2.6in]{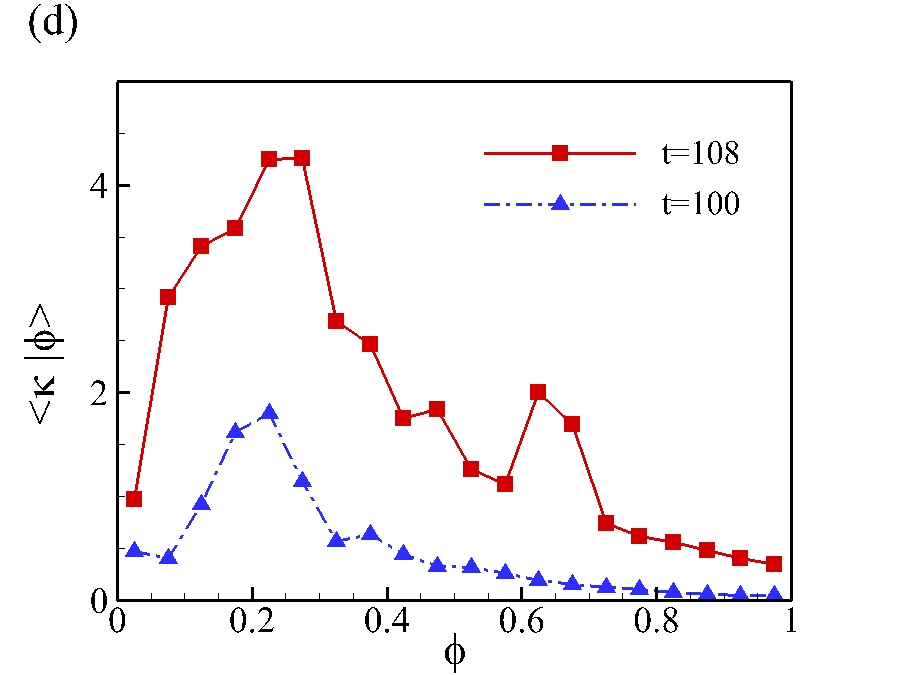}
\end{minipage}}%
  \caption{(Colour online) Vorticity components and the curvature of vortex lines averaged on material surfaces,  {(a) the absolute value of streamwise vorticity $|\omega_x|$; (b) the absolute value of wall-normal vorticity $|\omega_y|$; (c) the Lagrangian difference of absolute value of spanwise vorticity $\Delta_L|\omega_z|$; (d) the local curvature of vortex lines $\kappa$.}
}
\label{fig:wx}
\end{figure}

Although the Lagrangian change of the vorticity magnitude $\Delta_L|\bomega|$ averaged over the material surfaces around $\phi=0.2$ is small in figure~\ref{fig:sta_w},
all of the vorticity components have significant variations on the surfaces in figure~\ref{fig:wx}.
This implies that the deformation of the material surfaces around $\phi=0.2$ is accompanied by intensive stretching and folding of vortex lines with high curvatures.
The local curvature of vortex lines can be calculated as
\begin{equation}
   \kappa=\frac{|\boldsymbol{\omega}\times(\boldsymbol{\omega}\bcdot\bnabla\boldsymbol{\omega})|}{|\boldsymbol{\omega}|},
  \label{Eq:kappa}
\end{equation}
and the averaged curvatures $\<\kappa|\phi\>$ conditioned over material surfaces are shown in figure~\ref{fig:wx}(d).
The curvatures of vortex lines increase for all the surfaces, which indicates that the vortex lines evolve from nearly straight lines into curved lines. In particular, the peak of $\<\kappa|\phi\>$ is near $\phi=0.23$, suggesting that vortex lines are most curved on these labelled material surfaces during the transition.



\subsection{Brief summary and discussion}
From three different criteria and perspectives, the influential material surfaces are identified and characterized. 
First, the most deformed material surface near $\phi=0.225$ shows the signature shapes as the triangular bulge and the hairpin-like structure,
and its evolution provides a quantitative Lagrangian presentation of the peak-valley splitting.
Second, the maximum elevation of the material surface near $\phi=0.125$ is related to ejections, momentum transport and the formation of high-shear layers.
Finally, the material surface with the largest vorticity growth is around $\phi=0.325$, but the largest curvature of vortex lines and variations for three vorticity components, in general, occur near $\phi=0.2$, which implies that the vortex lines become highly curved on the surface with $\phi\approx0.2$.

 {Therefore, although the whole flow field is filled with an infinite number of material surfaces, the influential material surface of $\phi=0.217$ or $\phi^+=45$ with the maximum deformation as well as the maximum curvature of vortex lines and variation of vorticity components, is selected as a representative structure to study the structural evolution in the K-type transitional flow in the next section. It has been justified that the surfaces close to this influential surface show similar dynamical behaviours in this section, and the surfaces remote from the representative one, namely the surfaces that are close to the wall or near the central region, are only slightly disturbed in the evolution and provide little understanding on flow dynamics as discussed in \S\,\ref{sec:L-surface}.}

\section{Evolution of influential material surfaces}\label{sec:5}
In \S\S\,\ref{sec:3} and \ref{sec_influential}, the characterization of identified influential material surfaces can provide quantifications on the structural evolution and signature events in the transition from a Lagrangian perspective.
Although material surfaces with arbitrary initial conditions cannot reveal unambiguous turbulent dynamics, an explicit connection between Lagrangian kinematics and dynamics can be educed by the influential material surface with a physically interesting initial condition, presently the vortex sheet, in transitional wall flows.
In this section, we will use the material surface as a surrogate of the vortex surface to investigate dynamic mechanisms in the  {late transition}.

\subsection{Approximation of vortex surfaces}\label{sec:approx_vortex_surface}
As remarked in \citet[]{yang2010lagrangian},
the material surface can be considered as a good approximation for the vortex surface
before significant topological changes in high-Reynolds-number flows.
As mentioned in \S\,\ref{sec:initial_scalar},
the initial material surface of $\phi_0=y_0$ is a good approximation of the vortex surface as sketched in figure~\ref{fig:1}.

The cosine $\lambda_\omega\equiv\boldsymbol{t}_\omega\bcdot\boldsymbol{n}$ of the angle between the vorticity and the Lagrangian scalar gradient
can be used to quantify the local deviation between an iso-surface of $\phi$ and a vortex surface \citep[see][]{yang2010multi, yang2011vsf}.
As shown in figure~\ref{fig:diagram_n}, if $\boldsymbol{\omega}$ is perpendicular to $\boldsymbol{n}$ in the whole field,
the material surfaces completely coincide with the vortex surfaces with $\langle\lambda_\omega\rangle=0$.

\begin{figure}
\psfrag{a}{$t$} \psfrag{b}{$\langle\lambda_\omega\rangle$}
\begin{center}
 \includegraphics[width=0.6\textwidth]{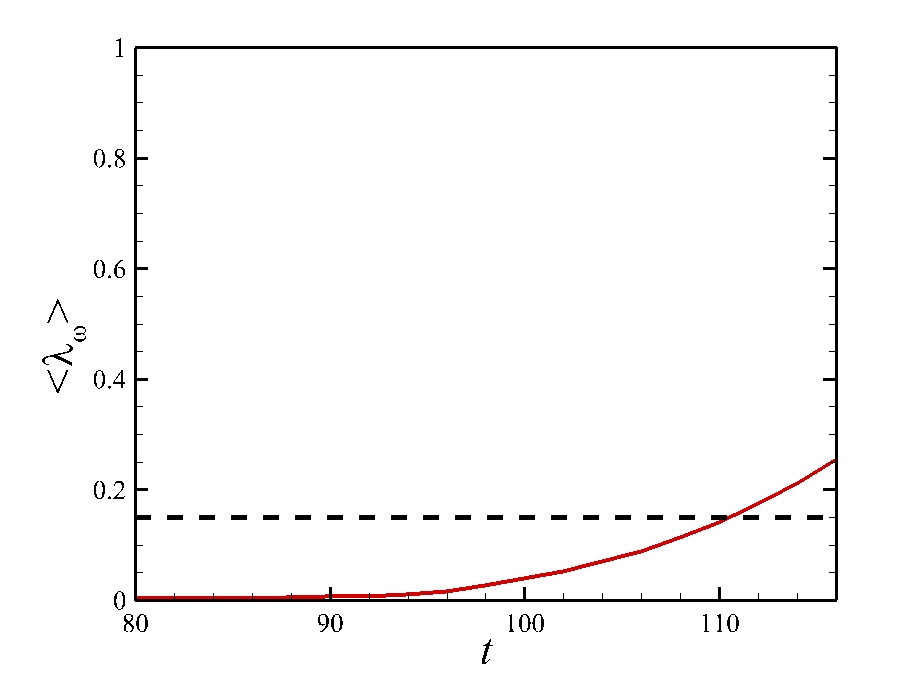}
  \caption{(Colour online) Temporal evolution of $\langle\lambda_\omega\rangle$, where $\lambda_\omega$ is the cosine of the angle
  between $\boldsymbol{\omega}$ and $\bnabla\phi$. The dashed line shows the $15\%$ level. \protect\\}
\label{fig:l_w}
\end{center}
\end{figure}

Temporal evolution of $\langle\lambda_\omega\rangle$ calculated from the present transitional channel flow is shown in figure~\ref{fig:l_w}.
For strictly inviscid incompressible flow with conservative body forces,
the Helmholtz vorticity theorem shows that the material surfaces that are vortex surfaces at $t=0$ remain so for $t>0$.
Thus $\langle\lambda_\omega\rangle=0$ can be satisfied for $t>0$ with $\langle\lambda_\omega\rangle=0$ at $t=0$ in inviscid flow.
In a viscous flow, $\langle\lambda_\omega\rangle$ slowly increases owing to the viscous effect and the breakdown of the Helmholtz theorem in figure~\ref{fig:l_w}.
The averaged deviation $\langle\lambda_\omega\rangle$, however, remains  { less than the $15\%$ level for $t<110$} in the present transitional flow. Therefore, we can reasonably assume that vortex surfaces are still well approximated by the material surfaces before significant topological changes of vortical structures in the transition.

It is noted that the quantitative results of material surfaces in the present study are restricted to the evolution during $80\leq t\leq108$ with $\langle\lambda_\omega\rangle\leq15\%$ to ensure the good approximation of vortex surfaces.
The whole process in the evolution of the influential material surface will be roughly divided into three stages as follows.


\subsection{Stage 1, triangular bulge}


 {In the evolution of vortical structures in transitional wall flows, the so-called `$\Lambda$-vortex' is frequently mentioned and often considered as a prior stage of hairpin-like strictures \citep[\eg][]{Sandham1992,Bake2002,Borodulin2002a,Guo2010,Sayadi2013}.}
 {Controversies, however, still exist about the definition and existence of the `$\Lambda$-vortex'. For example, \citet[]{Kleiser1985} found that the $\Lambda$-like structure is represented by the dislocation of vortex lines within a $\Lambda$-shaped region from the evolution of material lines in transitional plane Poiseuille flow. Similar observations were found in the transitional boundary layer, and \citet[]{Liu2014} concluded that the `$\Lambda$-vortex tubes' do not exist because the vortex lines can penetrate the $\Lambda$-shaped tubes visualized by the Eulerian vortex identification criterion.}
 {
\citet[]{Bernard2011,Bernard2013} pointed out that it is not reasonable to consider the coherent structures simply as rotational regions. Alternatively, the `vortex furrows' evolving from spanwise-aligned vortex filaments were proposed as the principal structural element in the K-type transition.
Therefore, the disagreement on the $\Lambda$-shaped vortical structure is partially caused by the different methods used to characterize the coherent structures.
In the present study, we seek the essence of the $\Lambda$-shaped structure using material surfaces along with almost attached vortex lines and the vorticity magnitude on the lines, which combines the strengths of different identification methods.}


\begin{figure}
\centering \subfigure{
\begin{minipage}[c]{0.5\textwidth}
\includegraphics[width=1\textwidth]{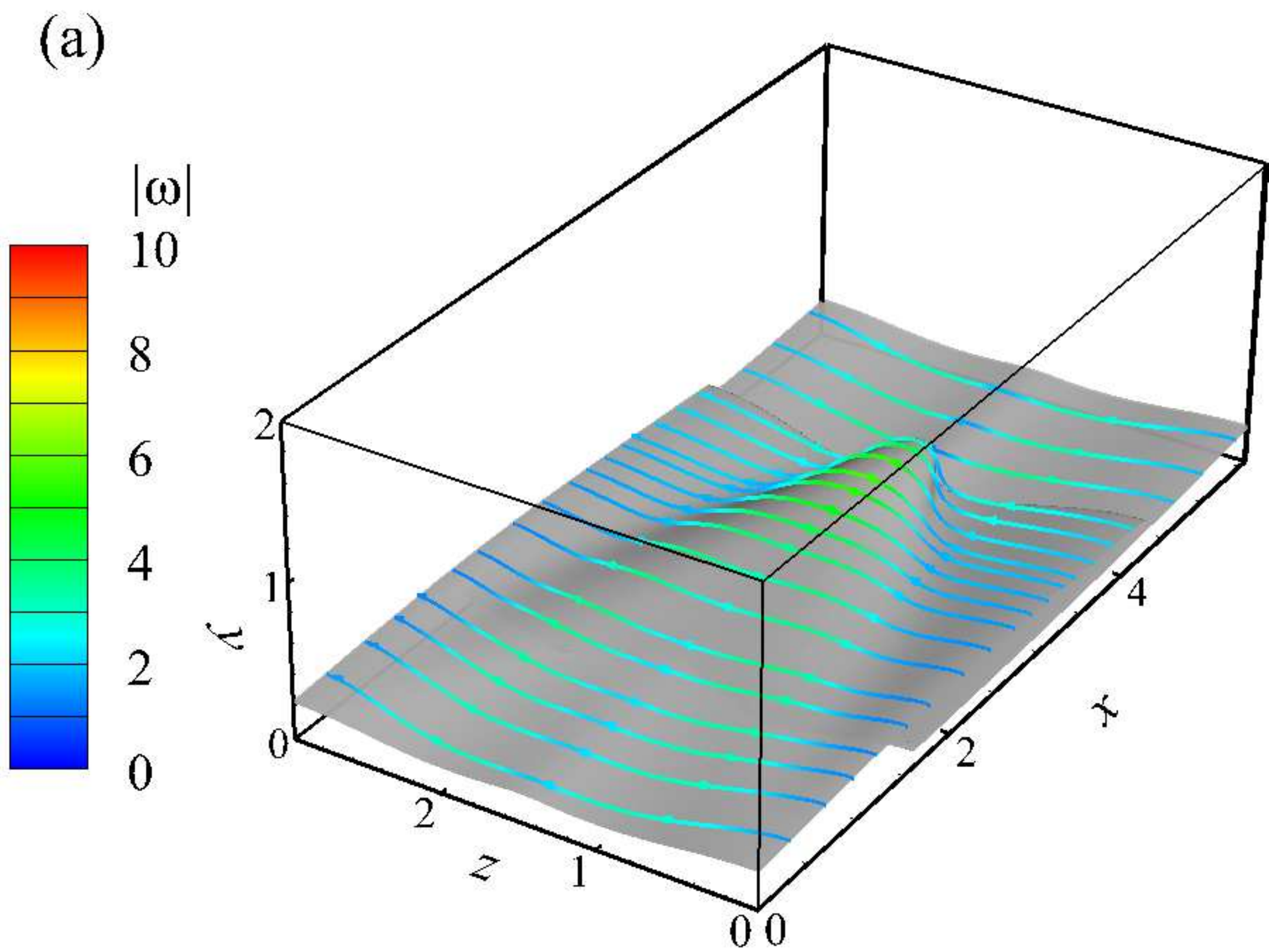}
\end{minipage}}%
  \centering \subfigure{
\begin{minipage}[c]{0.5\textwidth}
\includegraphics[width=1\textwidth]{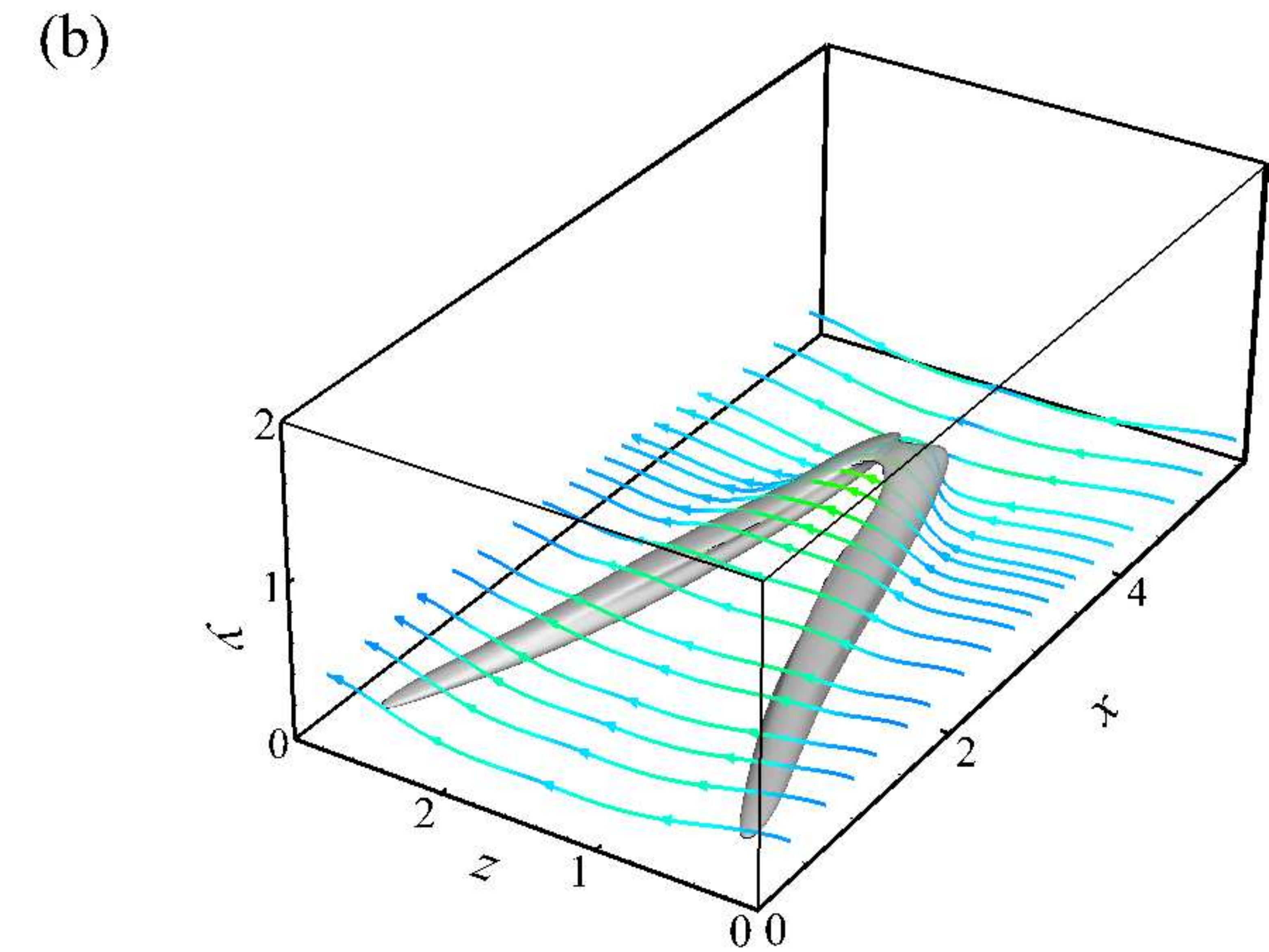}
\end{minipage}}%
  \caption{(Colour online) Comparison of Lagrangian and Eulerian vortical structures at $t=100$.
Vortex lines are integrated from the surfaces, and colour-coded by the magnitude of vorticity $|\bs \omega|$.
  (a) material surface of $\phi=0.217$, (b) iso-surface of the swirling strenth $\lambda_{ci}$ (50\% of the maximum value).\protect\\}
\label{fig:h100}
\end{figure}

The influential material surface of $\phi=0.217$ or $\phi^+=45$, which has the maximum deformation in a period of interest as discussed in \S\,\ref{sec:geometry_surface}, shows a triangular bulge at $t=100$ in figure~\ref{fig:h100}(a).
The formation of this bulge can be qualitatively analyzed  {from the evolution of material surfaces in \S\,\ref{sec:triangular} instead of the wave interactions \citep[see][]{Herbert1988,Kachanov1994}.}
In addition, the vortex lines, colour-coded by the magnitude of vorticity, are integrated from the material surface, and they are almost on the surface. This demonstrates that the material surface can be considered as a surrogate of the vortex surface.

For comparison,  {the Eulerian structures are extracted by the iso-surfaces of the swirling strength $\lambda_{ci}$ that is the imaginary part of the complex eigenvalue of the velocity gradient tensor \citep[]{Zhou1999}.
The $\Lambda$-like structure identified in the present transitional channel flow is shown in figure~\ref{fig:h100}(b).}
At the mean time, vortex lines are also drawn near the the iso-surface of $\lambda_{ci}$, but
tube-like vortical structures cannot be identified from the vortex lines. Other Eulerian vortex identification criteria based on the velocity gradient, such as the $Q$- and $\lambda_2$-criteria, show the similar result (not shown).
It is interesting to observe that the ridge of the Lagrangian triangular bulge  {covered with folding vortex lines} spatially coincides with the Eulerian $\Lambda$-like structure,  {and a high-shear layer with large magnitudes of vorticity is generated on the top of the bulge \citep[]{Rist1995}.}


 {Therefore, the conceptual $\Lambda$-shaped vortex tubes composed of vortex lines do not exist at this stage, and the $\Lambda$-like structure from the Eulerian identification criterion can be misinterpreted without caution.} Moreover, the generation mechanism of the $\Lambda$-like structure is not easy to be elucidated and traced from the initial laminar state using the Eulerian methods applied on the structures that are not uniquely defined at different time instants.

 {From the Lagrangian study, the triangular bulge, which consists of $\Lambda$-shaped swirling regions at two ridges and the triangular high-shear layer on the top, is identified as the representative structure in this stage. }
Furthermore, from $|\bs \omega|$ on the vortex lines drawn in figure~\ref{fig:h100}, there is no obvious vorticity  {concentration} region near the triangular bulge.
This indicates that the mean shear, rather than the vortical structures with obvious swirling motions, is still dominating in the evolution of material surfaces at this stage.

\subsection{Stage 2, vorticity intensification}\label{sec:stage2}

\begin{figure}
\centering \subfigure{
\begin{minipage}[c]{0.5\textwidth}
\includegraphics[width=2.6in]{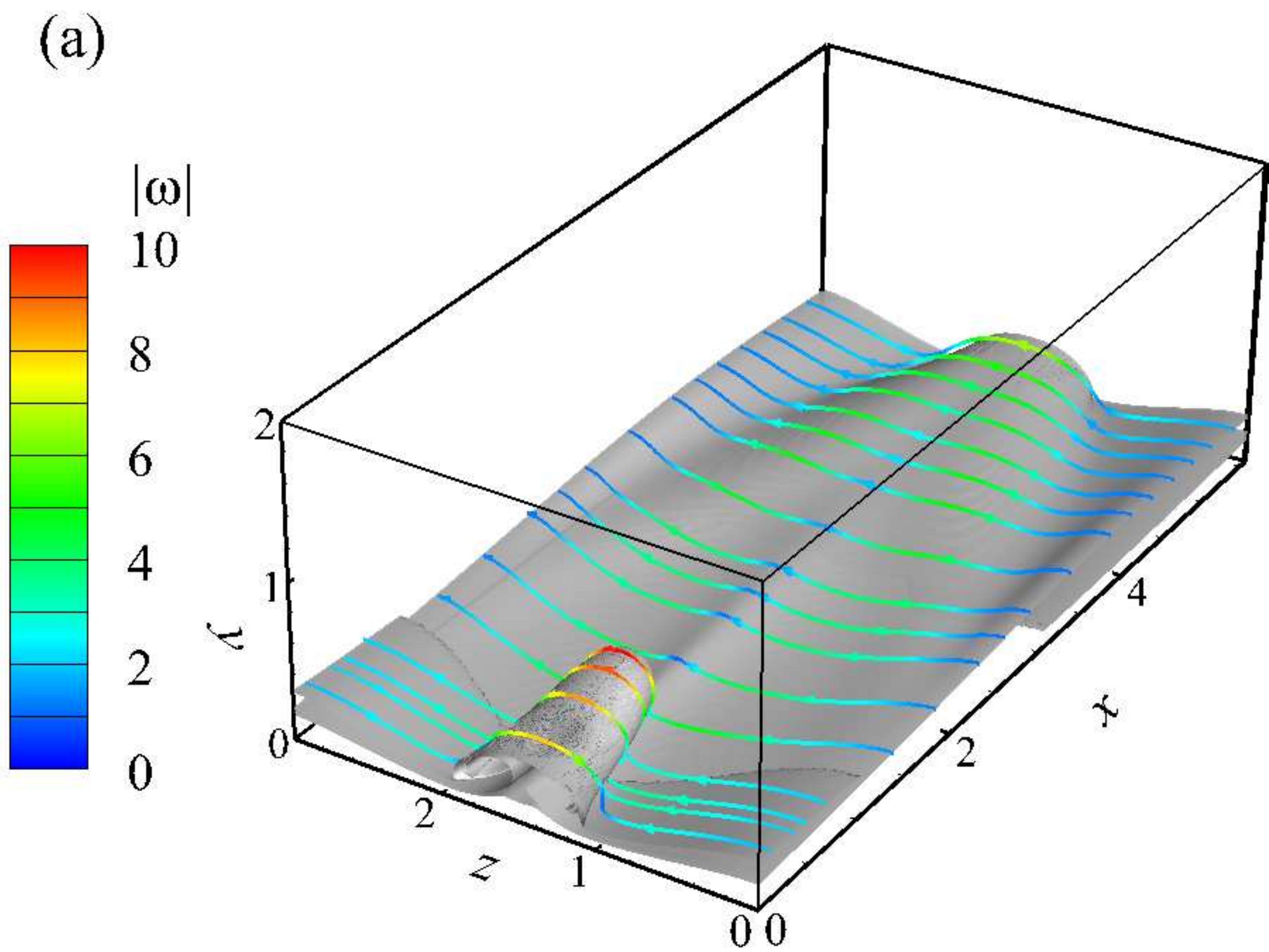}
\end{minipage}}%
  \centering \subfigure{
\begin{minipage}[c]{0.5\textwidth}
\includegraphics[width=2.6in]{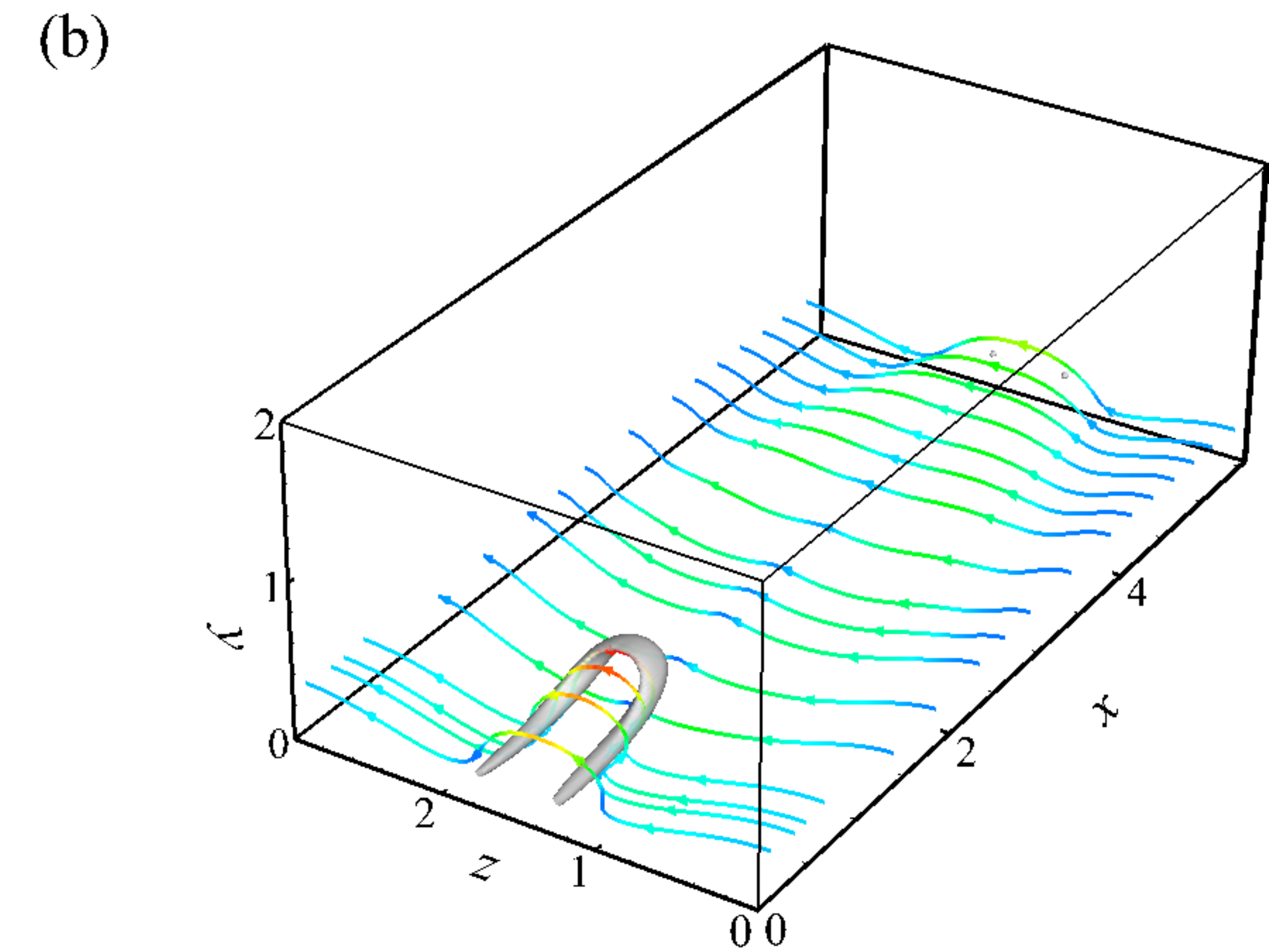}
\end{minipage}}%
  \caption{(Colour online) Comparison of Lagrangian and Eulerian vortex structures at $t=104$.
  Vortex lines are integrated from the surfaces, and colour-coded by the magnitude of vorticity $|\bs \omega|$.
  (a) material surface of $\phi=0.217$, (b) iso-surface of the swirling strenth $\lambda_{ci}$ (50\% of the maximum value).\protect\\}
\label{fig:h104}
\end{figure}

In figure~\ref{fig:all-2}, the material surface subsequently evolves into a hairpin-like structure after the formation of the triangular bulge.
In this stage, the vorticity begins to be intensified in local regions and the tube-like vortical structures are generated in figures~\ref{fig:h100} and \ref{fig:h104},
so the evolution of vorticity 
becomes important in this stage.
The governing equation for the vorticity in an incompressible flow can be written as
\begin{equation}
   \frac{\mathrm{D}\boldsymbol{\omega}}{\mathrm{D}t}=\boldsymbol{\omega}\bcdot\mathsfbi{S}+\nu\nabla^2\boldsymbol{\omega}.
  \label{Eq:ww1}
\end{equation}

For the flow at $t=100$ in figure~\ref{fig:h100}, $\partial u/\partial y$ is the largest component in the rate-of-strain tensor $\mathsfbi{S}$ owing to the domination of the mean shear at this time,
so (\ref{Eq:ww1}) can be simplified as
\begin{equation}
\left. \begin{array}{ll}
\displaystyle\
    \frac{\mathrm{D}\omega_x}{\mathrm{D}t}=\frac{\omega_y}{2}\frac{\partial u}{\partial y},\\[8pt]
\displaystyle\
    \frac{\mathrm{D}\omega_y}{\mathrm{D}t}=\frac{\omega_x}{2}\frac{\partial u}{\partial y}.\\[8pt]
 \end{array}\right\}
  \label{Eq:wxz}
\end{equation}
Here, the viscous term $\nu\nabla^2\boldsymbol{\omega}$ in (\ref{Eq:ww1}) is ignored before topological changes of the vorticity for high-Reynolds-number flows.

Assuming the mean shear ${\partial u}/{\partial y}=2S_0$ is a constant, the solution of (\ref{Eq:ww1}) is
\begin{equation}
\left. \begin{array}{ll}
\displaystyle\
    \omega_x=\frac{\omega_{x0}+\omega_{y0}}{2}\exp(S_0t)+\frac{\omega_{x0}-\omega_{y0}}{2}\exp(-S_0t),\\[8pt]
\displaystyle\
    \omega_y=\frac{\omega_{x0}+\omega_{y0}}{2}\exp(S_0t)-\frac{\omega_{x0}-\omega_{y0}}{2}\exp(-S_0t),\\[8pt]
 \end{array}\right\}
  \label{Eq:Awxz}
\end{equation}
where the initial values for $x$- and $y$-components of vorticity are $\omega_{x0}$ and $\omega_{y0}$, respectively.
Because of the symmetry, only the lower half of the channel with $S_0\ge0$ is considered.
From the first term in the right hand side of (\ref{Eq:Awxz}), the amplitudes of $\omega_x$ and $\omega_y$ increase exponentially if $\omega_{x0}$ and $\omega_{y0}$ are finite and have the same sign.

\begin{figure}
\begin{center}
  \includegraphics[width=0.65\textwidth]{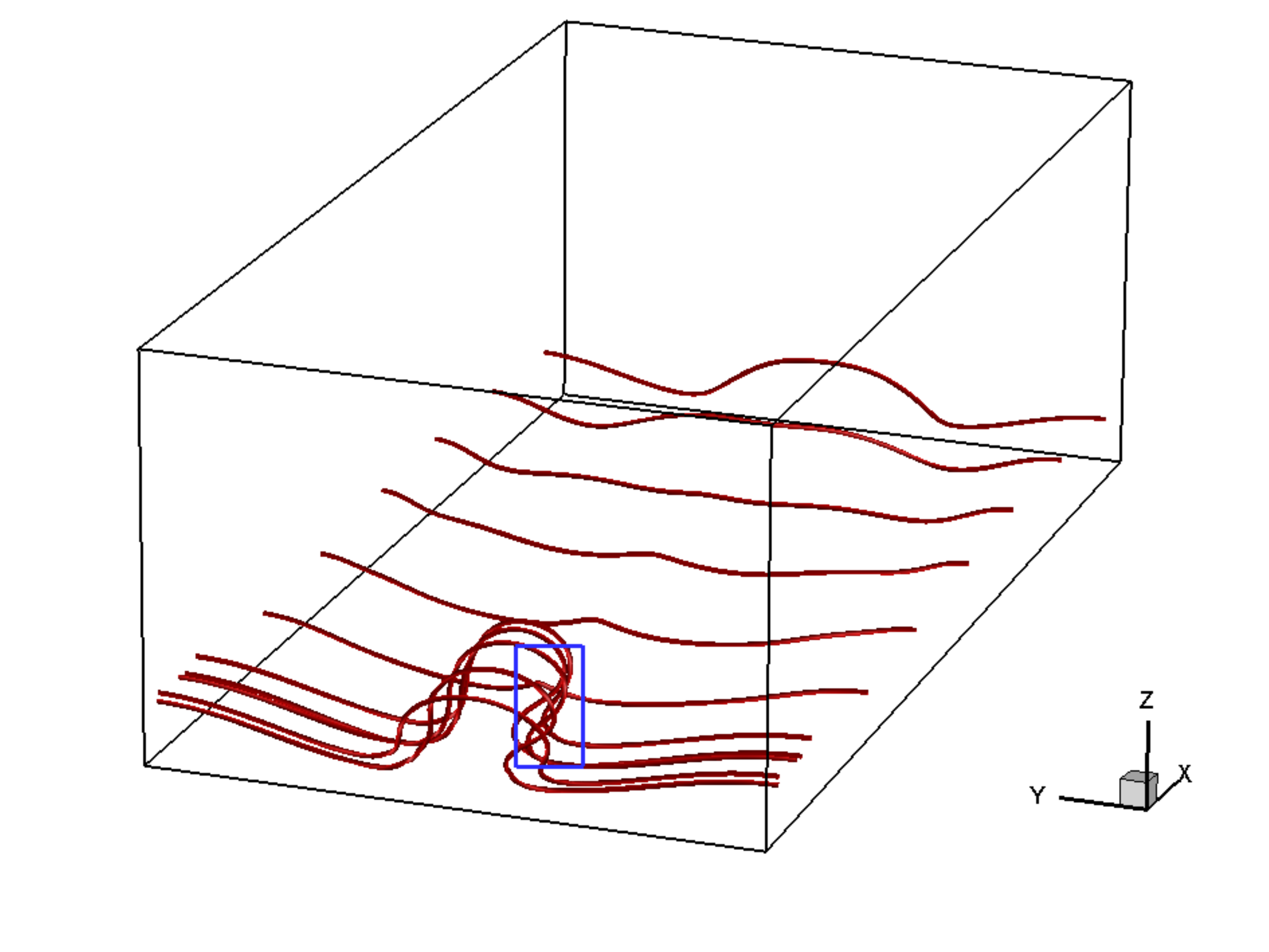}
  \caption{(Colour online) Vortex lines extracted near the material surface with $\phi=0.217$.
  The folding of vortex lines is highlighted in the blue rectangle.\protect\\}
\label{fig:lines104}
\end{center}
\end{figure}

As shown in figure~\ref{fig:h100}(a), the inclination angles of the vortex lines near the bulge ridges
imply that $\omega_y$ and $\omega_x$ are finite with $\omega_x\omega_y>0$.
As indicated in (\ref{Eq:Awxz}), $\omega_x$ and $\omega_y$ can increase rapidly in this region, which is also shown in figure~\ref{fig:wx}.
This implies that the vortex lines are intensively stretched
and the most intensified $|\bs \omega|$ occurs at the ridge of the triangular bulge, which is also observed in figure~\ref{fig:lines104} at $t=104$.

The structural evolution appears to be led by the tip of the triangular bulge where is the most distant from the wall on the material surface.
The vortex lines at the tip are advected by large mean shear and induced velocities, so they are highly curved as an $\Omega$-shape.
In the Lagrangian view, the bulge is lifted and elongated in figure~\ref{fig:h100}(a).
Then the surface rolls up in wall-normal and spanwise planes, and tube-like structures are generated beneath the bulge tip in figure~\ref{fig:h104}(a).
The vortex lines are stretched in the streamwise direction because of the mean flow shear,  {which was inferred as a possible mechanism for the formation of the hairpin-like structures in \citet[]{Kim1986}}, and then they are folded inward at the neck as shown in the blue rectangle in figure~\ref{fig:lines104}.
The folding vortex lines are concentrated and they are rapidly pulled out with intensified $|\bs \omega|$.
 {The accumulation of the streamwise vorticity in transitional boundary layer was also observed from the stretch and folding of material lines and surfaces in the K-type transition in \citet[]{Bernard2013}, but the material surface was discretized on a coarse mesh so that the possible rolling up of the tube-like structures was not observed.}
Finally, the arch (or head) of the hairpin-like structure are pulled out at this part of the material surface with two vortex tubes behind as `legs'. This procedure is generally consistent with the conceptional model qualitatively illustrating the generation of a hairpin-like vortex in \citet{Perry1982}.

 {
We remark that the vorticity intensification stage is skipped in some simplified model for the generation of the hairpin-like or ring-like vortical structures \citep[\eg][]{Hama1963,Moin1986}. In these models, the initial vorticity is assumed to be concentrated in thin tube-like structures and is much higher than the background vorticity produced by the mean shear, which can be considered as the beginning of the third stage below.}


\subsection{Stage 3, hairpin-like structures}\label{sec:stage3}
\begin{figure}
\centering \subfigure{
\begin{minipage}[c]{0.5\textwidth}
\includegraphics[width=2.6in]{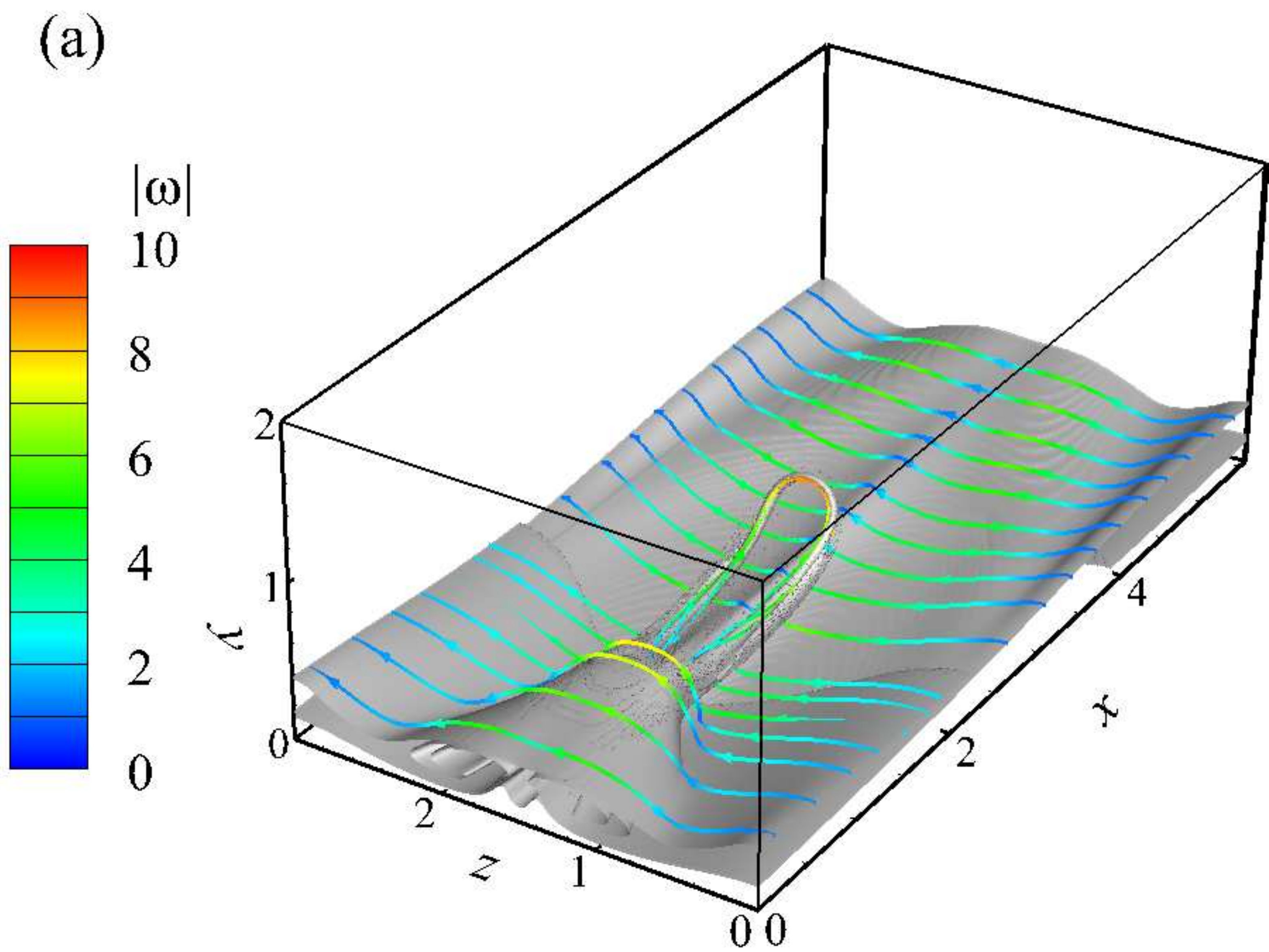}
\end{minipage}}%
  \centering \subfigure{
\begin{minipage}[c]{0.5\textwidth}
\includegraphics[width=2.6in]{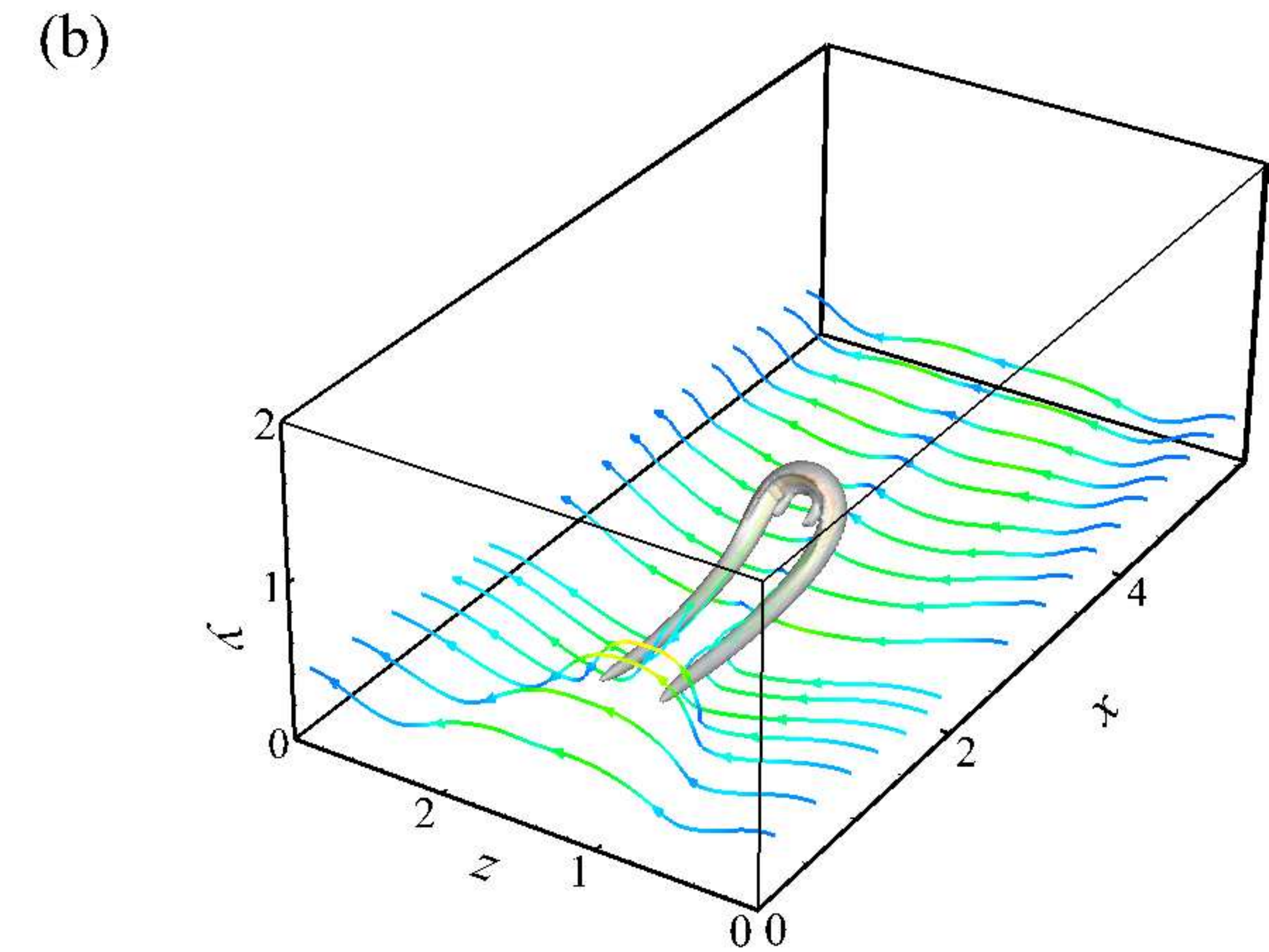}
\end{minipage}}%
  \caption{(Colour online) Comparison of Lagrangian and Eulerian vortex structures at $t=106$.
   Vortex lines are integrated from the surfaces, and colour-coded by the magnitude of vorticity $|\bs \omega|$.
  (a) material surface of $\phi=0.217$, (b) iso-surface of the swirling strenth $\lambda_{ci}$ (25\% of the maximum value).\protect\\}
\label{fig:h106}
\end{figure}

The `hairpin vortex' is often described as the hairpin-shaped, tube-like structures. In the evolution, the Lagrangian hairpin-like structures are stretched out from the triangular bulge at $t=106$ in figure~\ref{fig:h106}(a) after the vorticity intensification stage. At the stage after $t=106$, the vorticity distribution could be considered as isolated, thin and elongated vortex tubes or filaments.

It is noted that the skeleton of the hairpin-like structure can be also captured by the Eulerian $\lambda_{ci}$-criterion at $t=106$ in figure~\ref{fig:h106}(b), because the Eulerian criteria have the capability to identify the `vortex core' at a time instant \citep[]{Zhou1999}. The tracking of the hairpin-like structures, however, appears to be ambiguous from the Eulerian criterion in figures~\ref{fig:h100}(b), \ref{fig:h104}(b) and \ref{fig:h106}(b), because the iso-contour value is not uniquely defined and usually it has to be varied \emph{ad hoc} with time to keep the characteristic size of extracted structures similar at different times.

The vortex-induced velocity $\boldsymbol{u}_{i}$ at a point $\boldsymbol{r}$ in space can be calculated by the Biot-Savart law
\begin{equation}
   \boldsymbol{u}_{i}(\boldsymbol{r},t)=-\frac{\Gamma}{4\pi}\int\frac{(\boldsymbol{r}-\boldsymbol{r}')\times
   \boldsymbol{t}_{\omega}}
   {|\boldsymbol{r}-\boldsymbol{r}'|^3}G(|\boldsymbol{r}-\boldsymbol{r}'|)\mathrm{d}s'
  \label{Eq:B-S_l}
\end{equation}
in a Lagrangian viewpoint as discussed in \citet[]{Moin1986}, where $s'$ is the arclength of the vortex filament,
$\boldsymbol{r}'(s',t)$ is a curve representing the tube of concentrated vorticity,
$\Gamma$ is the circulation of the vortex tube, $\boldsymbol{t}_{\omega}$ is the tangential vector of the vortex filament,
and the function $G(|\boldsymbol{r}-\boldsymbol{r}'|)$ models the effect of distributed vorticity within the core of effective radius $\sigma$.
This equation is valid in the thin-filament approximation, i.e., $\sigma$ is very small and the structure of the vortex core does not change.

The self-induced velocity $\boldsymbol{u}_{si}$ of the vortex filament can be obtained by assuming the point $\boldsymbol{r}$ is also on the filament $\boldsymbol{r}'(s',t)$.
By only considering the local effect \citep[]{Batchelor1967},
\citet[]{Zhou1999} showed that $\boldsymbol{u}_{si}$ can be approximated by the leading term of (\ref{Eq:B-S_l}) as
\begin{equation}
   \boldsymbol{u}_{si}=K_c\frac{\Gamma}{4\pi}(\kappa\boldsymbol{b}_{\omega}),
  \label{Eq:B-S_s}
\end{equation}
where $K_c$ is a constant depending on the distribution of vorticity within the vortex core,
and $\boldsymbol{b}_\omega$ is the local bi-normal vector of the vortex filament $\boldsymbol{r}'(s',t)$.

As the vortex lines are concentrated within tube-like regions with growing vorticity,
the geometry of the vortex lines plays an increasingly important role in the evolution of material surfaces that are mostly composed of vortex lines.
From (\ref{Eq:B-S_s}), the self-induced velocity $\boldsymbol{u}_{si}$ of the vortex filament is aligned with $\bs b_\omega$ of the vortex line and proportional to $\kappa$.

\begin{figure}
\begin{center}
\includegraphics[width=0.97\textwidth]{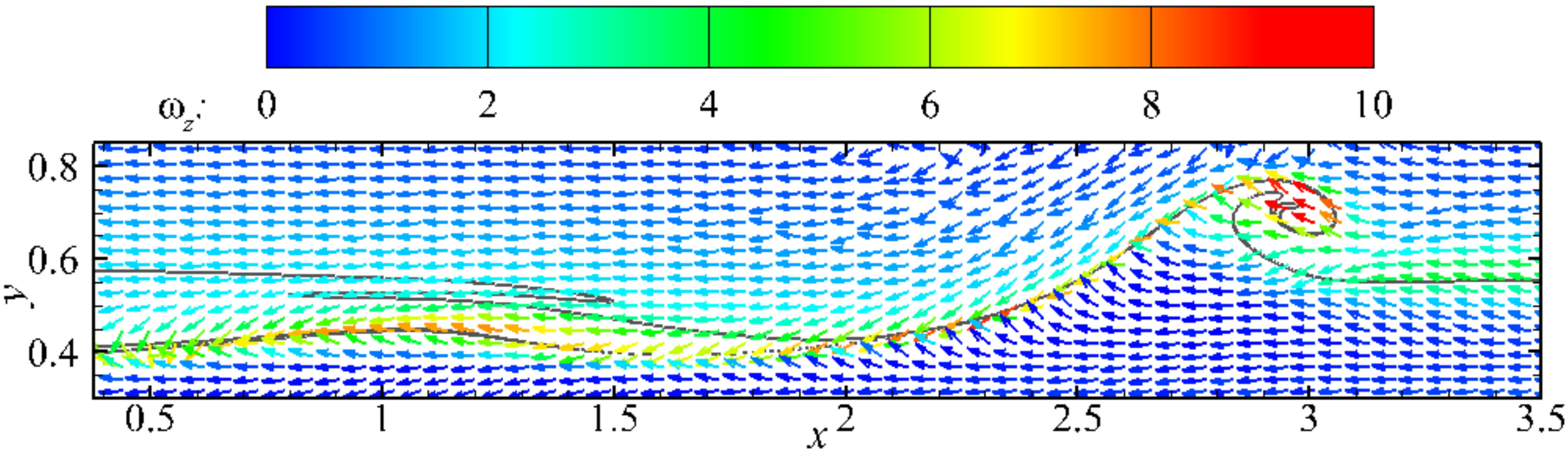}
  \caption{(Colour online)  The bi-normal vector $\bs b_\omega$ on the spanwise and wall-normal plane-cut at  {the peak position} $z=L_z/2$ at $t=106$.
  The vectors are colour-coded by the spanwise vorticity $\bs\omega_z$. The solid lines are the contour lines of the Lagrangian field with $\phi=0.217$ and $\phi=0.433$.}
\label{fig:binormal}
\end{center}
\end{figure}

In this stage, the geometry of vortex lines is important in dynamics because the motions of vortex lines and material surfaces are affected by the self-induced velocity $\bs u_{si}$.
The distribution of the bi-normal vector $\bs b_\omega$ on the $x$--$y$ plane at  {the peak position} $z=L_z/2$ is shown in figure~\ref{fig:binormal},
where $\bs b_\omega$ can be calculated as
\begin{equation}
   \boldsymbol{b}_{\omega}=\frac{\boldsymbol{\omega}\times(\boldsymbol{\omega}\bcdot\bnabla\boldsymbol{\omega})}
   {|\boldsymbol{\omega}\times(\boldsymbol{\omega}\bcdot\bnabla\boldsymbol{\omega})|}.
  \label{Eq:bi-normal}
\end{equation}
The vectors are colour-coded by the spanwise vorticity $\omega_z$ and overlayed by two typical contour lines of the Lagrangian field with $\phi=0.217$ and $\phi=0.433$ in table~\ref{tab:material}.
It is shown that the vorticity is most concentrated at the arch of the hairpin-like structure near $x=3$.
From (\ref{Eq:B-S_s}), $\boldsymbol{u}_{si}$ at the tip of the hairpin-like structure is upper left, so the hairpin is lifted under its self-induction.
At the mean time, the hairpin-like structure is elongated in the streamwise direction controlled by the mean shear.

\begin{figure}
\centering \subfigure{
\begin{minipage}[c]{0.5\textwidth}
\includegraphics[width=2.6in]{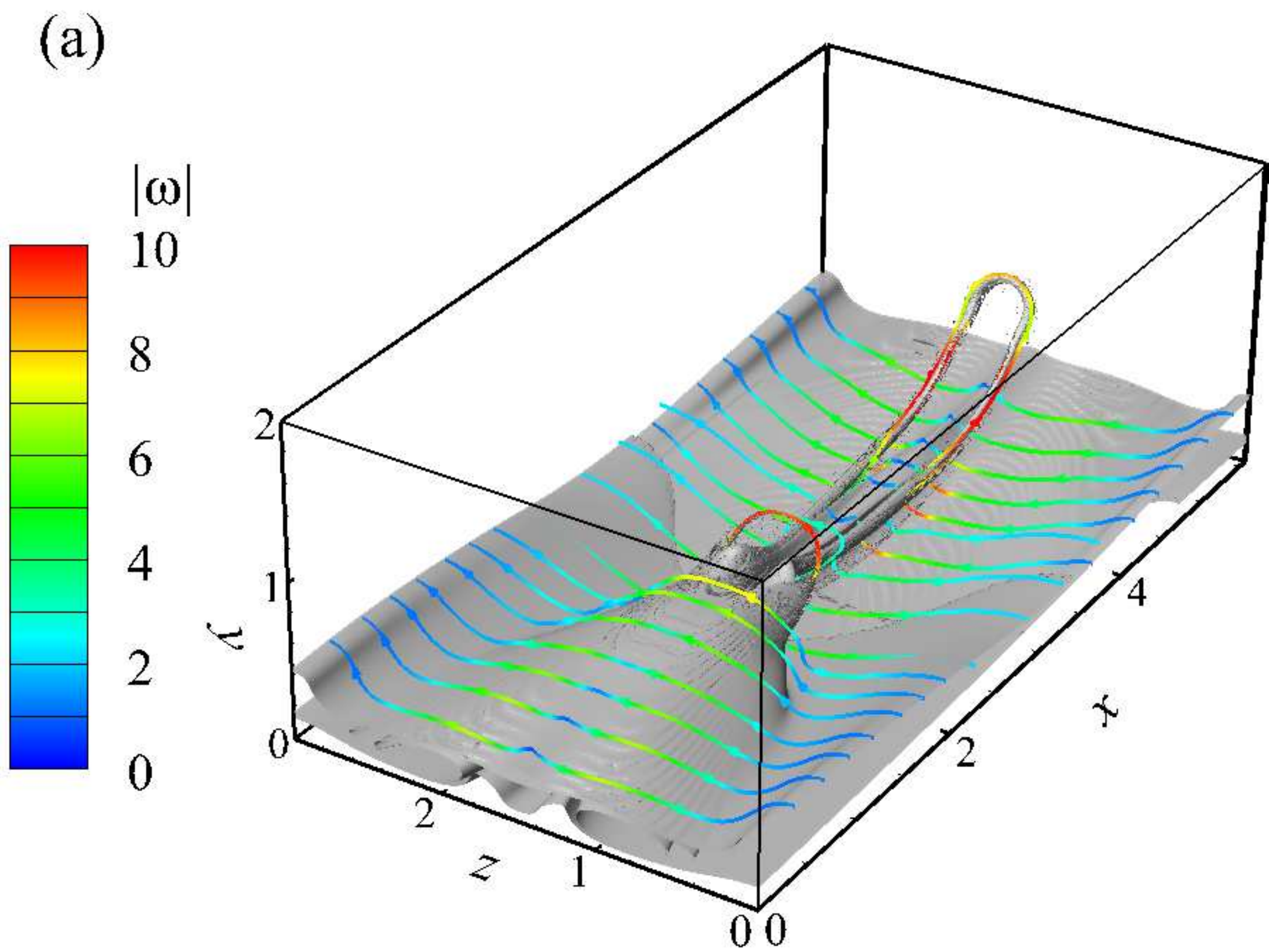}
\end{minipage}}%
  \centering \subfigure{
\begin{minipage}[c]{0.5\textwidth}
\includegraphics[width=2.6in]{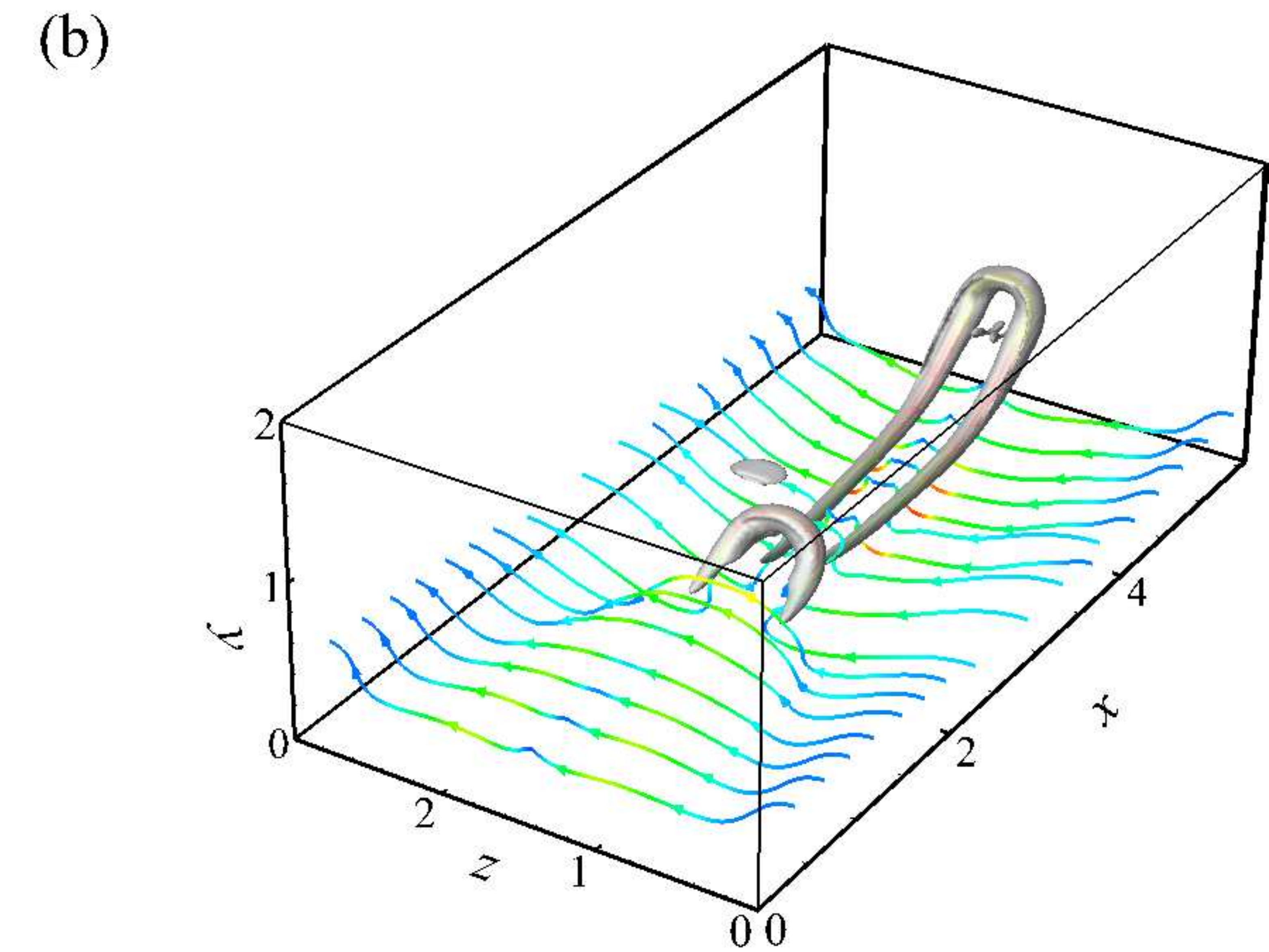}
\end{minipage}}%
  \caption{(Colour online) Comparison of Lagrangian and Eulerian vortex structures at $t=108$.
  Vortex lines are integrated from the surfaces, and colour-coded by the magnitude of vorticity $|\bs \omega|$.
  (a) material surface of $\phi=0.217$, (b) iso-surface of the swirling strenth $\lambda_{ci}$ (25\% of the maximum value).\protect\\}
\label{fig:h108}
\end{figure}

In figure~\ref{fig:binormal}, the secondary vorticity intensification region occurs near $x=1.1$ where $\bs u_{si}$ has a finite upward component at $t=106$.
After the stretching of the `legs' of the primary hairpin-like structure, vortex lines start to concentrate behind the stretched structure, and the secondary hairpin-like structure is generated in this region at $t=108$. This formation is shown in figure~\ref{fig:h108}, and its mechanism is similar to that for the primary one.

 {
Since the present study focuses on the period during the beginning of the late transitional stages, the ring-like vortical structures perhaps generated at later times have not been observed. The ring-like vortex in the very late stage of transition is usually considered to be generated by the reconnections of the `legs' of $\Lambda$-like or hairpin-like structures \citep[\eg][]{Borodulin2002a,Bake2002} with the Crow instability \citep{Crow1970}.
In this stage, the deviation between the material surfaces and vortex surfaces can be greatly amplified, so a correction method must be introduced to faithfully describe the evolution of the vortex surfaces with significant topological changes \citep[]{yang2010lagrangian}.
}

\subsection{Brief summary and discussion}
In this section, a Lagrangian perspective on the formation and evolution of the  {signiture} structures is provided by tracking the influential material surface.
This material surface with the maximum deformation during the evolution is uniquely defined to avoid the ambiguity, and is a reasonable surrogate of a vortex surface before significant topological changes of vortex surfaces.
Therefore, the evolution of this material surface can shed light on the dynamics of vortical structures prior to possible vortex reconnections around $t=110$.

\begin{figure}
\begin{center}
  \includegraphics[width=0.98 \textwidth]{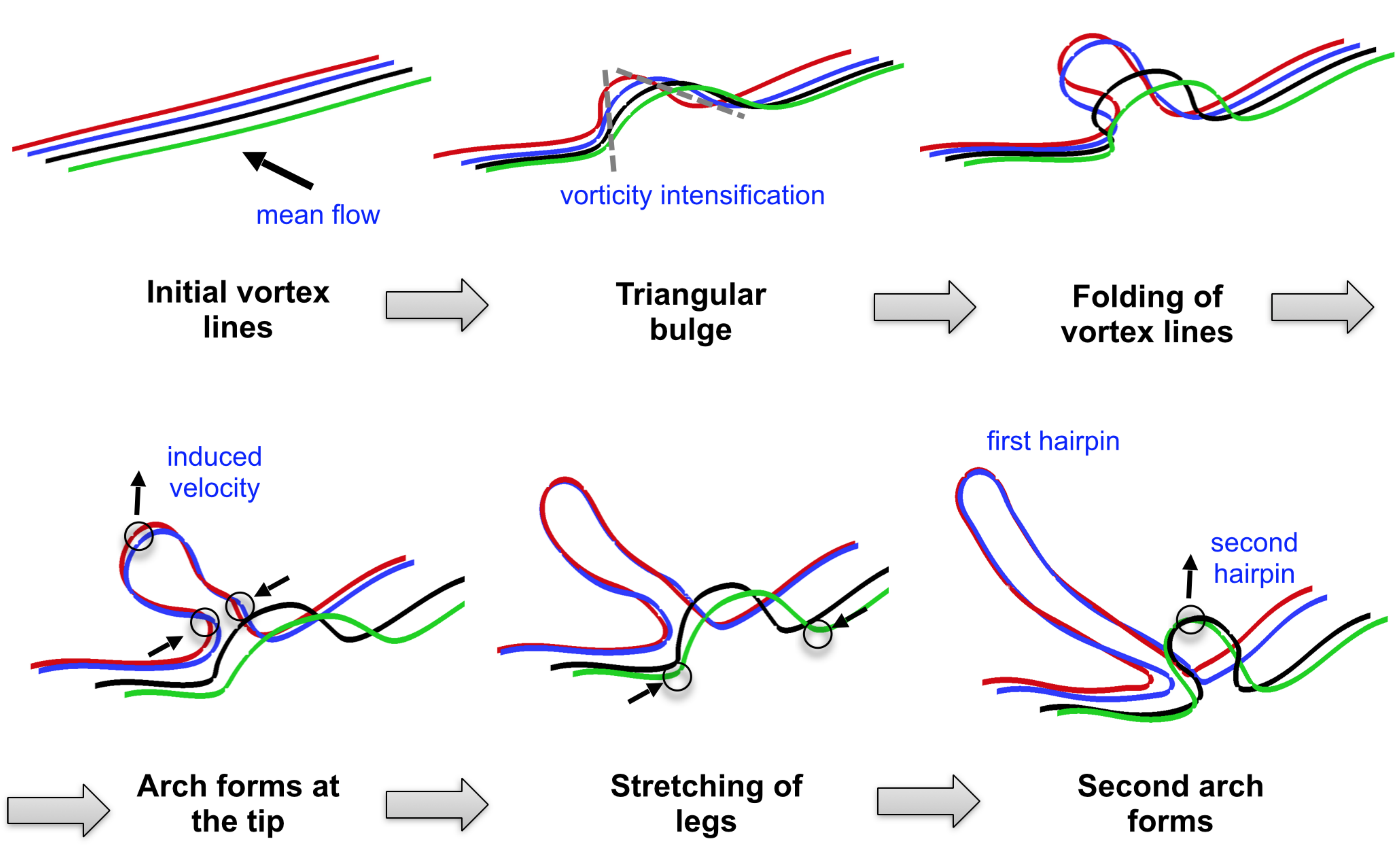}
  \caption{(Colour online) A schematic diagram of the evolution of the  {influential material surface} in transitional channel flow.
  The skeleton of the evolving material surface is presented by four typical vortex lines marked by different colours.
  \protect\\}
\label{fig:schem}
\end{center}
\end{figure}

A schematic diagram for the evolution of the Lagrangian hairpin-like structures, represented by typical vortex lines, in the  {late transition}al channel flow is summarized in figure~\ref{fig:schem}, which is also combined with previous findings focusing on the  {vortical structures in late transitional stages \citep[see e.g.][]{Kleiser1991,Kachanov1994,Borodulin2002a,Bake2002} and the generation of hairpin-like structure packets in turbulent channel flows \citep[]{Zhou1999,Green2007}.}
Since the amplitudes of the initial disturbances are small,
the flow is nearly laminar and the initial parallel vortex lines are attached on a planar material surface.
A triangular bulge is generated because the disturbed surface is streched by the mean shear,
and the vortex lines also bend at two ridges of the bulge.
The vorticity near the ridges begins to be intensified,
and the vortex lines are folded and elongated in this region.
The most intensified vorticity is at the bulge tip, and the vortex lines at this part are pulled out as the arch of the hairpin-like structure.
The hairpin arch is lifted and legs are stretched by the combination of the mean shear and the self-induced velocity, then the signature hairpin-like structure develops into a self-sustained state.
 {After the detachment of the primary hairpin-like structure from the triangular bulge}, the secondary hairpin arch is generated at  {the tip of the original} triangular bulge under the similar mechanism for the generation of the primary one.

 {
It is noted that the influential material surface is only used as a representative surface and its evolution is not isolated. This surface can interact with, or `influence' in some sense, the motion of neighbouring material surfaces. The surfaces below the influential one are elevated near the `head' of the hairpin-like structure, and the surfaces above the influential surface warp around the hairpin-like structure with high-shear layers generated between the `legs'.}


\section{Conclusions}\label{sec:conclusion}
The Lagrangian scalar field is applied in the K-type transitional channel flow.
By tracking the Lagrangian field, we are able to investigate the evolution of material surfaces represented by the iso-surfaces of Lagrangian fields.
Three critical issues for Lagrangian investigations on the evolution of coherent structures in transitional flows using material surfaces are addressed, including the uniqueness of the initial setup of material surfaces, significances of different evolving material surfaces, and implications of the evolving material surfaces with a physically interesting initial condition on vortex dynamics.

First, the initial scalar field is uniquely determined based on the criteria of the best approximation to vortex sheets and geometric invariance in the laminar state. The iso-surfaces of this initial field are both vortex surfaces and stream surfaces in the initial laminar state, and they are streamwise-spanwise planes.

Second, we seek the influential material surface that has significant kinematics. Since material surfaces with different initial wall distances can have different evolutionary geometries in the transition, the material surface with the maximum deformation is identified using the Lagrangian conditional mean of scalar gradient magnitude. It evolves from a streamwise-spanwise plane to a triangular bulge, and then into a signature hairpin-like structures. The evolution of this influential material surface not only affects the deformation of neighbouring ones, but also shows large variations of vorticity components and curvatures of vortex lines.
Additionally, the evolutionary geometry of the Lagrangian field is characterized by the wall-normal displacement of the surfaces, by which we define  Lagrangian elevation/descent event.
These Lagrangian events can quantify ejections and sweeps, and also have strong correlations to the momentum transport.

Finally, the material surface can be considered as a surrogate of the vortex surface before significant topological changes of vortical structures with the averaged deviation less than 15\%.
Therefore, the present study provides a Lagrangian perspective on the dynamics of the evolving vortical structures in transitional channel flow by focusing on the evolution of the influential material surface with the maximum deformation.


The evolution of the most deformed material surface is divided into three stages. In the first stage, the initially disturbed planar material surface in a weakly disturbed shear flow tends to evolve into a triangular bulge.
 {The triangular bulge consists of the swirling regions at the ridge and the high-shear layer on the top.}
In the second stage, the vorticity on the material surface is gradually intensified at the ridge of the triangular bulge  {owing to vortex stretching}.
With the folding of vortex lines, the material surface rolls up into the hairpin-shaped, tube-like structure.
 {Subsequently the hairpin-like structure is stretched and detached from the triangular bulge.}
In the third stage, the evolution of the Lagrangian hairpin-like structure with the large local vorticity intensification can be approximated as a vortex filament and described by the Biot-Savart law.
Under the effect of the mean shear and the self-induction,
the hairpin-like structure is lifted and elongated, and subsequently the secondary hairpin-like structure is generated behind the primary one.

 {The similarities and differences between our results and the previous findings using the Eulerian vortex identification criteria \citep[]{Kim1986,Zhou1999}, experimental visualizations \citep[]{Hama1963,Guo2010}, and Lagrangian tracking approaches \citep[]{Kleiser1985,Rist1995,Green2007,Bernard2013} are discussed.
It is noted that the continuous temporal evolution of vortical structures can be elucidated by tracking a single uniquely defined material surface at different times in the present study. Generation mechanisms of the triangular bulge and the hairpin-like structure are clarified as the preferential stretch of the unevenly disturbed vortex sheet under the mean shear and the rolling up of two ridges of the triangular bulge with the vorticity intensification, respectively. Moreover, the importance of the vorticity intensification stage that is often skipped in the vortex-filament models is stressed.}


In addition to the presented flow at $Re_\tau=208$, another transitional channel flow with a higher Reynolds number $Re_\tau=375$ has only been partially studied owing to the high computational cost.
The evolution of material surfaces at the higher Reynolds number shows more significant geometrical changes, e.g., the material surfaces and vortex lines with larger deformations, but qualitative differences have not been observed, so the results are omitted here.

We remark that the present work  {does not consider the very late transitional stage with the final flow breakdown}, because the deviation between the material surface and the vortex surface cannot be ignored after the significant topological changes or vortex reconnections.
 {The vortex-surface-field method proposed by \citet[]{yang2010lagrangian}, has been proved to be promising in characterizing the evolution of vortex surfaces with topological changes \citep[see][]{yang2011vsf,pullin2014vsf}.}
Thus, the vortex-surface-field method is expected to be applied to study the flow in the late transition in the near future. Moreover, the current methodology can be easily applied to other transitional flows with a sequence of time-resolved, three-dimensional velocity data sets from numerical simulations or experiments.

\begin{acknowledgments}
The authors thank C. B. Lee and J. Z. Wu for helpful comments. Numerical simulations were carried out on the TH-1A supercomputer in Tianjin and the TH-2A supercomputer in Guangzhou, China. This work has been supported in part by the National Natural Science Foundation of China (Grant Nos.~11342011, 11472015 and 11522215), and the Thousand Young Talents Program of China.
\end{acknowledgments}

\appendix
\section{Scalar gradient magnitude and the DLE field}\label{sec:LCS}
It was discussed in \citet[]{Wang2015} that the scalar gradient magnitude $g$ is related to the DLE field that is usually used to identify the hyperbolic LCS \citep[\eg][]{Green2007} in a flow evolution.
The DLE is defined as
\begin{equation}
  DLE_{\Delta T}(\bs x,t)=\frac{1}{2\Delta T}\log{\sigma_{\Delta T}(\bs x,t)},
  \label{eq:dle}
\end{equation}
where
\begin{eqnarray}
  \sigma_{\Delta T}(\bs x,t) & = & \lambda_{max} \left\{\left[\frac{\partial{\bs x_0}}{\partial{\bs x}}\right ]^\mathrm{T}
  \left[\frac{\partial{\bs x_0}}{\partial{\bs x}}\right] \right\} \nonumber\\
   & = & \lambda_{max}\left\{\left[\frac{\partial{\bs x_0}}{\partial{\bs X(\bs x_0,t_0|t)}}\right]^\mathrm{T}
  \left[\frac{\partial{\bs x_0}}{\partial{\bs X(\bs x_0,t_0|t)}}\right]\right\}.
\end{eqnarray}
Here, $\lambda_{max}\{\mathsfbi{M}\}$ denotes the maximum eigenvalue of the matrix $\mathsfbi{M}$, and the superscript $\mathrm{T}$ stands for the transpose of the matrix.
The fluid particles are tracked from the location $\bs X$ at $t$ to the initial location $\bs x_0$ at $t_0$,
and the tracking interval $\Delta T=t_0-t$ is negative for backward tracking.
The Cauchy--Green deformation tensor is represented by the real symmetric matrix  $\mathsfbi{P}$ as
\begin{equation}
  \mathsfbi{P}\equiv\left[\frac{\partial{\bs x_0}}{\partial{\bs x}}\right ]^\mathrm{T}
  \left[\frac{\partial{\bs x_0}}{\partial{\bs x}}\right].
\end{equation}

Similarly, the scalar gradient vector can be expressed as
\begin{equation}
\left[\frac{\partial{\phi}}{\partial{\bs x}}\right] =
\left[\frac{\partial{\bs x_0}}{\partial{\bs x}}\right]
\left[\frac{\partial{\phi}}{\partial{\bs x_0}}\right],
\end{equation}
where the initial scalar gradient $\left[{\partial{\phi}}/{\partial{\bs x_0}}\right]^\mathrm{T}=[0,1,0]$ is a constant unit vector in the whole field.
Then the scalar gradient magnitude can be written as
\begin{eqnarray}
g & = & \left(  \left[\frac{\partial{\phi}}{\partial{\bs x}}\right]^\mathrm{T}
\left[\frac{\partial{\phi}}{\partial{\bs x}}\right] \right)^{1/2} \nonumber\\
  & = & \left( \left[\frac{\partial{\phi}}{\partial{\bs x_0}}\right]^\mathrm{T}
\left(\left[\frac{\partial{\bs x_0}}{\partial{\bs x}}\right]^\mathrm{T}
\left[\frac{\partial{\bs x_0}}{\partial{\bs x}}\right]\right)
\left[\frac{\partial{\phi}}{\partial{\bs x_0}}\right] \right)^{1/2}.\label{eq:g_matrix}
\end{eqnarray}

The real symmetric matrix $\mathsfbi{P}$ can be represented as $\mathsfbi{P}=\mathsfbi{U}^\mathrm{T}\mathsfbi{\Lambda} \mathsfbi{U}$, where the orthogonal matrix $\mathsfbi{U}$ satisfies $\mathsfbi{U}^\mathrm{T} \mathsfbi{U}=\mathsfbi{U}\mathsfbi{U}^\mathrm{T} =\mathsfbi{I}$, and $\mathsfbi{\Lambda}=\textrm{diag}(\lambda_1,\lambda_2,\lambda_3)$ is a diagonal matrix with the eigenvalues $\lambda_1$, $\lambda_2$ and $\lambda_3$ of $\mathsfbi{P}$.
Then (\ref{eq:g_matrix}) can be re-expressed as $g=(\bs q^\mathrm{T}\mathsfbi{\Lambda}\bs q)^{1/2}$, where
$\bs q\equiv\mathsfbi{U}
 \left[{\partial{\phi}}/{\partial{\bs x_0}}\right]$.
It is noted that $\bs q^\mathrm{T}\bs q =1$, because $\mathsfbi{U}$ is orthogonal and the transform does not change the magnitude of the vector $\left[{\partial{\phi}}/{\partial{\bs x_0}}\right]$.
Thus we can derive the inequality
\begin{equation}
g=( \lambda_1 q_1^2+\lambda_2 q_2^2+\lambda_3 q_3^2)^{1/2}
\leq (\lambda_{max})^{1/2}.
\label{eq:g_lambda}
\end{equation}
Finally the relation between the DLE and $g$ in the present work can be derived from the (\ref{eq:dle}) and (\ref{eq:g_lambda}) as
\begin{equation}
 DLE_{\Delta T} \geq \frac{1}{\Delta T}\log{g}.
\label{eq:g_dle}
\end{equation}

Since $\lambda_{max}\{\mathsfbi{P}\}$ gives the maximum amplification of $g$, the DLE field has an explicit relation to $g$ or $\log g$ \citep[see][]{Wang2015}.
Hence, the contour plots of $g$ with the simple initial scalar gradient and the DLE field can be qualitatively similar (not shown), and their contours with large values tend to concentrate in ribbon-like regions.

\section{Analysis for the formation of the triangular bulge}\label{sec:triangular}
The evolutionary geometry of material surfaces shown in \S\,\ref{sec:L-surface} can be statistically quantified using the normal $\boldsymbol{n}$ of the material surfaces discussed in \S\,\ref{sec:geometry_surface}.
The transport equation for $\boldsymbol{n}$ can be derived from (\ref{Eq:phi}) and (\ref{Eq:grad_phi}) as \citep[see][]{Brethouwer2003}
\begin{equation}
   \frac{\mathrm{D}\boldsymbol{n}}{\mathrm{D}t}=-\boldsymbol{n}\bcdot\mathsfbi{S}
   \bcdot(\mathsfbi{I}-\boldsymbol{nn})-\frac{1}{2}\boldsymbol{n}\times\boldsymbol{\omega}.
  \label{Eq:vec_n}
\end{equation}
The evolution of $\boldsymbol{n}$ is influenced by the rate-of-strain tensor $\mathsfbi{S}$,
vorticity $\boldsymbol{\omega}$ and $\boldsymbol{n}$ itself, so it is determined by both the Eulerian velocity field
and the evolutionary geometry of the material surfaces.
It is challenging to analyze (\ref{Eq:vec_n}) owing to the nonlinear coupling among the three normal components, and $\mathsfbi{S}$ and $\boldsymbol{\omega}$ also become highly nonlinear in the laminar-turbulent transition.
Nevertheless, we can reveal the qualitative behavior of $\bs n$ for early times, along with some simplifications before the formation of the triangular bulge at $t\le100$ when the disturbances are still small.

In the early transitional stage, we consider the Eulerian velocity field as a constant laminar shear flow without any disturbances in the evolution.
This assumption is based on that the disturbances from (\ref{Eq:TS}) are much smaller than the basic Poiseuille flow velocity (\ref{Poiseuille}) in the very early transition stage.
Alternatively, we assume the initial material surface is disturbed, and the deformation of the surface is induced by the constant flow shear.


With this assumption, (\ref{Eq:vec_n}) can be simplified in the laminar shear flow (\ref{Poiseuille}) as
\begin{equation}
\left. \begin{array}{ll}
\displaystyle\
    \frac{\mathrm{D}n_x(t)}{\mathrm{D}t}=n_x(t)^2n_y(t)\zeta,\\[8pt]
\displaystyle\
    \frac{\mathrm{D}n_y(t)}{\mathrm{D}t}=n_x(t)(n_y(t)^2-1)\zeta,\\[8pt]
\displaystyle\
    \frac{\mathrm{D}n_z(t)}{\mathrm{D}t}=n_x(t)n_y(t)n_z(t)\zeta,
 \end{array}\right\}
  \label{Eq:dn}
\end{equation}
where $\zeta=2 U_0 (1-y)$ is a scaled coordinate. Considering the symmetry of the channel,
only the lower half of the channel is considered, which implies $0<\zeta\leq2U_0$.

\begin{figure}
\centering \subfigure{
\begin{minipage}[c]{1\textwidth}
\begin{center}
\includegraphics[width=4.5in]{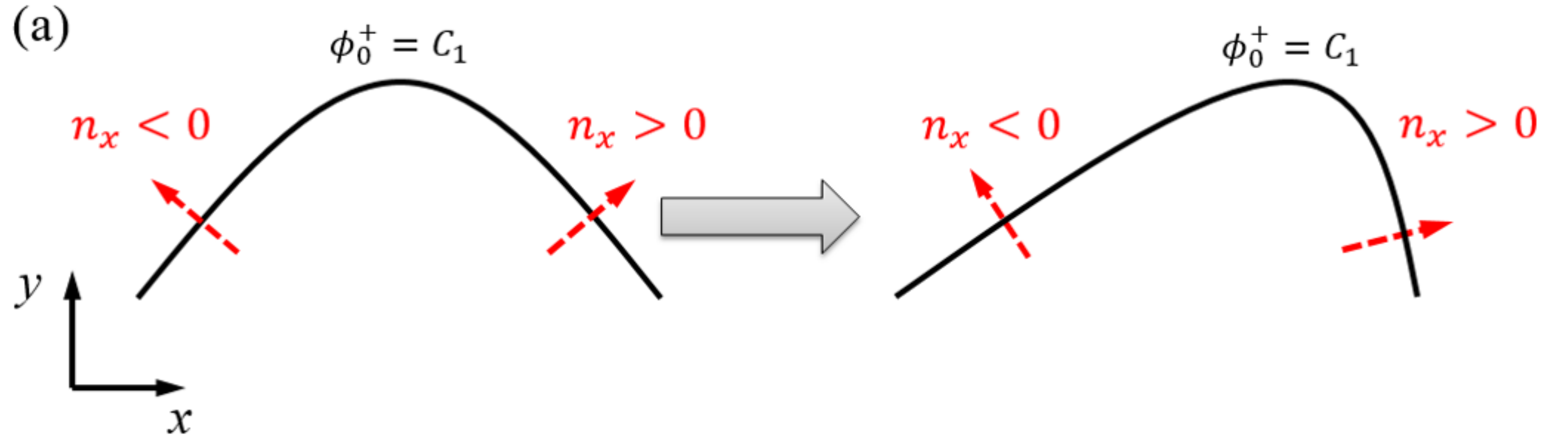}
\end{center}
\end{minipage}}
\centering \subfigure{
\begin{minipage}[c]{1\textwidth}
\begin{center}
\includegraphics[width=4.5in]{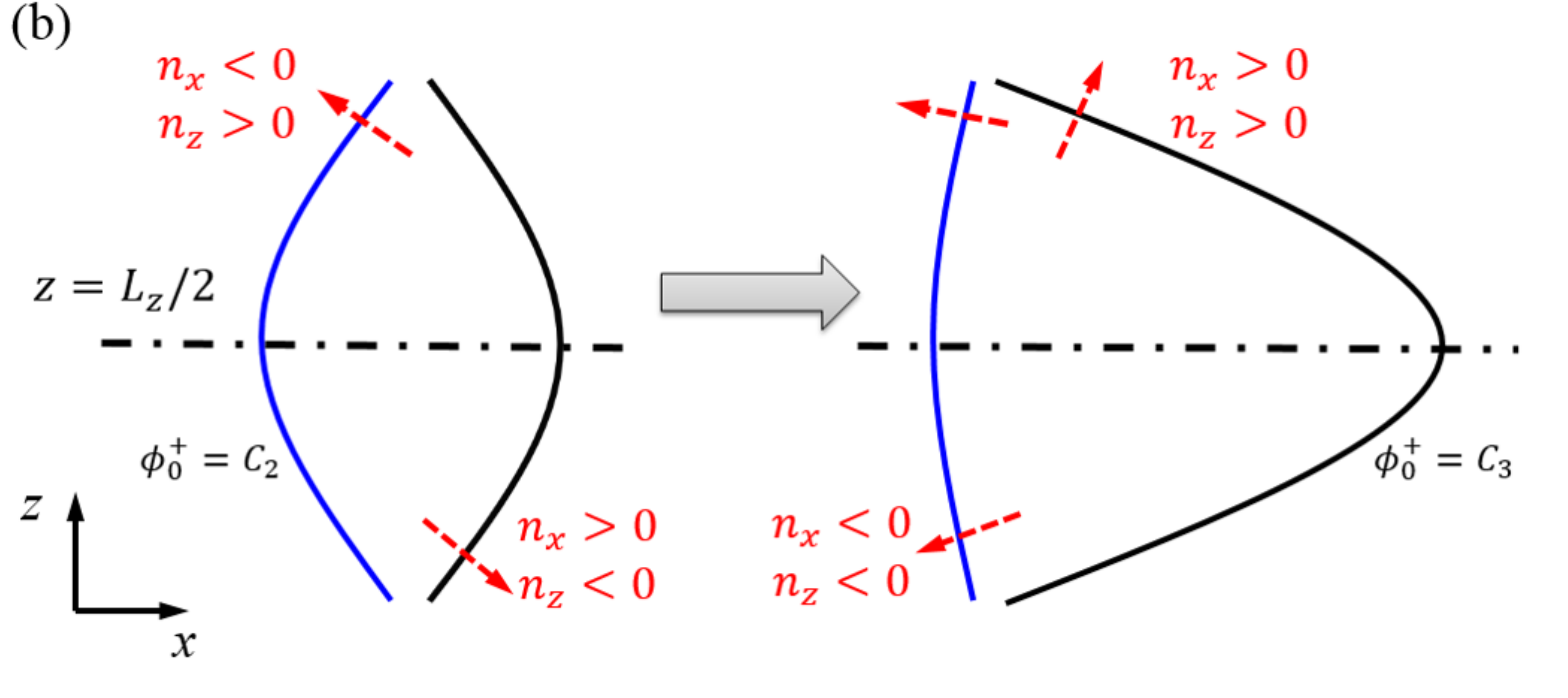}
\end{center}
\end{minipage}}
\caption{(Colour online) A diagram of the evolution of the normal $\boldsymbol{n}$ (dashed arrows) of the material surface in a simple shear flow with the mean velocity from left to right on two-dimensional plane-cuts
 from (\ref{Eq:dn}), where $C_1$, $C_2$ and $C_3$ are different constants.
(a) $x$--$y$ plane-cut, (b) $x$--$z$ plane-cut. \protect\\}
\label{fig:diagram-cut}
\end{figure}

Since the normal (0,1,0) of initial planar surfaces stay invariant from (\ref{Eq:dn}), the initial perturbations of the surfaces caused by the velocity disturbances are imposed for analysis on the evolution.
The eigenfunction of the two-dimensional TS wave in (\ref{Eq:TS}) is $\boldsymbol{u}_{2\textrm{D}}(y)=(u_{2\textrm{D}},v_{2\textrm{D}},0)$, so this two-component disturbance only cause folding of the material surface in two-dimensional $x$--$y$ planes in figure~\ref{fig:all-2}(a).
The amplitudes of the three-dimensional TS waves then amplify, and the wall-normal component $w_{3\textrm{D}}\exp[i(\alpha x+\beta z)]$ of $\boldsymbol{u}_{3\textrm{D}}$ in (\ref{Eq:TS}) induce periodic sinuous disturbances on the initially planar material surfaces in the streamwise and spanwise directions.
It is noted that the maximum disturbance occurs at $L_z/2$ or on the `peak plane'.
As shown in figure~\ref{fig:diagram-cut}, without loss of generality, we only consider the contour line of the disturbed surface as a convex arch on the $x$--$y$ plane, and $n_y$ is positive everywhere without rolling up of the surface for the weak disturbances.
For the contour lines of the initial material surface on the $x$--$z$ plane, $n_x$ and $n_z$ can be positive or negative at different regions.

A diagram of the generation of the triangular bulge caused by the TS waves
is presented in figure~\ref{fig:diagram-cut}, and it can be explained qualitatively by (\ref{Eq:dn}).
Since $n_x(t)^2n_y(t)\zeta>0$, $n_x$ increases on the arch with time, 
which implies that
the arch leans forward in figure~\ref{fig:diagram-cut}(a).
On the $x$--$z$ planes, the evolution of $n_z$ depends on the sign of $n_xn_z$. 
From (\ref{Eq:dn}), $n_z$ increases for the branch with $n_xn_z>0$ and decreases for $n_xn_z<0$.
The contour lines in figure~\ref{fig:diagram-cut}(b) can be divided into four quadrants depending on the signs of $n_x$ and $n_z$.
As shown in figure~\ref{fig:diagram-cut}(b), it is straightforward to find that the downstream contour line is stretched in the streamwise direction and the upstream line is flattened with time.
Therefore, a convex bulge generated from the initial disturbance tends to evolve into a triangular shape from the top view (also refer to figure~\ref{fig:all-2}(b)).

It is noted that the simplified analysis here cannot be valid for long times, because the amplification of the disturbances for the velocity and material surfaces cannot be ignored when the material surfaces exhibit notable three-dimensional geometrical characteristics.

\bibliographystyle{jfm}

\bibliography{JFM_draft_zym_version_3-10}


\clearpage


\end{document}